\documentclass[10pt]{article}
\usepackage[top=3.2cm, bottom=3cm, left=3.4cm, right=3.4cm]{geometry}
\usepackage{graphicx}
\usepackage{calc}

\usepackage{amsmath}
\usepackage{bbm}
\usepackage{amsfonts}
\usepackage{amssymb}
\usepackage{amsbsy}
\usepackage[T1]{fontenc}
\usepackage{stmaryrd}
\usepackage{textcomp}
\usepackage{subfigure}
\usepackage{epstopdf}
\usepackage{lmodern}
\usepackage[overload]{empheq}
\usepackage{framed}
\usepackage[subfigure]{tocloft}
\usepackage{subeqnarray}
\usepackage{alphalph}
\usepackage[refpage,cfg,prefix]{nomencl}
\usepackage{multirow}
\usepackage{xifthen}
\usepackage{enumerate}
\usepackage[bottom]{footmisc}
\usepackage{natbib}  
\usepackage{wasysym}
\usepackage{bibentry}
\usepackage{verbatim}
\usepackage{url}
\usepackage{float}
\usepackage{setspace}

\usepackage{lipsum}                     % Dummytext
\usepackage{xargs}                      % Use more than one optional parameter in a new commands
\usepackage[pdftex,dvipsnames]{xcolor}  % Coloured text etc.
\usepackage[colorinlistoftodos,prependcaption,textsize=tiny]{todonotes}
\newcommandx{\todoadd}[2][1=]{\todo[linecolor=blue,backgroundcolor=blue!25,bordercolor=blue,#1]{#2}}
\newcommandx{\todoinf}[2][1=]{\todo[linecolor=yellow,backgroundcolor=yellow!25,bordercolor=OliveGreen,#1]{#2}}
\newcommandx{\seen}{\todo[linecolor=OliveGreen,backgroundcolor=OliveGreen!25,bordercolor=OliveGreen]{Review already checked}}
\newcommandx{\notseen}{\todo[linecolor=red,backgroundcolor=red!25,bordercolor=red]{Review NOT checked}}
\newcommandx{\todoimp}[2][1=]{\todo[linecolor=Plum,backgroundcolor=Plum!25,bordercolor=Plum,#1]{#2}}
\newcommandx{\todohid}[2][1=]{\todo[disable,#1]{#2}}
\newcommandx{\notekit}[2][1=]{\todo[linecolor=red,backgroundcolor=red!25,bordercolor=red,#1]{#2}}
\usepackage{color,colortbl}
\usepackage{relsize}

\makenomenclature

\newcommand{\D}[2]{\frac{\partial #1}{\partial #2}}
\newcommand{\DD}[2]{\frac{\partial^2 #1}{\partial #2^2}}
\newcommand{\DDdiff}[3]{\frac{\partial^2 #1}{\partial #2 \partial #3}}

\newcommand{\N}{\ensuremath{\mathbb{N}}}

\newcommand{\R}{\ensuremath{\mathbb{R}}}
\newcommand{\cN}{\ensuremath{\mathcal{N}}}
\newcommand{\cT}{\ensuremath{\mathcal{T}}}
\newcommand{\cI}{\ensuremath{\mathcal{I}}}

\newcommand{\LR}[1]{\left(#1\right)}
\newcommand{\LRb}[1]{\left\lbrace#1\right\rbrace}

\newcommand{\LRs}[1]{\left [#1\right ]}
\newcommand{\st}{\ensuremath \, | \,}
\newcommand{\C}[2]{\ensuremath C^{(m)}({#1},#2)}
\newcommand{\Cb}[2]{\ensuremath \bar{C}({#1},#2)}

\newif\ifletter

\begin{document}

%--------------------------------------------------------------------
% Title
\title{\textbf{Stochastic and deterministic modelling of cell migration}}

%--------------------------------------------------------------------
% Authors
\author{\textbf{Enrico Gavagnin\footnote{Corresponding author: e.gavagnin@bath.ac.uk}   \,and Christian A. Yates} \\ \small{\textit{Department of Mathematical Sciences},}\\ \small{\textit{University of Bath, Claverton Down, Bath, BA2 7AY, UK}}}
	
\date{}
\maketitle

%--------------------------------------------------------------------
% Abstract
\begin{abstract}

Mathematical models are vital interpretive and predictive tools used to assist in the understanding of cell migration. There are typically two approaches to modelling cell migration: either micro-scale, discrete or macro-scale, continuum. The discrete approach, using agent-based models (ABMs), is typically stochastic and accounts for properties at the cell-scale. Conversely, the continuum approach, in which cell density is often modelled as a system of deterministic partial differential equations (PDEs), provides a global description of the migration at the population level.
Deterministic models have the advantage that they are generally more amenable to mathematical analysis. They can lead to significant insights for situations in which the system comprises a large number of cells, at which point simulating a stochastic ABM becomes computationally expensive. However, finding an appropriate continuum model to describe the collective behaviour of a system of individual cells can be a difficult task. Deterministic models are often specified on a phenomenological basis, which reduces their predictive power. Stochastic ABMs have advantages over their deterministic continuum counterparts. In particular,  ABMs can represent individual-level behaviours (such as cell proliferation and cell-cell interaction) appropriately and are amenable to direct parameterisation  using experimental data. It is essential, therefore, to establish direct connections between stochastic micro-scale behaviours and deterministic macro-scale dynamics. 

In this Chapter we describe how, in some situations, these two distinct modelling approaches can be unified into a discrete-continuum equivalence framework. We carry out detailed examinations of a range of fundamental models of cell movement in one dimension. We then extend the discussion to more general models, which focus on incorporating other important factors that affect the migration of cells including cell proliferation and cell-cell interactions. We provide an overview of some of the more recent advances in this field and we point out some of the relevant questions that remain unanswered.\\

    
\end{abstract}

%====================================================================
%--------------------------------------------------------------------
% Introduction
\vspace{0.5cm}
\section{Introduction}
\label{sec:introduction}
%-------------------------------------------------------------------- 
%Why bother             
The process of cell migration plays an essential role in several developmental and pathological mechanisms. For example, cell migration orchestrates morphogenesis throughout the development of the embryo \citep{gilbert2003med, keller2005cmd}, and plays a crucial role in wound-healing \citep{maini2004twm, deng2006rma} and immune responses \citep{madri2000cmi}. Migration also contributes to many pathological processes, including vascular disease \citep{raines2000emr} and cancer \citep{hanahan2000hoc}. 

Many of these phenomena are highly complex and the collective behaviour of a set of interacting cells. At the individual-cell-level, a variety of mechanisms can be involved, including cell-cell adhesion  \citep{niessen2007tja,trepat2009pfd}, attraction \citep{yamanaka2014vas} and repulsion \citep{carmona2008cil}. In other cases, cells can respond to a chemical gradient which regulates and guides their motility (\textit{chemotaxis}) \citep{ward2003dbd, keynes1992rca}. In addition, when the migration occurs over a sufficiently long time, cell proliferation and death can also play important roles in the process \citep{mort2016rdm}. An understanding the impact that these (and other) individual-cell mechanisms have on global collective migration is important, since it can illuminate the origin of major diseases and suggest effective therapeutic approaches.

Over the past few decades, great progress has been made in understanding many aspects of cell migration \citep{ridley2003cmi,friedl2009ccm}. However, we still lack a proper understanding of many of its underlying mechanisms. In particular, there are aspects which remain impenetrable to experimental biologists \citep{staton2004cma}. One of the major issues when studying cell migration is obtaining and interpreting experimental data. For example, a comprehensive controlled experiment \textit{in vivo} or a realistic \textit{in vitro} set up can be difficult or impossible to obtain. In other cases, it may be extremely complicated to investigate the microscopic origin of a macroscopic phenomenon, given the number of different actions that a single cell can perform \citep{westermann2003aml,dworkin1985cim,trepat2009pfd,tambe2011ccg}. In this context, mathematical modelling has become a necessary interpretive and predictive tool to assist in the understanding of such complex phenomena \citep{noble2002orc,simpson2006lii,maini2004twm}. 

Mathematical models have been employed as tools for validation of the experimentally generated hypotheses. Moreover, due to the advances in computation of the last few decades, mathematical models are becoming more detailed and accurately parametrised which makes them capable of generating experimentally testable hypotheses and exploring new questions which are still not approachable from an experimental prospective \citep{tomlin2007bnm}. For example, the use of stochastic mathematical models facilitates the investigation of biological systems in which randomness plays an important role and for which viewing only a single instance of the evolution of a system can be inconclusive \citep{lee2011ccc}. In a mathematical modelling framework, the system can be repeatedly simulated using the same deterministic or randomised initial conditions. Equally, repeats with slightly altered initial conditions or parameters are useful for quantifying the sensitivity of systems in which a small fluctuation in the initial state of the system can have a significant effect on the outcome of the experiment, or in understanding the sensitivity of the system to certain parameter choices, respectively. In addition, the outcome of a mathematical simulation can be examined easily since all the variables of the system are explicitly accessible \citep{flaherty2007mmc}.\\

There are typically two approaches to modelling cell migration, depending on the scale of interest. At the level of an individual cell, stochastic, discrete agent-based models (ABMs) are popular. Each cell is modelled as a single individual (agent) with its own rules of movement, proliferation and death. Alternatively, to model collective cell migration at the population level, a deterministic, continuum partial differential equation (PDE) for the population density is normally used\footnote{Although deterministic ABMs \citep{kurhekar2015abd} and stochastic continuum models \citep{schienbein1993lef,dickinson1993sma} have also be considered, the most common pairing in the literature involves stochastic ABMs and population-based deterministic models. Therefore, these are the two modelling sub-types that we will consider in this review.}. These two modelling paradigms have complementary advantages and disadvantages which we summarise in Table \ref{table:pro_cons_contiuum}. 

Generally speaking,  ABMs are attractive because their macro-scale behaviour is  completely self-induced, rather than being superimposed on a phenomenological basis \citep{ben2000cso, shapiro1988bmo}. This is particularly interesting when a complex structure emerges at the population-level \citep{othmer1997abc,thompson2011lmn,thompson2012mcm}. In fact, although the agents represent the driving force of such structure, typically they are unaware, as individuals, of the macroscopic configuration of the system, since their behaviour is dictated only by their local environment.

Another advantage of ABMs is that their formulation is generally more intuitive than continuum models. Each cell's actions can be easily incorporated in the model by implementing appropriate rules for the corresponding agent, which mimic specific known biological behaviours. The parameters regulating of each of these rules can then be inferred directly from real observations which makes ABMs more easily relatable to experimental data. 

\begin{figure}[h!!]
\begin{center}
\includegraphics[width= 0.8\columnwidth]{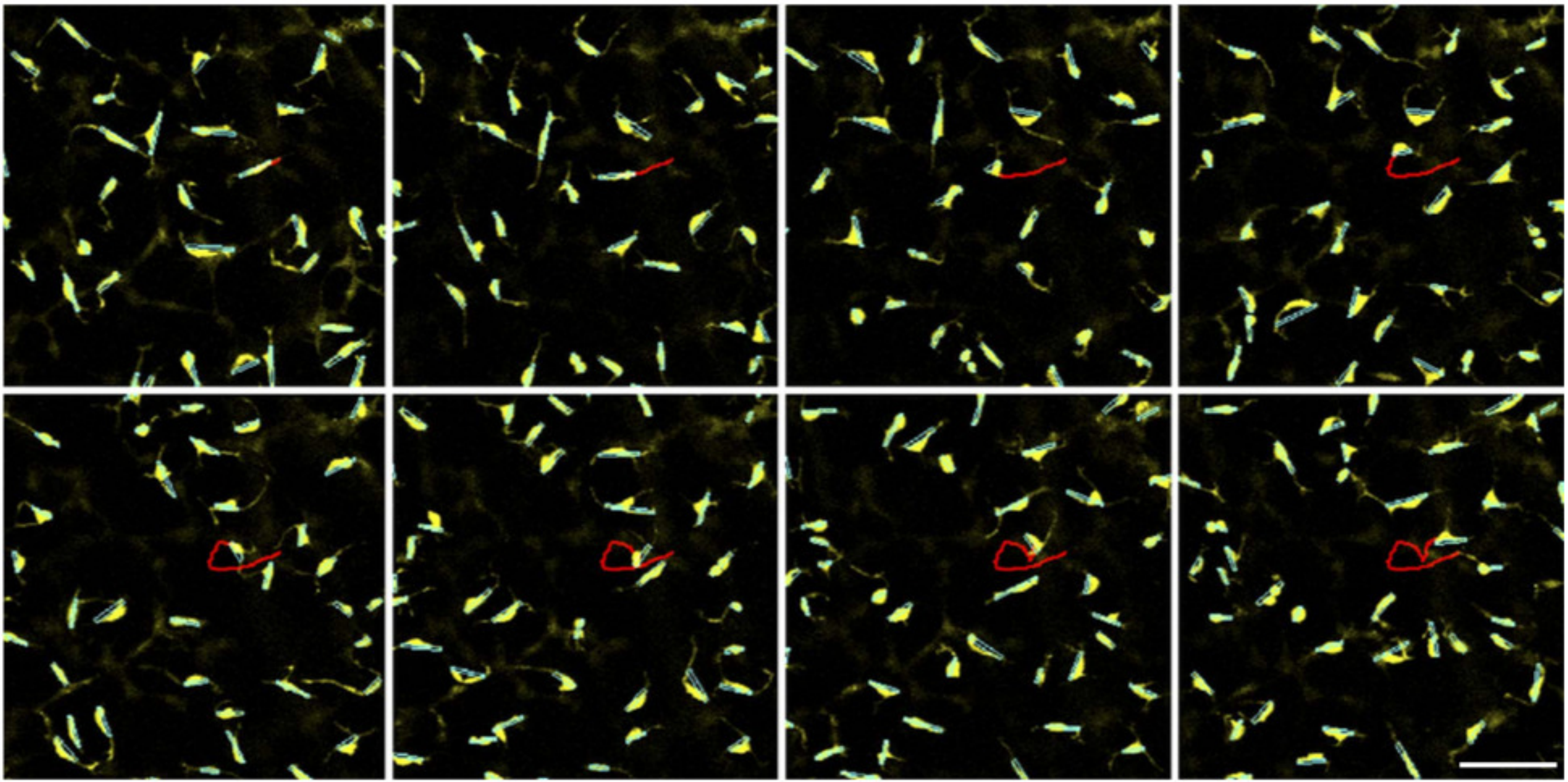}
\end{center}
\caption{A time-lapse sequence of melanoblasts migrating in \textit{ex vivo} culture of E14.5 (14.5 days post fertilisation) mouse skin. The Feret's diameter of each cell body is indicated in cyan. The path of a single migrating cell is indicated in red. Melanoblasts migrate along their Feret's diameter - the longest distance between any two points along a given cell boundary. Reproduced from \citet{mort2016rdm} with the permission of Nature Communications.}
\label{fig:mort2016} 
\end{figure}

\citet{mort2016rdm}, for example, used a simple ABM to study the migration of mouse melanoblasts, the embryonic precursors of melanocytes, responsible for pigmentation (see Figure \ref{fig:mort2016}). The authors studied the effects of a family of mutations to the receptor tyrosine kinase, Kit. These mutations affect the success of the colonisation of the growing epidermis by the melanoblasts, leading to unpigmented regions of hair and skin. \citeauthor{mort2016rdm} employed an ABM to study the interplay of movement and proliferation of melanoblasts and they carried out a statistical analysis on single cell trajectories in order to parametrise the model against experimental data. By simulating their ABM and comparing the results with experimental data, \citeauthor{mort2016rdm} were able to show that belly-spot formation in Kit mutants is likely to be induced by a reduction in proliferation rate, rather than motility, a result which was contrary to the received wisdom in the experimental literature. \\

Deterministic models also have their advantages. For example, when the population size becomes large, a PDE model is generally desirable, since there exist a wide range of well-developed numerical tools for their rapid solution. This is in direct contrast to agent-based models which are typically coded up \textit{ad hoc} and require multiple computationally intensive repeats in order to gather reliable ensemble statistics. Moreover, continuum models have the advantage that they are generally more amenable to mathematical analysis than ABMs, and this analysis can often lead to significant and general insights. For example, PDEs can be used to carry out stability analysis to determine the conditions which lead to pattern formation \citep{anguige2009odm}. In other scenarios, one can use travelling wave analysis to obtain expressions for the speed of the cell invasion in terms of the model parameters \citep{murray2007mbi}. 

Depending on the biological questions of relevance and available experimental data, either deterministic-continuum or stochastic-discrete models (or some combination of both) may be appropriate. Ideally, the individual-level representation can be used to parametrise the model, and the continuum level-description should link the population-level results back to the parameters of interest. 
The problem of connecting the parameters of the individual-level model to those of a representative population-level description represents, therefore, a crucial step of the multi-scale modelling process.\\

\begin{table}
\begin{center}
\def\arraystretch{2}
\begin{tabular}{ c|c|c| } 
& Advantages & Disadvantages\\
\hline
&  & \\[-22pt]
\hline
Discrete/Stochastic & \begin{tabular}{@{}c@{}}
Detailed structure and \\[-12pt] explicit implementation\\
Direct connection to  \\[-12pt] experimental data \\ Incorporation of randomness
\end{tabular}  & \begin{tabular}{@{}c@{}}
High computational cost \\
\\[-12pt] Multiple simulations required \\ Relatively inaccessible to \\[-12pt] mathematical analysis
\end{tabular}   \\
\hline
&  & \\[-22pt]
\hline
Continuum/Deterministic & \begin{tabular}{@{}c@{}}
Fast to simulate \\ Amenable \\[-12pt] to mathematical analysis \\ Suitable for systems of \\[-12pt] large numbers of cells \end{tabular}  &  \begin{tabular}{@{}c@{}}
Lack of fine detailed\\[-12pt] structure \\ Difficult to link \\[-12pt] experimental data \\ Ignore the effects \\[-12pt] of randomness \end{tabular} \\
\hline
\end{tabular}

\end{center}
\caption{Summary of the advantages and disadvantages of adopting a discrete/stochastic approach, such as an ABM, as opposed to a continuum/deterministic approach, typically represented by a PDE.}
\label{table:pro_cons_contiuum}
\end{table}

In this Chapter, we provide an overview of a range of techniques which can be used to connect ABMs of cell migration to macroscopic PDEs for average cell density. We review a series of stochastic and deterministic models which are capable of reproducing some of the key features of the behaviour of cells and describe how these two modelling regimes can be linked to form a multi-scale mathematical framework. 

In Section \ref{sec:MF} we carry out a detailed derivation of deterministic models from ABMs which focus on cell movement. We present derivations in two different scenarios, depending on whether the spatial domain in which the ABM is defined is partitioned into a finite lattice (\textit{the on-lattice case}) or not (\textit{the off-lattice case}). In each of these situations, we first study the case in which cells move independently of others (\textit{non-interacting cells}) and then we introduce the ability of cells to sense the occupancy of neighbouring regions of space and to avoid overlapping with other cells (\textit{interacting cells}). All the derivations in Section \ref{sec:MF} are carried out in one dimension and assuming a \textit{mean-field} moment closure approximation. We discuss generalisations to higher dimensions and other closure approximations in Sections \ref{sec:HD} and \ref{sec:HOA}, respectively.

The remaining part of the chapter is devoted to reviewing a series of biologically relevant features that can be incorporated in the models of cell migration. In each case we  highlight the implications of introducing a given behaviour both at the individual- and population-level. More precisely, in Section \ref{sec:proliferation} we present models which incorporate the ability of cells to reproduce by division. In Section \ref{sec:external_signalling} we present models in which cells can interact indirectly through an external signal, e.g. \textit{chemotaxis} or slime following. Some examples of direct forms of cell-cell interactions, such as adhesion-repulsion, pushing and pulling, are discussed in Section \ref{sec:cell-cell_interaction}. In Section \ref{sec:growing_domain} we review a series of recent papers which model cells migrating on a growing domain. Finally, we discuss how to derive a macroscopic limit for ABMs which are not based on simple random walks and which are capable of representing persistence of motion, in Section \ref{sec:persistence_of_motion}. 
We conclude the Chapter with a short discussion and final considerations in Section \ref{sec:conclusion}.

%--------------------------------------------------------------------
% MORT

%--------------------------------------------------------------------
% Methods
\section{Cell motility}
\label{sec:methods}
In this Section we present a detailed description of the basic models of cell motility, which represent the fundamental basis for the majority of the spatially extended representations of the remaining part of the Chapter. We use the case of these models to illustrate standard approaches to deriving diffusive, deterministic, continuum representation from ABMs, based on occupancy master equations. We carry out the explicit derivations for one-dimensional versions of the models in Section \ref{sec:MF}, and we discuss the generalisation to higher dimensions in Section \ref{sec:HD}. We consider two distinct cases, depending on whether the motility mechanism of the ABM is implemented on- or off-lattice. In each case, we focus on two variants of the model, both with and without crowding effects. We explain the standard techniques for deriving the corresponding deterministic descriptions at the population-level in each case. Note that, throughout the Chapter, we adapt the notation of models taken from the literature for consistency.

%--------------------------------------------------------------------
% The Mean Field Approximaiton
\subsection{Connecting stochastic and deterministic models of cell movement}
\label{sec:MF}

%--------------------------------------------------------------------
% The on-lattice case
\subsubsection{On-lattice models}
\label{sec:on_lattice}
Broadly speaking, on-lattice ABMs can be classified as cellular automata in which a set of agents occupy some or all sites of a lattice. In general, these agents have a number of  state variables associated with them and a set of rules prescribing the evolution of their state and position. There exists a great variety of forms of ABMs. The appropriateness of each representation depends on the phenomenon that is being modelled. For example, cellular Potts models have been used by \citet{turner2002iac} and \citet{turner2004dcm} in the context of cancer modelling and by \citet{graner1992sbc} in the context of cell sorting via differential adhesion. \citet{othmer1997abc} modelled bacterial aggregation by using a position-jump process and \citet{painter1999sfj} adopted a similar approach to study the interplay of chemotaxis and volume exclusion. One of the advantages of using these on-lattice models is that their formulation and analysis tend to be straightforward in comparison to their off-lattice counterparts. Moreover, the presence of the grid considerably reduces the computational cost of  simulations when a large number of agents is involved. However, since in the majority of scenarios the assumption that cells move on a discrete grid is not appropriate, the implementation of cell behaviours is usually phenomenological. \\

Consider a one-dimensional domain, $[0,L],$ with periodic boundary conditions. An ABM on the domain $[0,L]$ comprises a set of $\cN \in \N$ agents positioned in $[0,L]$ and a range of stochastic rules which govern their evolution in time. The ABMs that we consider throughout this Section are all based on continuous-time \textit{position-jump processes} \citep{othmer1988mdb,othmer1997abc}. This means that the position of the agents undergoes series of sequential Markovian jumps \textit{i.e.}, at any given time, the evolution of the position of each agent depends only on its current position. An alternative approach is to use \textit{velocity-jump processes}, in which the Markov property applies to the velocity of the agents, instead of their position. We discuss this approach in Section \ref{sec:persistence_of_motion}. 

Discrete-time ABMs are popular in the literature \citep{simpson2007sic, simpson2009mss,treloar2011vjm,treloar2012vjp}, although, for the examples presented here, we treat time as a continuous variable\footnote{In general, when the time discretisation step is sufficiently small, the behaviour of the discrete-time ABMs is similar to its continuous-time counterpart.} \citep{othmer1988mdb,othmer1997abc}.  As time evolves, agents can attempt movement events which occur as a Poisson process with rate $\alpha$. In other words, each agent attempts to move after an independent exponentially distributed waiting time with parameter $\alpha$. When such attempts take place, we say that the agent has been selected to attempt a movement.\\

\begin{figure}
\begin{center}
{\includegraphics[width=0.7 \columnwidth ]{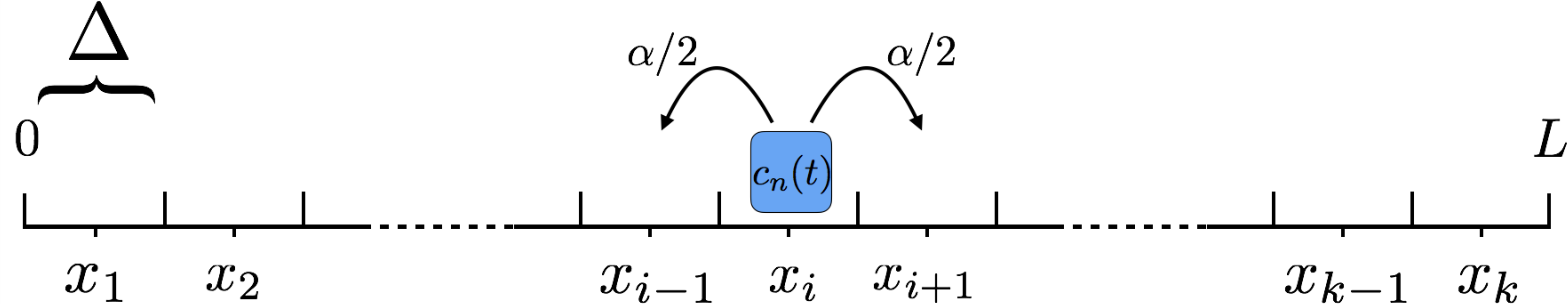}
\caption{Schematic of the on-lattice ABM in one dimension. The domain is partitioned in $k$ intervals of length $\Delta$. A single agent, $n$, is represented as a blue square and it occupies interval $i$, \textit{i.e.} $c_n(t)=x_i$. The agent attempts to move, with rate $\alpha$, to one of the two adjacent sites.}
\label{fig:schematics}}
\end{center}
\end{figure}

Consider the on-lattice scenario, in which the domain is partitioned in $k$ intervals, each of length $\Delta =L/k$, whose centres are denoted by $x_1,\dots , x_k$, respectively (see Figure \ref{fig:schematics} for a schematic illustration).
We denote the position of the $n$-th agent at time $t$ as $c_n (t)$, hence $c_n(t)\in \LRb{x_1, \dots, x_k}$ for every $t\in \R^+$ and $n=1, \dots, \cN$. 

\begin{figure}
\begin{center} 
\subfigure[][]{\includegraphics[scale=0.12]{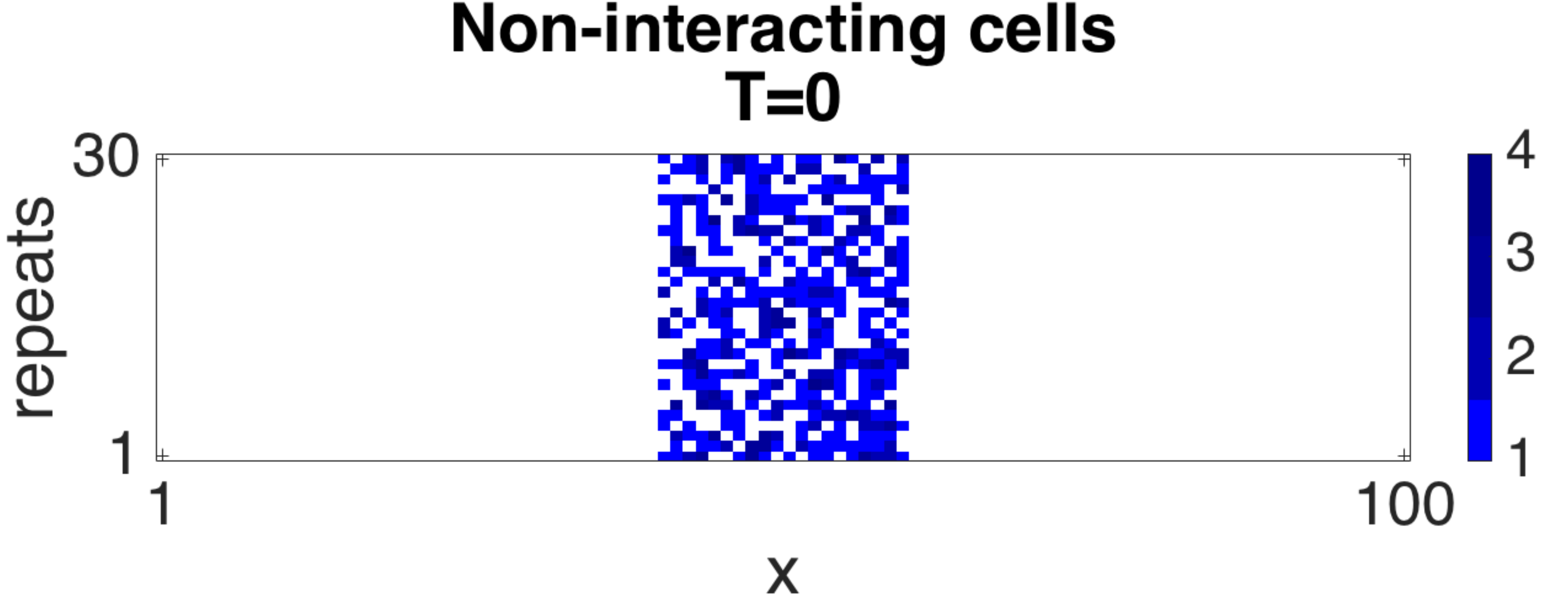}} \quad
\subfigure[][]{\includegraphics[scale=0.12]{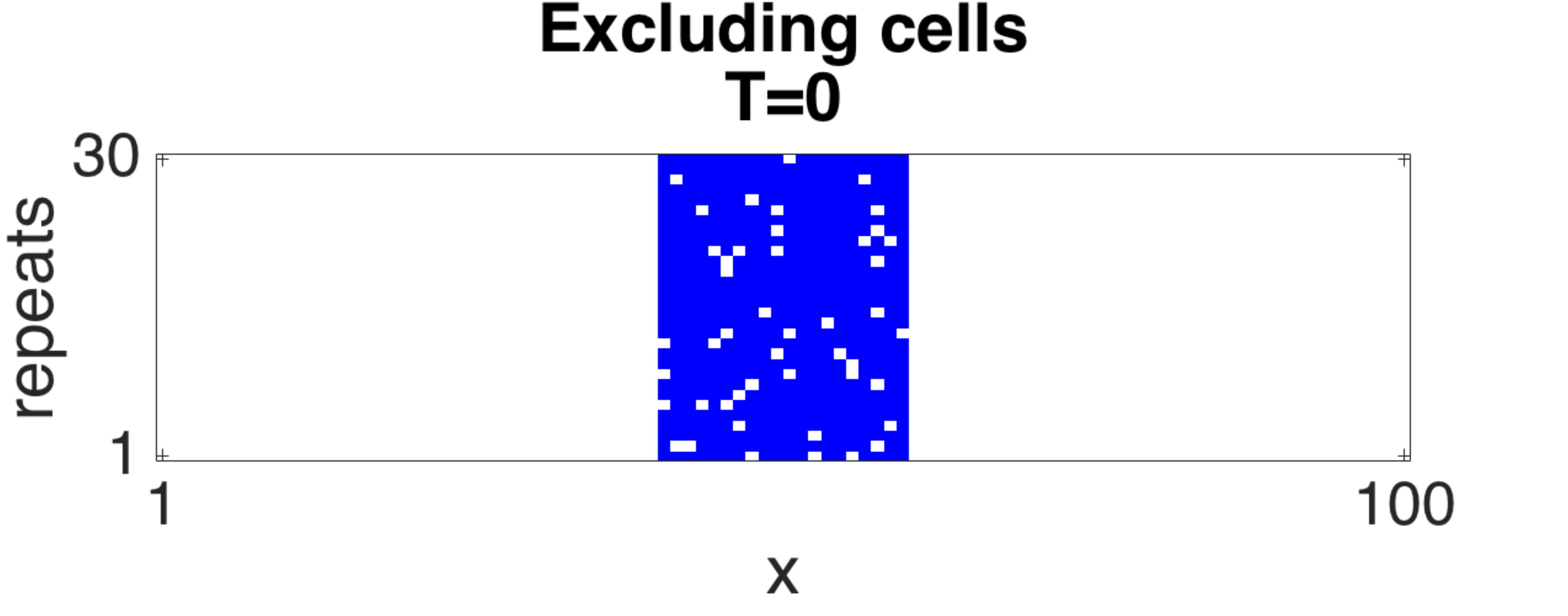}}\\[-5pt]
\subfigure[][]{\includegraphics[scale=0.12]{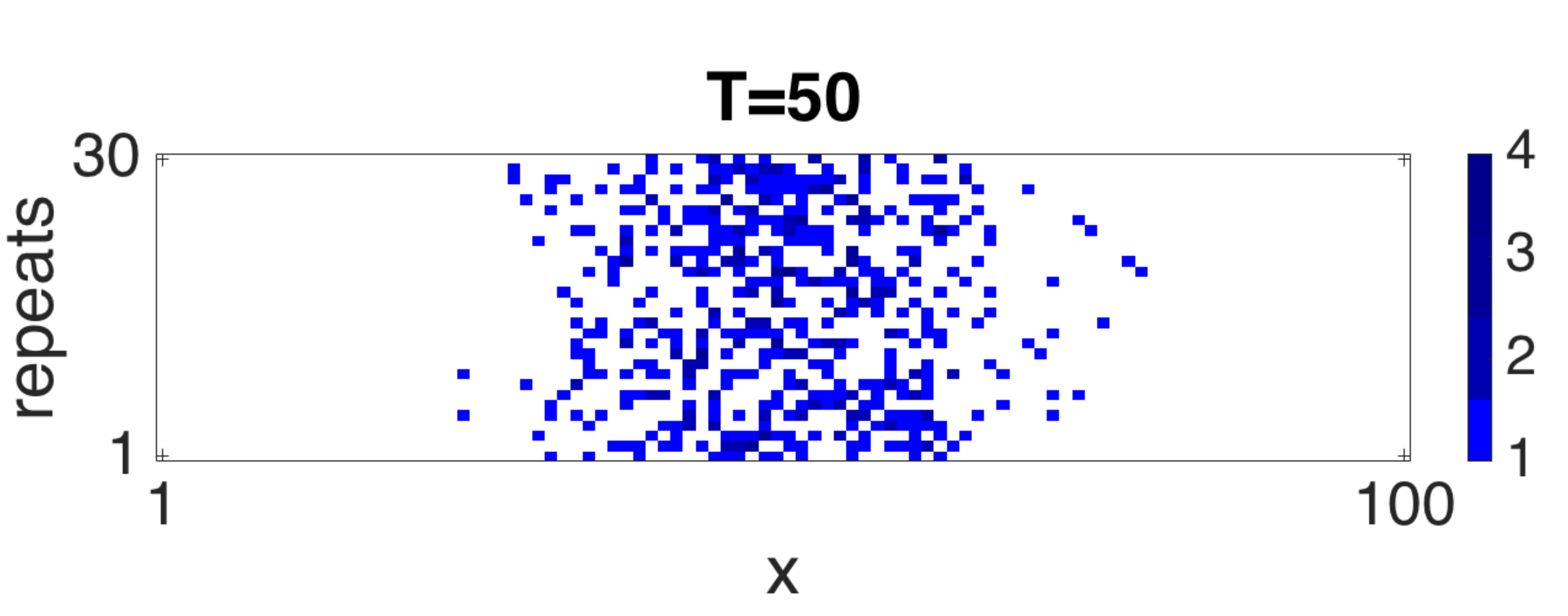}} \quad
\subfigure[][]{\includegraphics[scale=0.12]{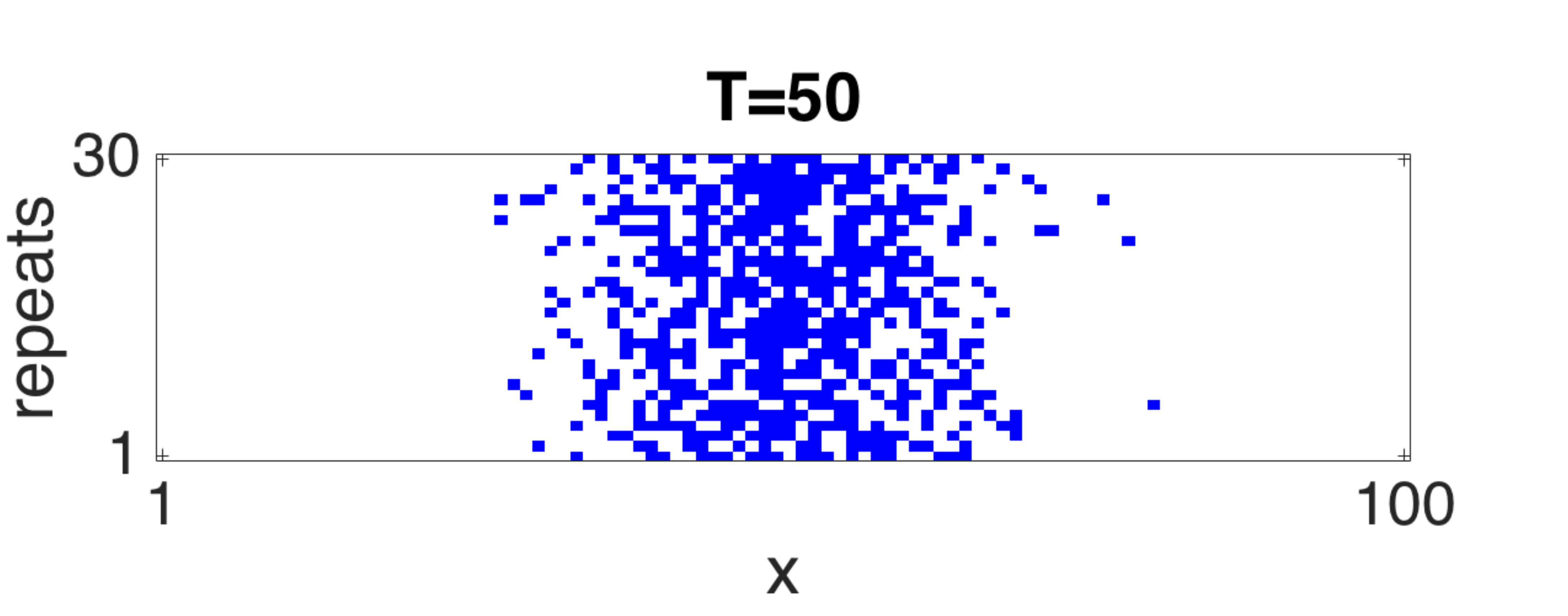}}\\[-5pt]
\subfigure[][]{\includegraphics[scale=0.12]{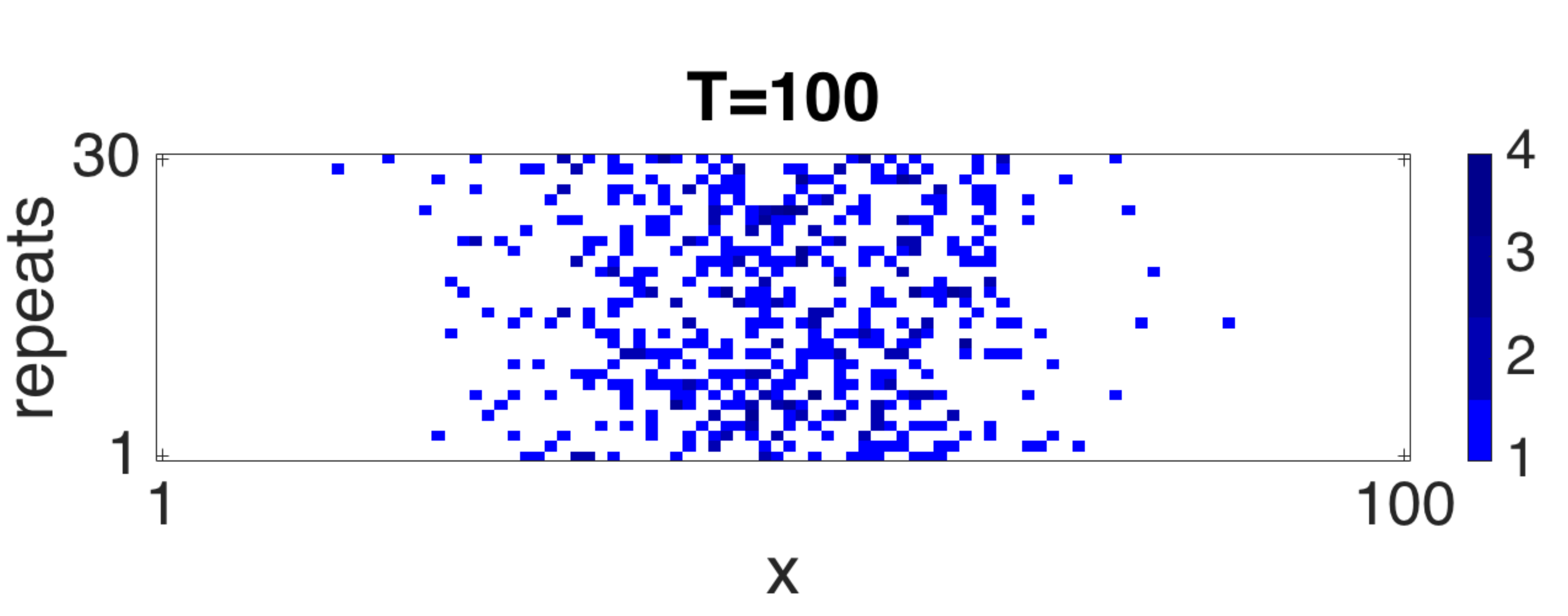}}\quad
\subfigure[][]{\includegraphics[scale=0.12]{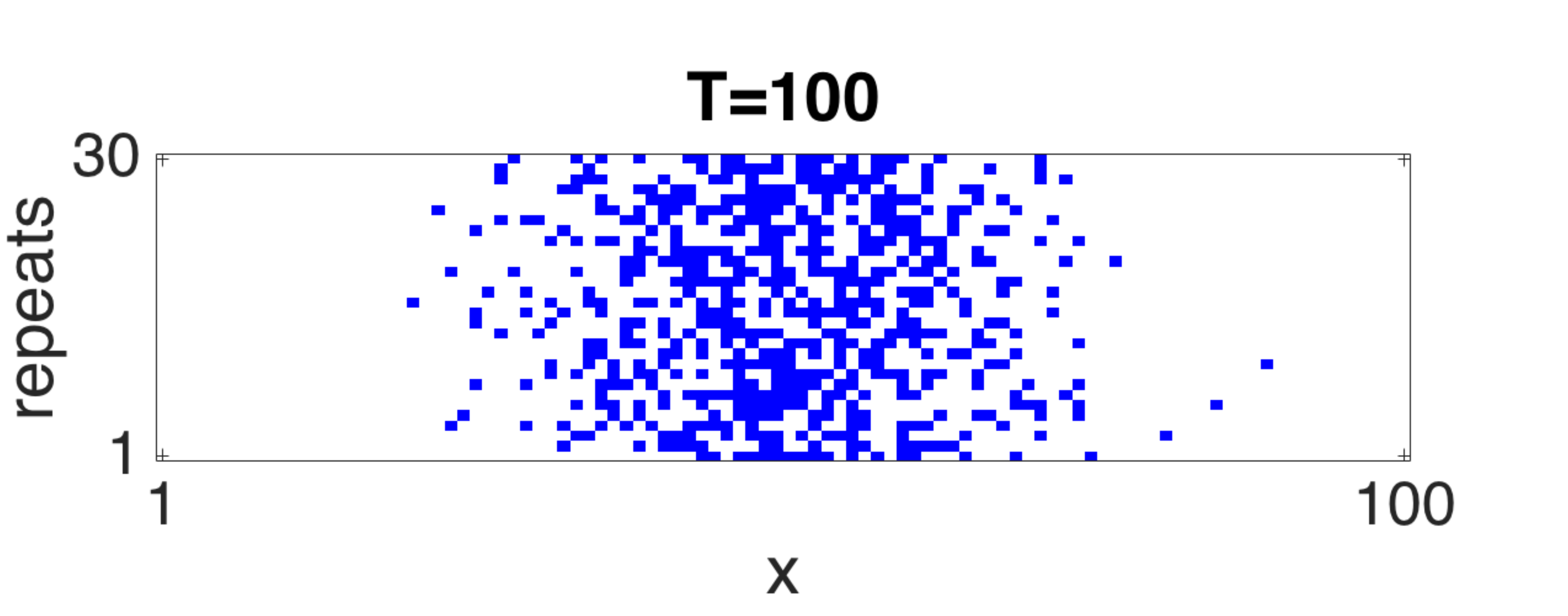}}\\[-5pt]
\subfigure[][]{\includegraphics[scale=0.12]{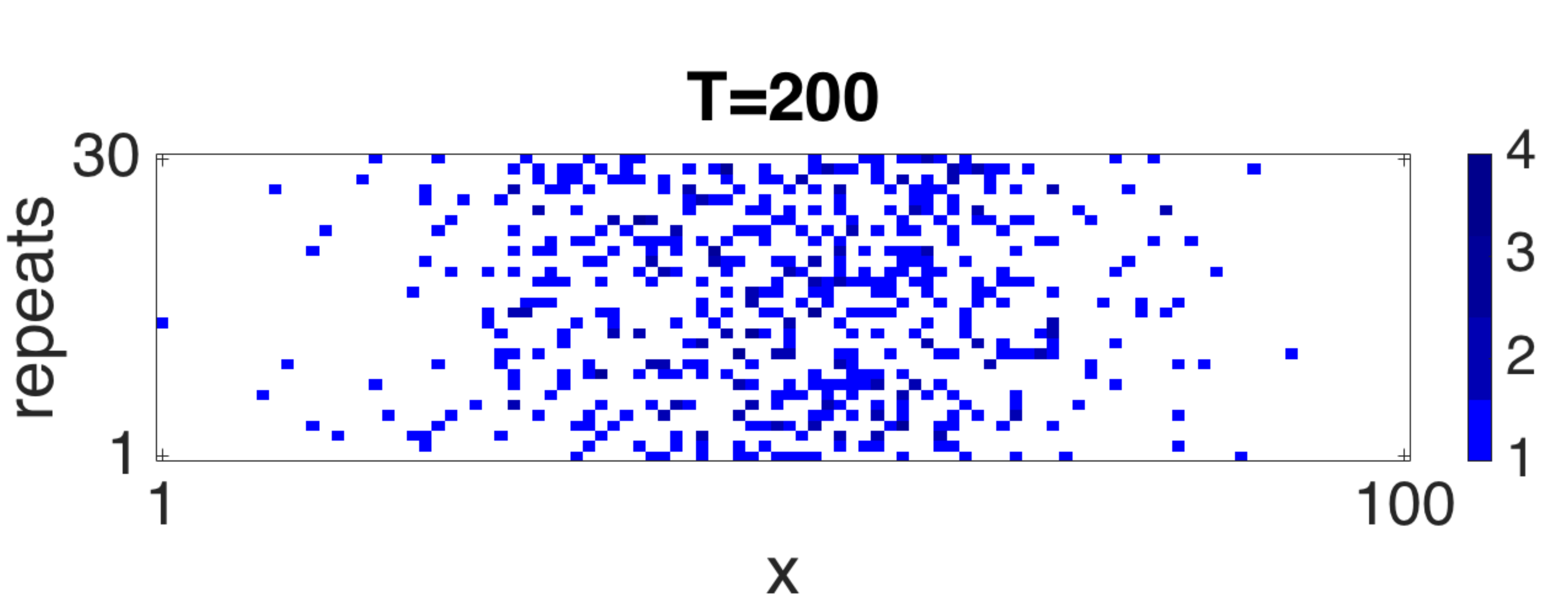}}\quad
\subfigure[][]{\includegraphics[scale=0.12]{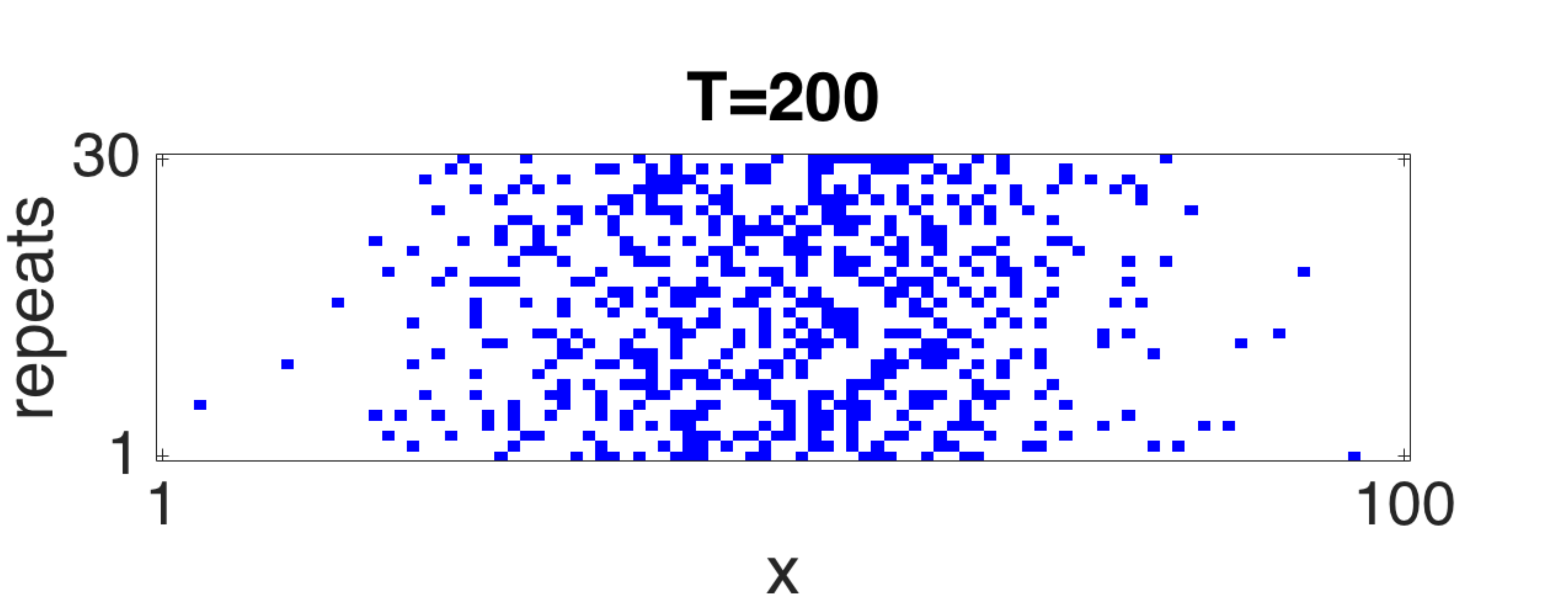}}\\
\subfigure[][]{\includegraphics[scale=0.3]{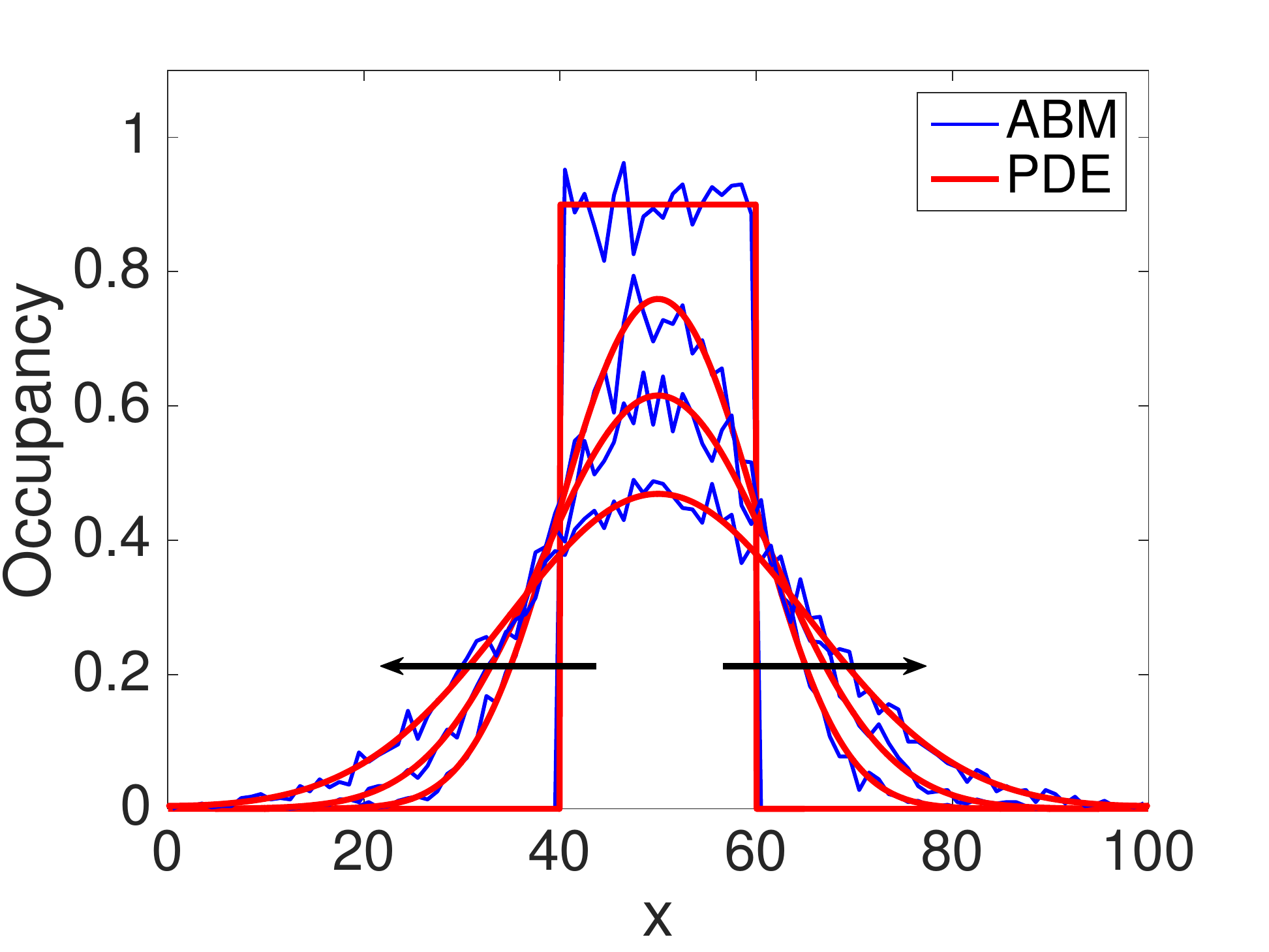}}
\subfigure[][]{\includegraphics[scale=0.3]{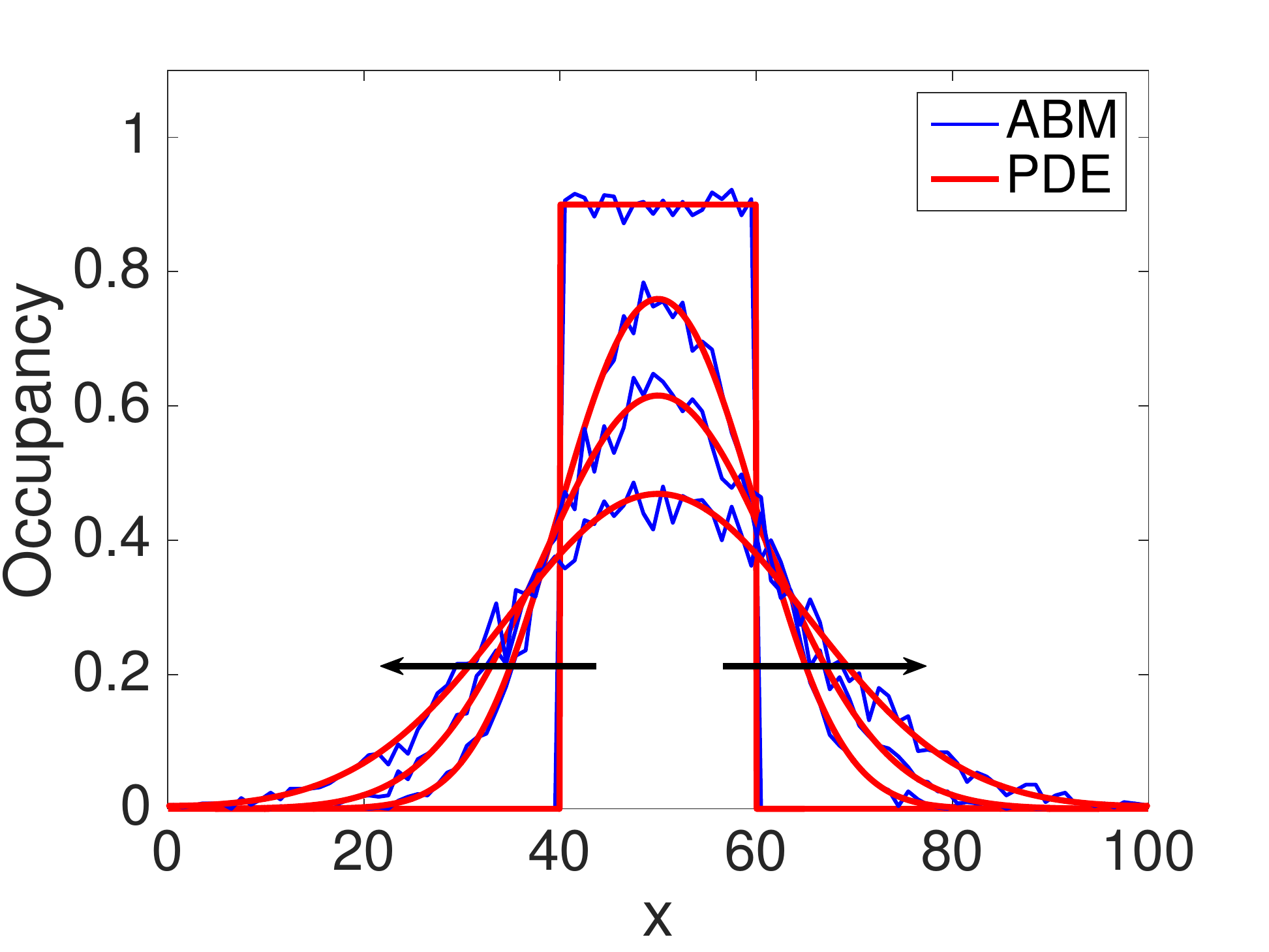}}
\end{center}
\caption{Comparison between stochastic and deterministic models of cell motility in on a one-dimensional lattice . Panels (a) to (h) show four snapshots of 30  simulations of the ABM with non-interacting agents as described in Section \ref{sec:MF} (panels (a), (c), (e) and (g)) and with excluding agents ((panels (b), (d), (f) and (h)). The motility rate is $\alpha=1$ and the simulations are shown at times $T=0,50,100$ and $200$. In each panel $30$ independent repeats of the simulations are displayed on top each other. Empty sites are represented in white and occupied sites are represented in blue. For the non-interacting model, the occupied sites are coloured with a graded intensity of blue corresponding to the number of agents which occupy the site, as indicated in the colour bar. In panels (i) and (j) we show a comparison between the occupancy of the ABM (blue line), averaged over $500$ repeats, and the numerical solution of corresponding deterministic PDE (red line). Panel (i) is for the non-interacting model and panel (j) is for the model with volume exclusion. In both cases, the profiles are displayed at the times $T = 0, 50, 100, 200$, with the direction of the black arrows indicating increasing time. }
\label{fig:simulations_lattice}
\end{figure}

%--------------------------------------------------------------------
\paragraph{Non-interacting cells undergoing unbiased movement.\\}
In this Section, we consider the basic case of non-interacting agents. In particular, the movement rates of each agent are independent of the other agents' positions and unbiased. If, at time $t$, an agent, $n$, is selected to move, it attempts to move with equal probability, $1/2$, to either one of its adjacent sites, $c_n(t)\pm\Delta $. Equivalently, we can say that each agent moves to the right and to the left with rates $\cT^{\pm}=\alpha/2$, respectively. Notice that multiple agents can occupy the same lattice site. We aim to obtain a deterministic representation of the mean evolution of the agent density at a given time $t$. Therefore, we assume the model is simulated until time $t$ for a large number, $M$, of independent realisations, with each realisation identically prepared.
%  We might initialise every simulation displacing $\cN$ agents  uniformly at random within a central interval, for example. 
In Figure \ref{fig:simulations_lattice} (a) we consider 30 one-dimensional lattices with $L=100$ and $\Delta=1$. In each one-dimensional lattice  we populated the 20 central sites, $i=41,
\dots , 60$, at random with $18$ agents on average (with multiple agents per site possible). In Figures \ref{fig:simulations_lattice} (c),(e) and (g) we show three snapshots of these 30 identically prepared simulations of the ABM, as time evolves. We denote by $\C{x_i}{t}$ the number of agents which lie in the interval $i$ at time $t$ of the $m$-th repeat simulation, with $m=1, \dots , M  $. Namely 
\begin{equation}
\label{eq:C_def}
\C{x_i}{t}=|\LRb{n \st c_n(t)=x_i} |\, .
\end{equation}
If $C(x_i,t)=0$, we say that the site $i$ is \textit{empty} and if $C(x_i,t)>0$ we say it is \textit{occupied}. We define the mean occupancy of site $i$ at time $t$, averaged over the number of realisations, as
\begin{equation}
	\label{eq:C_bar_def_on_lattice}
	\Cb{x_{i}}{t}=\frac{1}{ M}\sum_{m=1}^{M}\C{x_{i}}{t} \, .
\end{equation}
By considering all the possible agents movements we can write down a conservation law for the average occupancy at time $t+\delta t$, where $\delta t$ is a sufficiently small that the probability that two or more movements take place in the interval $[t,t+\delta t)$ is $o(\delta t)$. This reads
\begin{equation}
	\label{eq:MA_simple_lattice}
	\Cb{x_{i}}{t+\delta t}=\Cb{x_{i}}{t}-\overbrace{\alpha \delta t\Cb{x_{i}}{t}}^{\text{moving out of site $i$}} + \underbrace{\frac{\alpha \delta t}{2} \LRs{\Cb{x_{i-1}}{t}+\Cb{x_{i+1}}{t}}}_{\text{moving into site $i$}} +\mathcal{O}(\delta t ^2)\,.
\end{equation}
The right-hand side of equation \eqref{eq:MA_simple_lattice} comprises three parts: the first term, which accounts for the occupancy of site $i$ at time $t$, and two terms which determine the expected loss of average occupancy due to agents moving out of site $i$ and the gain due to agents moving into site $i$, respectively. 

If we rearrange equation \eqref{eq:MA_simple_lattice}, divide through by $\delta t$ and take the limit as $\delta t \rightarrow 0$, we obtain the Kolmogorov equation (or continuous-time master equation) of the process which reads
\begin{equation}
\label{eq:semi_MA_simple_lattice}
\D{\Cb{x_{i}}{t}}{t}= \frac{\alpha}{2} \LRs{\Cb{x_{i-1}}{t}-2 \Cb{x_{i}}{t}+\Cb{x_{i+1}}{t}} \, .
\end{equation}

Now we Taylor expand the terms $\Cb{x_{i \pm 1}}{t}$ about the point $x_i$ to the second order, \textit{i.e.} we use the following approximations:

	\begin{subequations}
	\label{eq:taylor_lattice}
	\begin{align}
		\Cb{x_{i + 1}}{t}&=\Cb{x_{i}}{t}+\Delta \D{\Cb{x_{i}}{t}}{x}+\frac{\Delta^2}{2} \DD{\bar{C}}{x}+\mathcal{O}(\Delta^2) \, , \\
		\Cb{x_{i - 1}}{t}&=\Cb{x_{i}}{t}-\Delta \D{\Cb{x_{i}}{t}}{x}+\frac{\Delta^2}{2} \DD{\bar{C}}{x}+\mathcal{O}(\Delta^2) \, .
	\end{align}
	\end{subequations}
By substituting equations \eqref{eq:taylor_lattice} into equation \eqref{eq:semi_MA_simple_lattice} and taking the limit as $\Delta \rightarrow 0$ while holding $\alpha \Delta^2$ constant, we recover the diffusion equation for the continuous approximation of the average occupancy function, $\bar{C}$:
\begin{equation}
	\label{eq:diffusion}
	\D{\bar{C}}{t}=D \DD{\bar{C}}{x} \, ,
\end{equation}
where $D$ is the \textit{diffusion coefficient} defined as
\begin{equation}
	\label{eq:D_def_lattice}
	D=\lim_{\Delta\rightarrow 0} \frac{\alpha \Delta^2}{2} \, .
\end{equation}
We refer to Figure \ref{fig:simulations_lattice} (i) for a comparison between the average occupancy of the ABM and the diffusion equation \eqref{eq:diffusion}. The results confirm the good agreement between the two models.

The derivation of equation \eqref{eq:diffusion} from a simple random walk, as outlined above, is a well known result \citep{murray2007mbi,deutsch2007cam,codling2008rwm}. Alternatively, we could have carried out the derivation for the probability density function, $P(x_i,t)$, of finding a single agent at position $x_i$ at time $t$. This would result in a macroscopic PDE of the same form as equation \eqref{eq:diffusion}. The equation for $P$ can be obtained by dividing both sides of equation \eqref{eq:diffusion} by the total number of agents, $\cN$.

The diffusion equation \eqref{eq:diffusion} is a classical PDE, also  known as Fick's diffusion equation or the heat equation depending on the application. Extensive discussions on this type of equation can be found in \citet{crank1979mod} or \citet{welty2009fmh}. The existence of an analytic solution of equation \eqref{eq:diffusion} depends on the imposed initial and boundary conditions. For example, if we assume an infinite domain and that all agents are initialised in the same position:
\begin{equation}
	\Cb{x}{0}=\begin{cases}
		\cN & x=x^*\, , \\ 
		0 & \text{otherwise}\, ,
	\end{cases}
\end{equation}
equation \eqref{eq:diffusion} admits the fundamental solution given by
\begin{equation}
	\Cb{x}{t}=\frac{\cN}{\sqrt{4 \pi D t}} e^{-\frac{(x-x^*)^2}{4 D t}} \, .
\end{equation}

In the basic model outlined above, agents can only move to their nearest neighbour sites. A generalisation of this model, in which agents have the ability to perform non-local jumps is studied by \citet{taylor2015dab}. Specifically, in the model of \citet{taylor2015dab} agents are allowed to jump up to $Q$ sites away from their current position. For $q=1,\dots , Q$, each length-$q$ jump occurs with rate $\cT^{(\pm q)}$ in the right- and left-directions, respectively. When the motility is unbiased, \textit{i.e.} $\cT^{(q)}=\cT^{(\pm q)}$ for every $q$, the authors show that the  behaviour of their agents in the continuum limit evolves according to the diffusion equation \eqref{eq:diffusion}, but with diffusion coefficient given by 
\begin{equation}
	\label{eq:non_local_D}
	D^Q=\lim_{\Delta \rightarrow 0} \Delta^2 \sum_{i=1}^Q q^2 \cT^{(-q)} \, .
\end{equation}
This implies that, when the following condition is satisfied,
\begin{equation}
\label{eq:non_local_condition}
\alpha=\sum_{i=1}^Q q^2 \cT^{(q)}	\, ,
\end{equation}
%where $\cT$..
the local ABM and the non-local ABM are described by the same macroscopic PDE. There are an infinite number of possible combinations of $\cT^{(q)}$s which satisfy condition \eqref{eq:non_local_condition}. However, \citet{taylor2015dab} suggest that choosing 
 \begin{equation}
 \label{eq:best_non_local_condition}
\cT^{(q)}=\frac{\alpha}{q^2Q}\,,	
 \end{equation}
 is particularly appropriate, since it preserves the well-known property of diffusive processes, that the mean-squared displacement scales linearly with time \citep{codling2008rwm,othmer1988mdb}. A comparison of the simulations of the two AMBs confirms the accuracy with which the non-local models typically correspond to their local equivalent. The results also reveal a significant reduction in the average simulation time of the non-local ABM compared to the local ABM. This time-saving potential, given that condition \eqref{eq:non_local_condition} is satisfied, is highlighted by \citeauthor{taylor2015dab} as an important application of the non-local models in order to reduce the computational cost of stochastic simulations.
 
  However, it should be noted that, when dealing with steep gradients in agent numbers, the non-local model loses accuracy in comparison to the local model. This inaccuracy stem from the choice of the transitional rates \eqref{eq:best_non_local_condition} which match the terms of the local model only up to the second order of the expansion. In order to address this issue, the authors propose a spatially extended hybrid method in which regions containing steep gradients in agent numbers are dealt with using the local representation and regions in which gradients are more shallow using the non-local representation. This allows the acceleration of simulations afforded by the non-local representation, whilst maintaining the accuracy associated with the local method.

In the second part of the paper, \citeauthor{taylor2015dab} derive a general class of boundary conditions for local and non-local ABMs, corresponding to classical boundary conditions in the deterministic, continuum, diffusive limit. 

For a general non-local ABM, the authors study a class of first-order reactive boundary condition known as Robin conditions which, for the left-boundary, can be written as 
\begin{equation}
\label{eq:robin_BC}
	D\D{C(0,t)}{x}=C(0,t)B \,.
\end{equation}
Here $B=0$ corresponds to a purely reflective boundary and $B=\infty$ corresponds to a purely absorbing boundary. In order to obtain the corresponding stochastic boundary condition at the individual-level, \citeauthor{taylor2015dab} allow agents which attempt length-$q$ jumps that result in them hitting the boundary, to be absorbed (\textit{i.e.} removed from the domain) with absorption probability $a_{q,Q}$. If no absorption occurs, the agents reach the boundary and are then reflected in the opposite direction for the remaining number of steps of the jump. By taking a diffusive limit from the corresponding occupancy master equation, the authors determine the expression of the absorption rates in terms of the parameter, $B$, of the Robin boundary condition. The absorption rates are given by \eqref{eq:robin_BC}:
\begin{equation}
	a_{q,Q}=\Delta \frac{B}{D^Q}\LR{1+q^2 \sum_{i=q+1}^Q\frac{2}{i^2}} \,\quad \text{for}\quad q=1\dots Q\,,
\end{equation} 
where $\Delta$ is the lattice step, $Q$ is the maximum jump length and $D^Q$ is defined as in equation \eqref{eq:non_local_D}. In particular, when a local ABM is considered, the absorption rate, $a_{1,1}$, for agents at the site adjacent to the boundary is given by $$a_{1,1}=\Delta \frac{B}{D} \, . $$

Finally, \citet{taylor2015dab} extend their study to the case in which Neumann boundary conditions are imposed at the population level. For the left boundary this condition is given as
\begin{equation}
	\label{eq:neumann_BC}
	\D{C(0,t)}{x}= F \, ,
\end{equation}
where $F<0$ represents the influx into the system at this boundary. The corresponding ABM implementation requires new agents to enter the domain. For a general non-local model with maximum jump length $Q$, these agents are positioned in the $Q$ nearest sites to the boundary. By postulating a discrete master equation, which respects conservation of the total influx, $F$, and taking a diffusive limit as $\delta t, \Delta \rightarrow 0$, the authors derive an expression for the rate of introduction of agents into the $k$-th nearest site:
\begin{equation}
	f_{k,Q}=- \frac{F D^Q}{Q \Delta}\LR{\sum_{i=k}^Q\frac{2i-2 k +1}{k^2}} \,\quad \text{for}\quad k=1\dots Q\,,
\end{equation} 
which in the local case ($Q=1$), becomes 
$$f_{1,1}=- \frac{F D}{\Delta} \,.$$

%--------------------------------------------------------------------
\paragraph{Excluding cells\\}

We now incorporate volume exclusion in the model following the approach of \citet{simpson2009mss}. We modify the basic ABM of the previous Section, by introducing a specific form of agent-agent interaction which prevents two or more agents from occupying the same lattice site. We initialise the domain by populating the central interval, with $\cN$ excluding agents, \textit{i.e.} with the property that any given site can be occupied by at most one agent. In the example in Figure \ref{fig:simulations_lattice} (b) in each row we populated sites $i=41,\dots, 60$ with $18$ agents, on average, with a maximum of one agent per site.  When an agent, $n$, is selected for a movement event, one of its two neighbouring sites, $C_n(t)\pm \Delta $, is selected with equal probability, $1/2$. In order for the movement to be successful, however, the new selected site has to be empty. When the selected site is occupied, the movement is aborted and the selected agent remains in its position. This movement rule prevents agents from moving into occupied sites. Therefore, the excluding property of the initial condition is preserved as the time evolves. 
In particular, this implies that $C^{(m)}(x_i,t)$, defined in equation \eqref{eq:C_def}, takes only two values, $0$ and $1$. See Figures \ref{fig:simulations_lattice} (d),(f) and (h) for snapshots of simulations of the ABM.

For a given a realisation of the model, $m$, we make the usual moment-closure approximation that the occupancies of different sites, $i$ and $j$ with $i\ne j$, are independent, \textit{i.e.}:
\begin{equation}
\label{eq:MFA} 
\mathbb{P}\LRs{\C{x_i}{t}=1 \, ,\C{x_j}{t}=1} =\mathbb{P}\LRs{\C{x_i}{t}=1}\mathbb{P}\LRs{\C{x_j}{t}=1} \, ,
\end{equation} 
for every $t\in \R^+$. Equation \eqref{eq:MFA} is know as the \textit{mean-field assumption}, and it provides a good approximation in many scenarios, however it becomes invalid when spatial correlations play an important role in evolution, for example when proliferation or attractive forces between agents are involved (see Sections \ref{sec:proliferation} and \ref{sec:cell-cell_interaction}, respectively).  We discuss the validity of this approximation and possible alternative approaches to the mean field assumption in Section \ref{sec:HOA}. \\

By assuming the independence captured by equation \eqref{eq:MFA}, we can write down the transition rates from a given site $x_i$ as 

\begin{align*}
	\cT^+(x_i,t)&=\frac{\alpha \delta t}{2}\LR{1-\C{x_{i + 1}}{t}} \, ,\\
	\cT^-(x_i,t)&=\frac{\alpha \delta t}{2}\LR{1-\C{x_{i - 1}}{t}} \, .
\end{align*}	
Note that the terms $\LR{1-\C{x_{i \pm 1}}{t}}$ keep into account the possibility of abortion of the movement due to volume exclusion. Hence the occupancy master equation at position $x_i$ reads:
\begin{equation*}
	\label{eq:MA_exclusion_lattice}
	\begin{split}
	\C{x_{i}}{t+\delta t}=&\C{x_{i}}{t}-\overbrace{\frac{\alpha \delta t}{2} \C{x_{i}}{t} \LRs{(1-\C{x_{i-1}}{t})+(1-\C{x_{i+1}}{t})}}^{\text{moving out of $x_i$}} \\ & \underbrace{+\frac{\alpha \delta t}{2} (1-\C{x_{i}}{t})\LRs{\C{x_{i-1}}{t}+\C{x_{i+1}}{t}}}_{\text{moving into $x_i$}} +\mathcal{O}(\delta t ^2)\,.
	\end{split}
\end{equation*}
If we divide by $\delta t$, rearrange  and let $\delta t \rightarrow 0$, we obtain
\begin{equation}
\label{eq:semi_MA_exclusion_lattice}
\D{\C{x_{i}}{t}}{t}= \frac{\alpha}{2} \LRs{\C{x_{i-1}}{t}-2 \Cb{x_{i}}{t}+\C{x_{i+1}}{t}} \, ,
\end{equation}
for $m=1,\dots, M$. By summing equations \eqref{eq:semi_MA_exclusion_lattice} over each repeat and dividing by the total number of realisations, $M$, we recover equation \eqref{eq:semi_MA_simple_lattice} for the average occupancy, $\Cb{x_{i}}{t}$. Remarkably, we obtain exactly the same macroscopic representation as in the case of non-interacting agents, which is given by the canonical diffusion equation \eqref{eq:diffusion}. In Figure \ref{fig:simulations_lattice} (j) we show a comparison between the average occupancy of the ABM with excluding agents and the corresponding diffusion equation at increasing times.

Notice that this inclusion of volume exclusion on a lattice does not affect the population-level dynamics. In other words, since the diffusion equation \eqref{eq:diffusion} arises as the limit equation of the standard random walks of non-interacting agents, the effect of the local interaction between the agent in the simple exclusion process disappears as we let $\Delta, \delta t \rightarrow 0$. Since agents are indistinguishable, the situation in which two agents occupy adjacent positions and block each other's movement is equivalent to the scenario in which the two neighbouring agents swap their positions in the non-interacting random walk. If multi-species agents are considered, such equivalence no longer holds and the exclusion property leads to different continuous equations \citep{simpson2009mss}.

\begin{figure}
\begin{center}
{\includegraphics[width=0.7 \columnwidth]{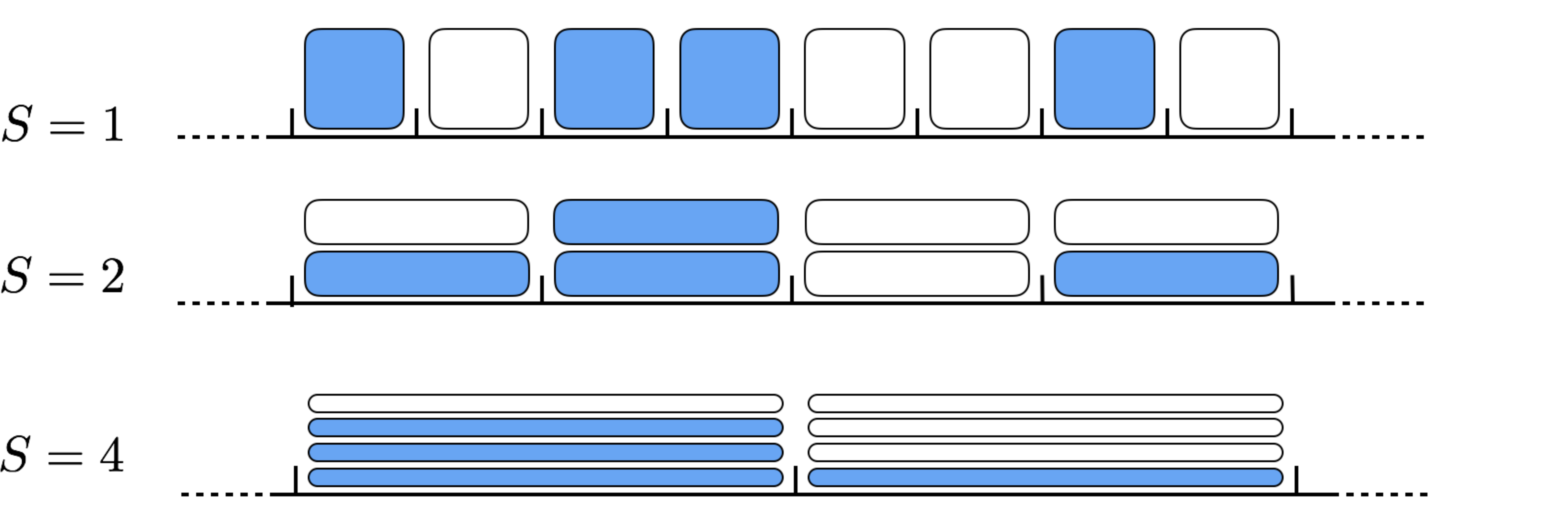}
\caption{Schematics of the partially excluding lattice with different values of the carrying capacity, $S=1,2,4$. Blue rectangle represent agents and white rectangles represent unused spaces. For unit carrying capacity, $S=1$, the fully excluding model is recovered, but as $S$ increases the spatial resolution coarsens.}
\label{fig:schematics_compartment}}
\end{center}
\end{figure}

\citet{taylor2015rtm,taylor2016cve} study alternative approaches to implementing volume exclusion in compartment-based models at different spatial scales. They consider a \textit{coarse-grained} representation of volume exclusion (as opposed to \textit{fine-grained} representation described above, in which at most one agent can occupy a given site) in which $S$ fine-grained sites are amalgamated together into a single coarse-grained site with capacity $S$ and length $S\Delta$. This coarse-grained representation is referred to as a partially-excluding ABM. See the schematic in Figure \ref{fig:schematics_compartment} for an illustration. Agents can perform jumps between neighbouring compartments with a rates proportional to $1/S^2$ and which scales linearly with the proportion of available space in the target compartment.  

\citet{taylor2015rtm} consider a uniform regular lattice. By comparing the fully-excluding model, $S=1$, and the partially-excluding model, $S>1$, the authors show that the mean and variance of the number of agents in each compartment is the same in both representations. In other words, it is possible to provide a consistent description of the effects of volume exclusion across different spatial scales with these coarse and fine representations. While the usage of a coarse-grained, partially-excluding approach leads to a significant time saving in their example simulations, it also requires movement events to occur across a wide range of spatial and temporal scales \citep{walpole2013mcm}.

In a more recent work, \citet{taylor2016cve} extend the study of these partially-excluding models to non-uniform lattices in which the sites' carrying capacities can vary across the domain. In particular, they develop a set of hybrid methods which allow the interfacing of regions of partially-excluding sites with regions of fully-excluding sites. The advantage of these hybrid methods is that they allow for the study of complicated scenarios, in which the accuracy of the fully excluding model is required in some regions of space, but the partially-excluding model can be exploited in other regions,  allowing a considerable reduction of computational cost in comparison to the ubiquitous fully-excluding model \citep{taylor2016cve}.\\

\citet{simpson2009mss} study other important extensions of the fundamental volume-excluding ABM outlined towards the start of this Section. In particular, the authors incorporate the possibility of having multiple subpopulations of agents with different motility parameters and a deterministic directional bias.

The ABM of \citet{simpson2009mss} is defined in discrete time and on a two-dimensional lattice with lattice step $\Delta$. An exclusion property is implemented, requiring that a lattice site can be occupied by at most one agent at the time. Agents move according to a biased random walk with the bias modulated by a parameter $\phi\in [-1,1]$.

In the first part of the paper, a single population of identical cells is considered. \citet{simpson2009mss} derived an advection-diffusion continuum approximation of the model by writing down the occupancy master equation and letting the bias parameter scale with the spatial step $\phi\sim \mathcal{O}(\Delta)$. The simulations of the ABM are averaged over the columns of the lattice and over many realisations. The resulting density profiles are compared with continuum descriptions showing good agreement with the corresponding PDE.

In the second part of the paper, the model is extended to describe the migration of $U$ sub-populations or species of cells. The cells of each species move according to a biased random walk, with the motility rate $\alpha^{(u)}$ and  bias intensity $\phi^{(u)}$, with $u=1,\dots U$, depending on the species. Using a similar derivation as in the one-species case, the authors derive a system of $U$ advection-diffusion equations describing the evolution of the occupancy of the different species, $\bar{C}^{(u)}$. When the bias is turned off, the system reads 
\begin{equation}
	\D{\bar{C}^{(u)}}{t}=D^{(u)} \D{}{x}\LRs{\LR{1-\sum_{u=1}^U \bar{C}^{(u)}}\D{\bar{C}^{(u)}}{x} +\bar{C}^{(u)} \D{}{x} \sum_{u=1}^U \bar{C}^{(u)}}, \quad u=1,\dots, U 
\end{equation}
where 
$$D^{(u)}=\lim_{\Delta\rightarrow 0} \frac{\alpha^{(u)} \Delta^2}{2} \, .$$ 
 In this case, the exclusion property leads to non-linear diffusivity for the individual species. However, in the case in which all species have the same motility parameters as each other, by summing all the equations, unsurprisingly, simple diffusion is recovered for the total population.
 
To compare the two levels of description, the authors consider a population of agents formed by two species initialised in adjacent regions with different initial densities. The results highlight a spontaneous aggregation in one of the species' density profiles in the continuous model. From this observation, \citet{simpson2009mss} conclude that the single species densities do not obey any maximum principle, i.e. the monotonicity of the density profile is not preserved in time, although this is true for the total population.

%--------------------------------------------------------------------
% The off-lattice case
\subsubsection{Off-lattice models}
In off-lattice ABMs the positions of the agents are represented in continuous space. This improvement adds realism to the model, since the movement of real cells is not constrained to a discrete grid. Moreover, the continuous framework introduces a larger variety of possible actions which cells can perform. For example, when modelling the migration of cells in two dimensions, the off-lattice framework allows both the distance and the direction of the movement to be a continuous variable, rather than being restricted to a discrete set of values, as in the on-lattice case. This extension is consistent with many biological observations of cell migration in which cells are not restricted to a lattice \citep{stokes1991arm,plank2004lnl}. However it has the disadvantage that it makes the mathematical analysis more complicated and sometimes intractable. 

%--------------------------------------------------------------------
\paragraph{Non-interacting cells undergoing unbiased movement.\\}

\begin{figure}
{\begin{center} 
\includegraphics[width=0.7 \columnwidth]{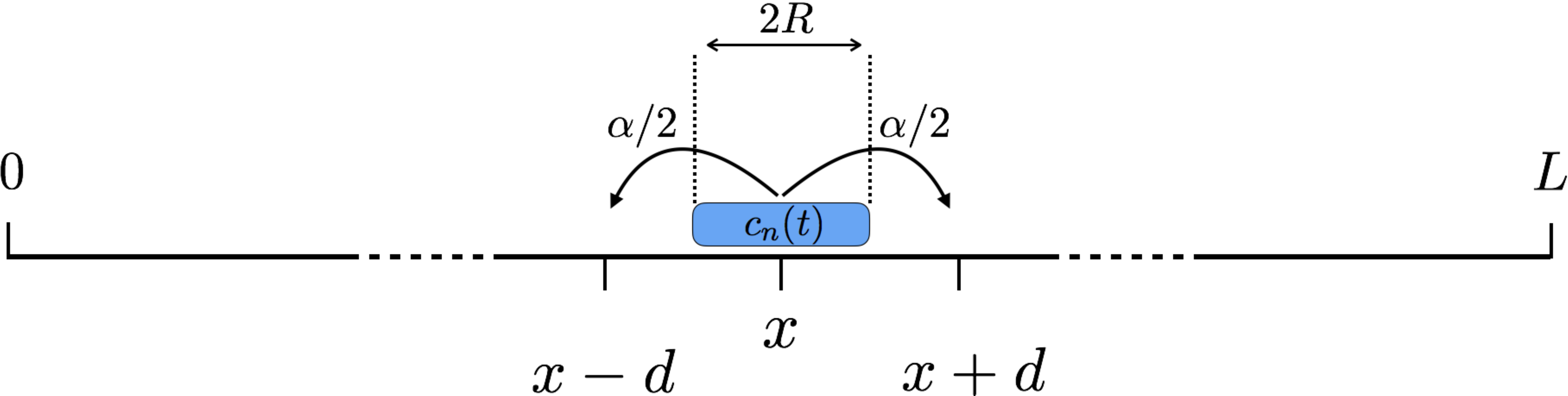}
\end{center}
\caption{Schematics of the off-lattice ABM in one dimension. A single agent, $n$, is represented by an blue interval of length $R$ and whose centre, $c_n(t)$, is defined as the position of the agent. The agent attempts to move with rate $\alpha$ a distance $d$ in one of the two directions.}
\label{fig:schematics_off_lattice}}
\end{figure}

\begin{figure}
\begin{center} 
\subfigure[][]{\includegraphics[scale=0.12]{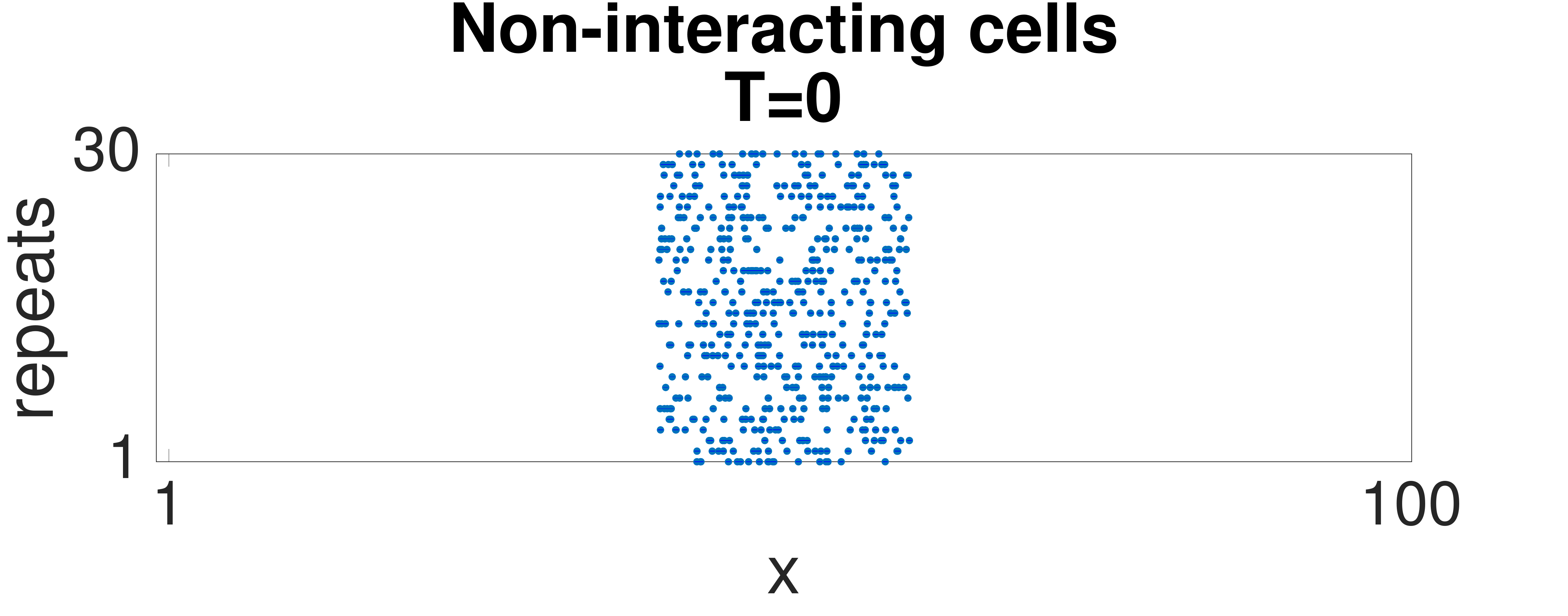}} \quad
\subfigure[][]{\includegraphics[scale=0.12]{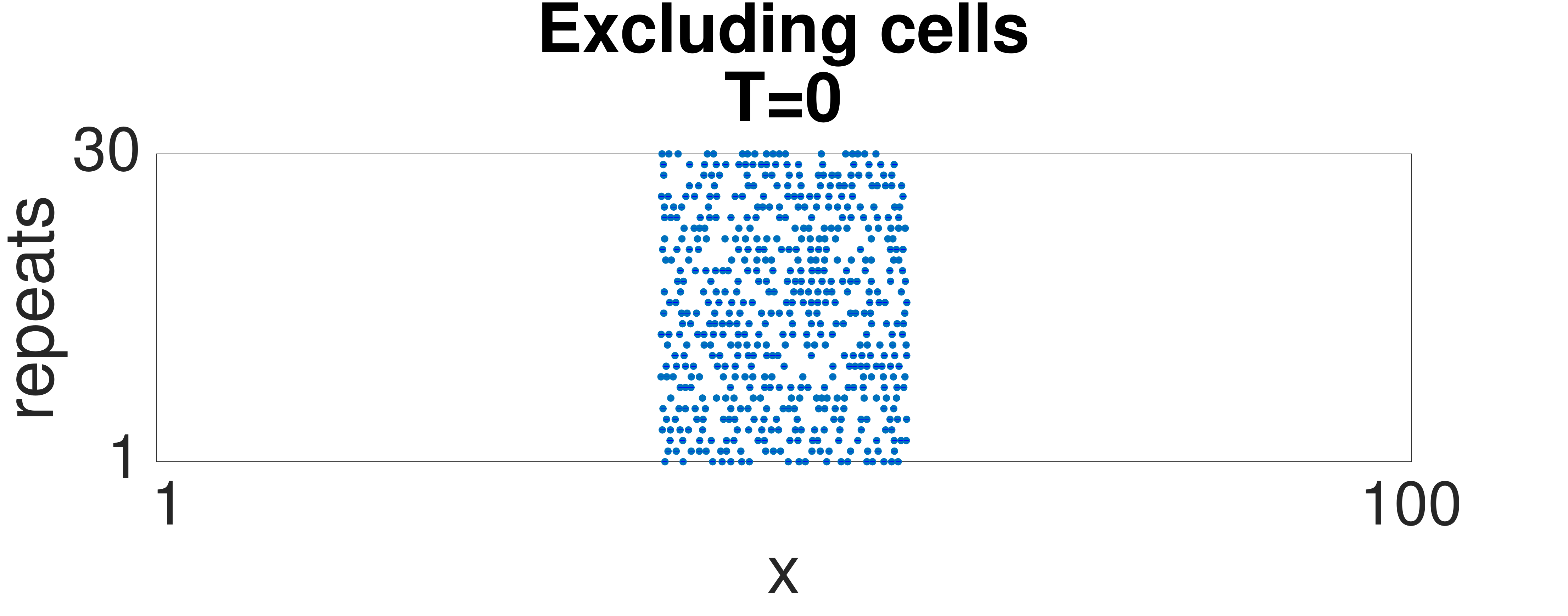}}\\[-5pt]
\subfigure[][]{\includegraphics[scale=0.12]{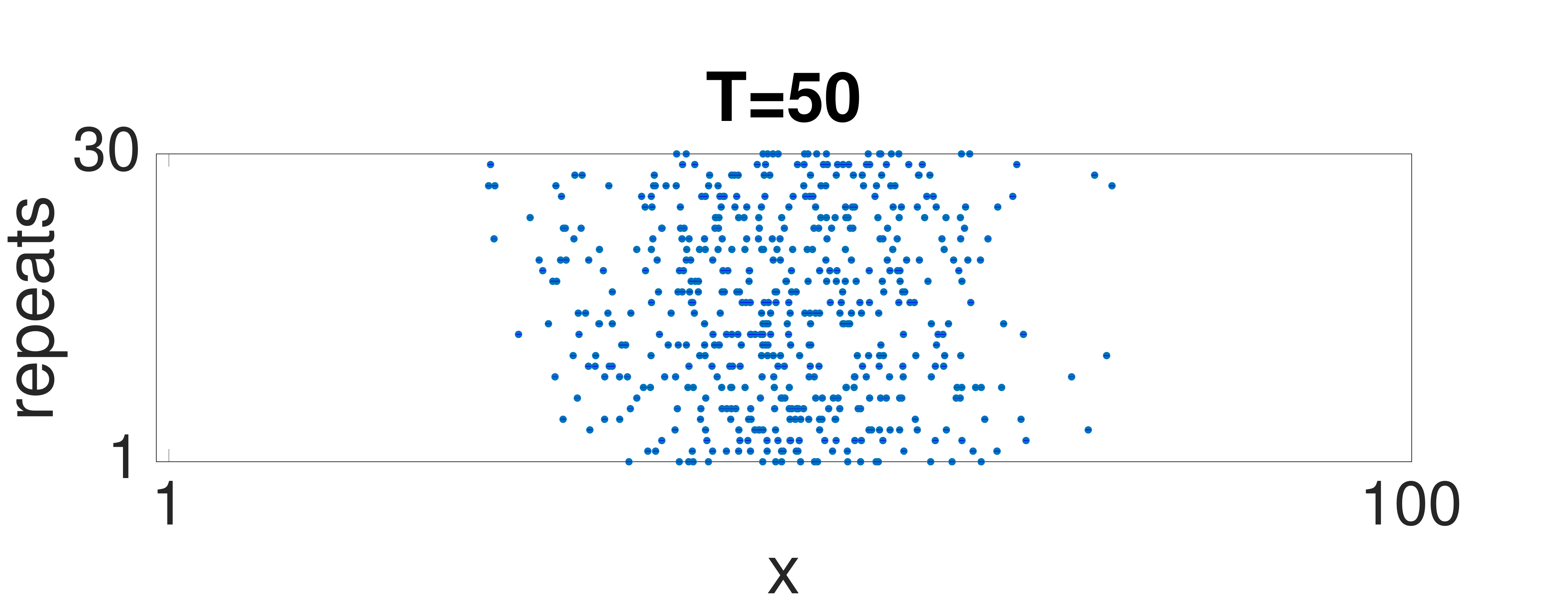}} \quad
\subfigure[][]{\includegraphics[scale=0.12]{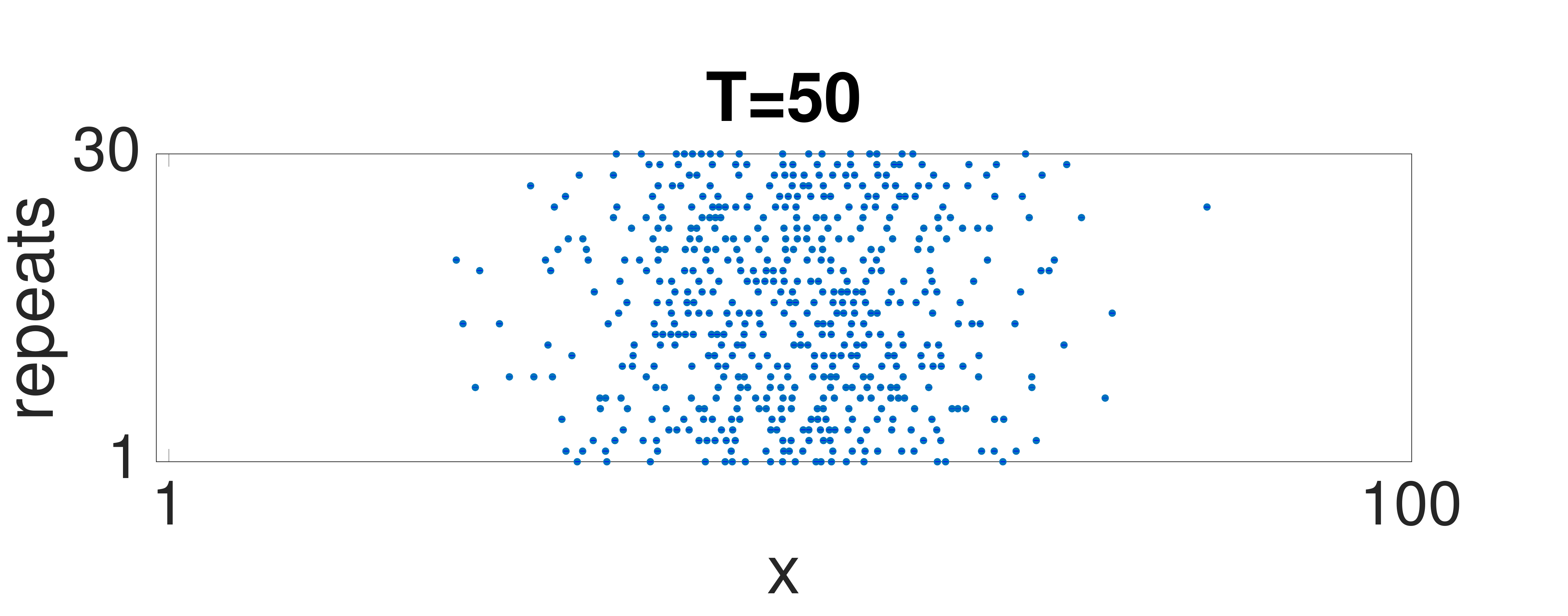}}\\[-5pt]
\subfigure[][]{\includegraphics[scale=0.12]{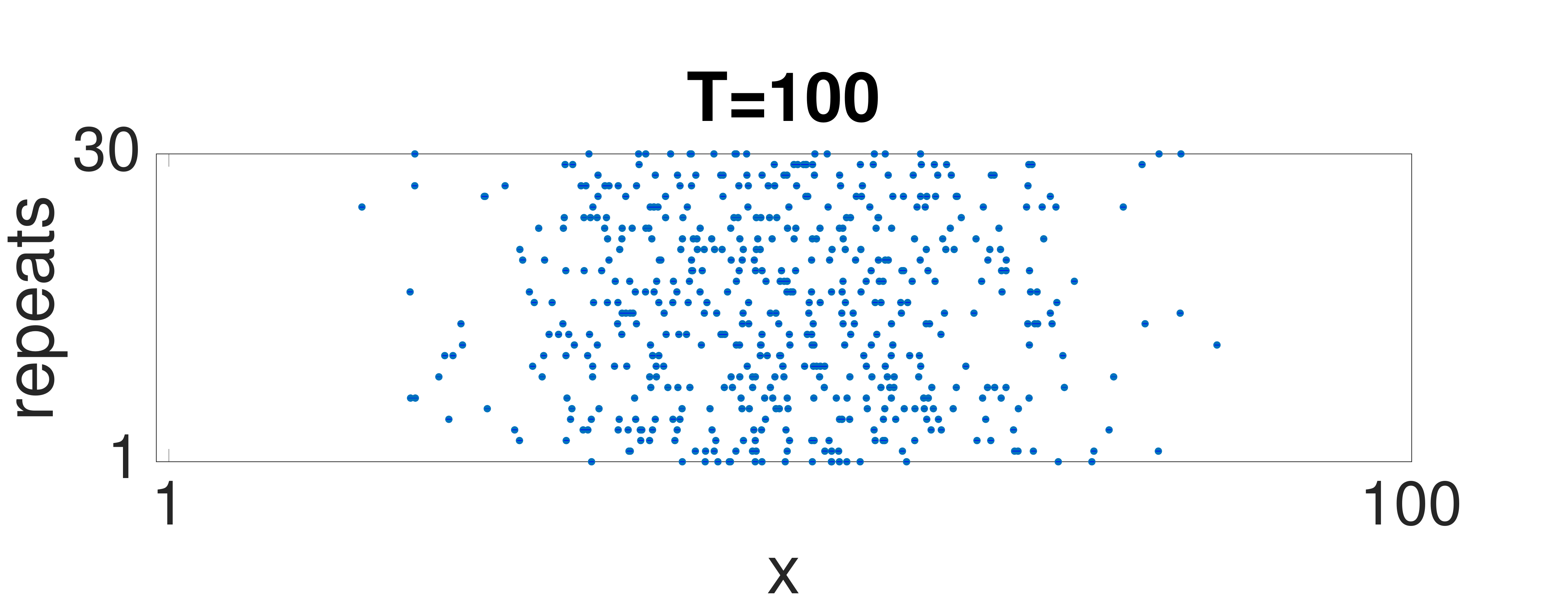}}\quad
\subfigure[][]{\includegraphics[scale=0.12]{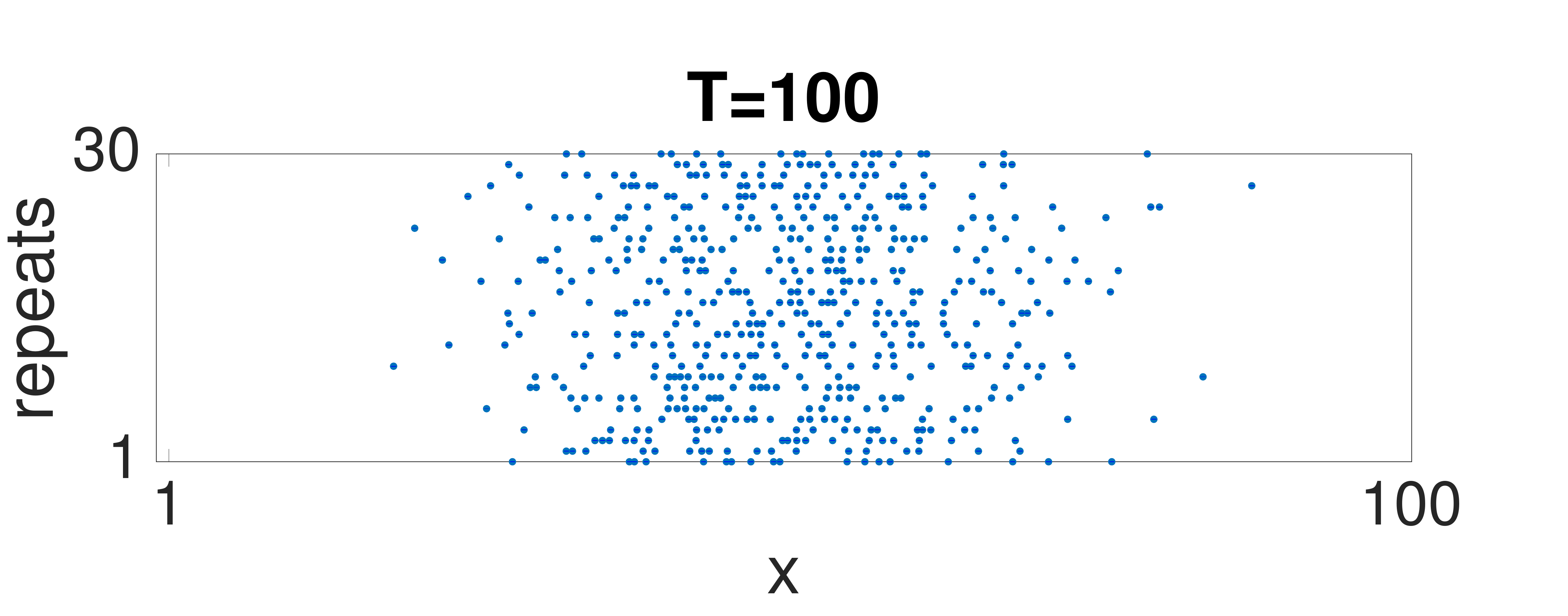}}\\[-5pt]
\subfigure[][]{\includegraphics[scale=0.12]{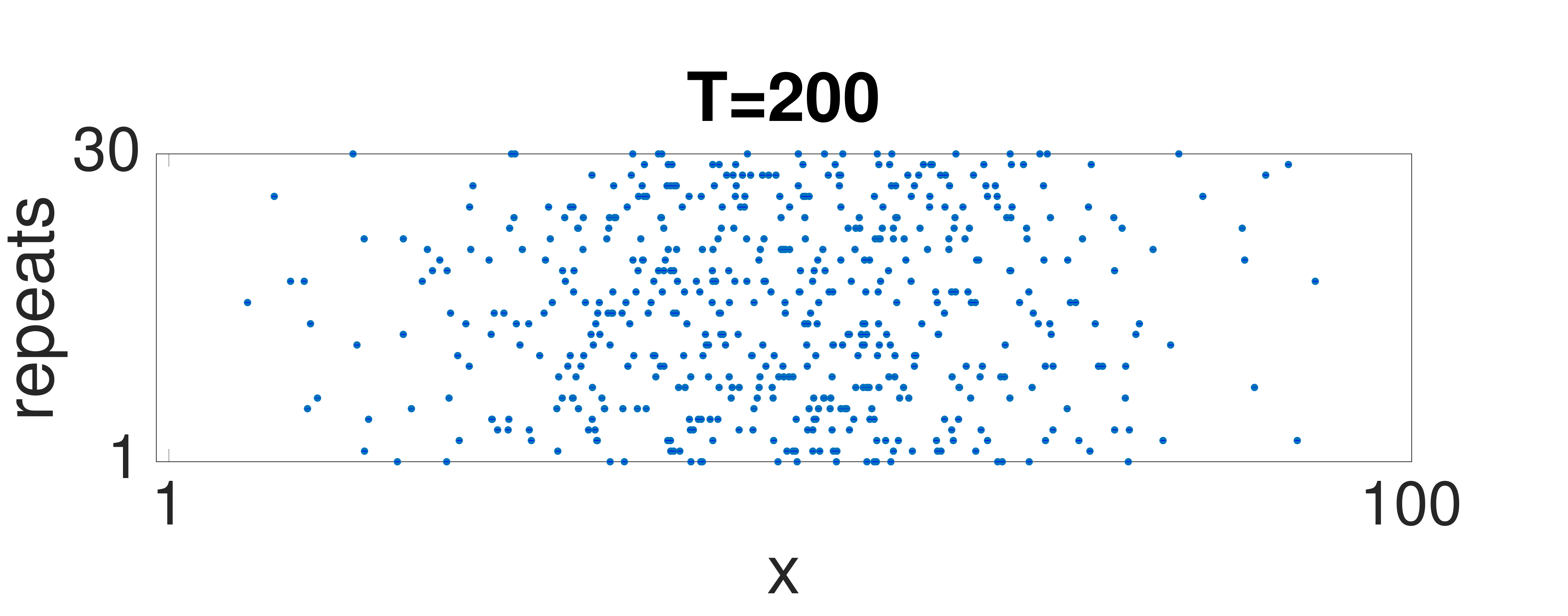}}\quad
\subfigure[][]{\includegraphics[scale=0.12]{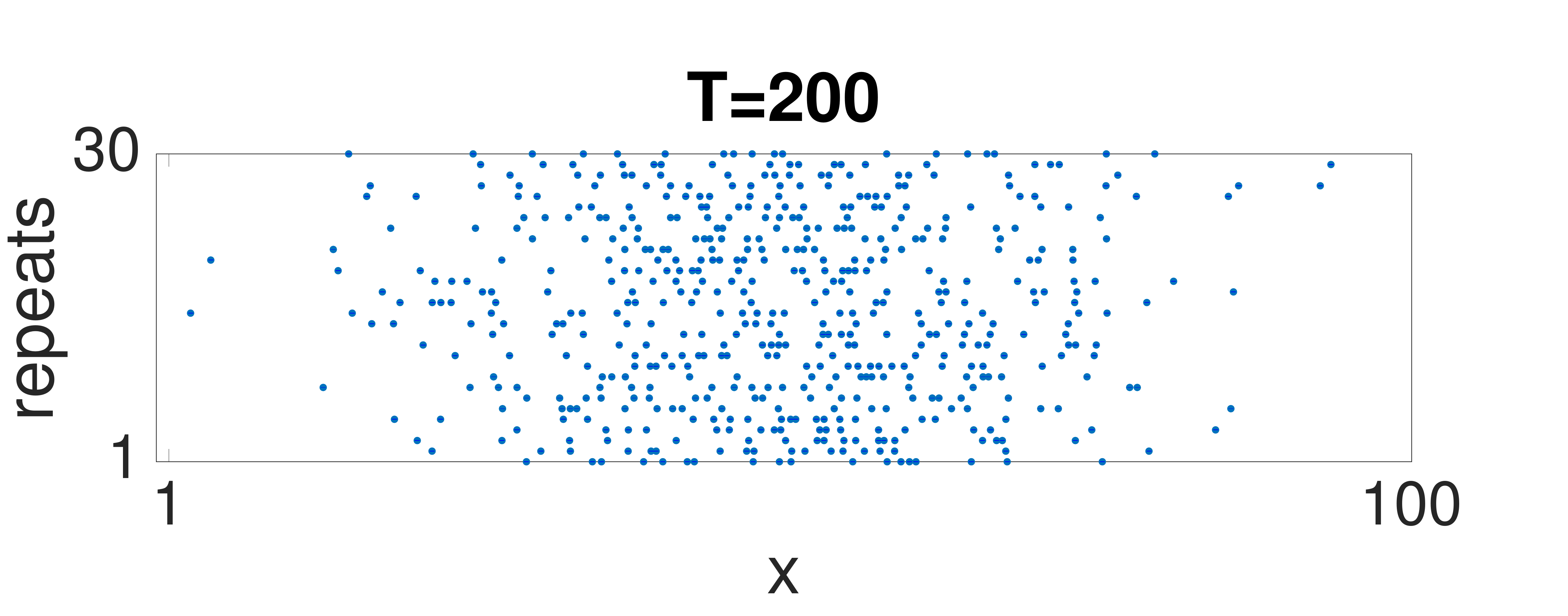}}\\
\subfigure[][]{\includegraphics[scale=0.3]{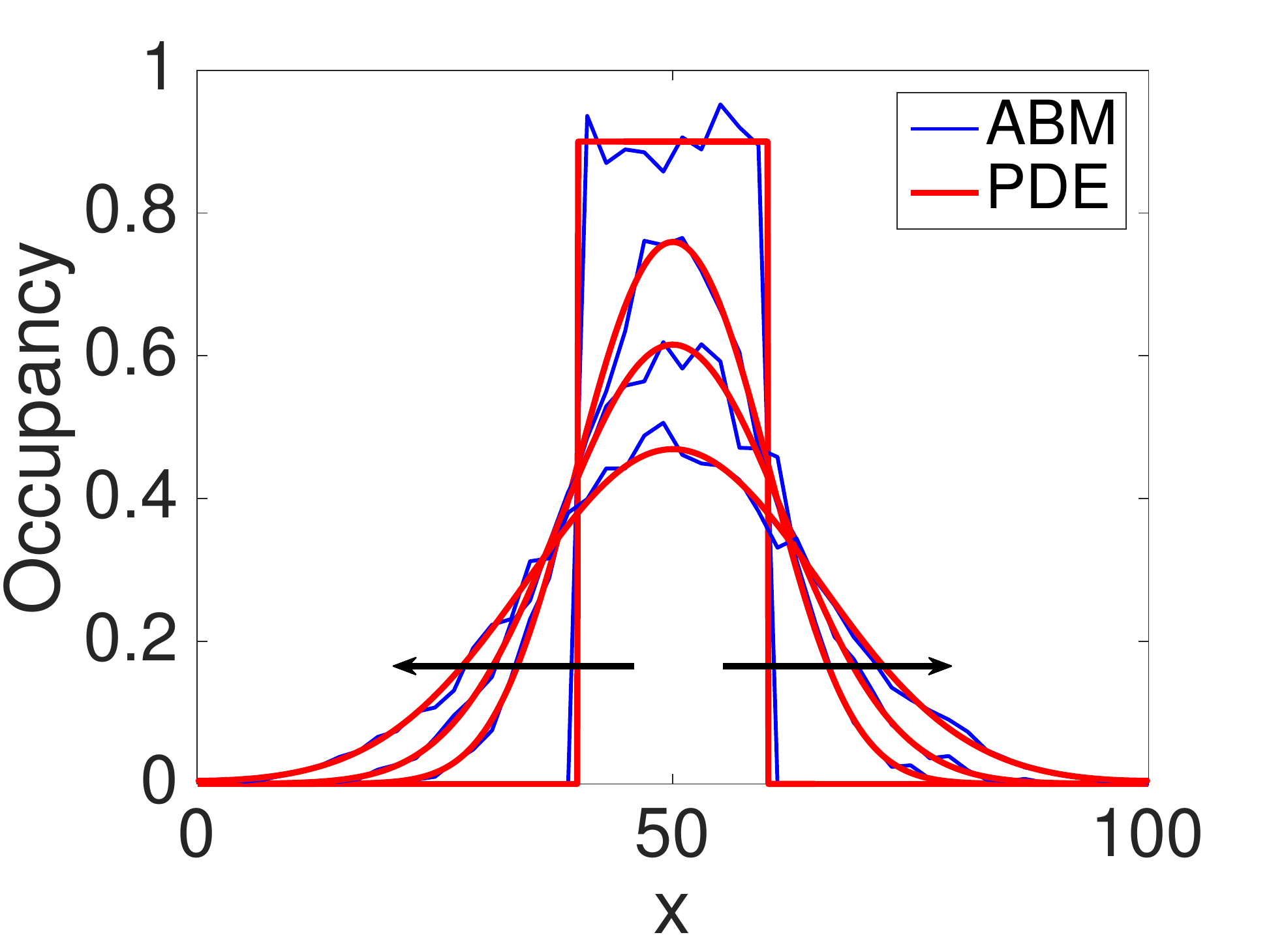}}
\subfigure[][]{\includegraphics[scale=0.3]{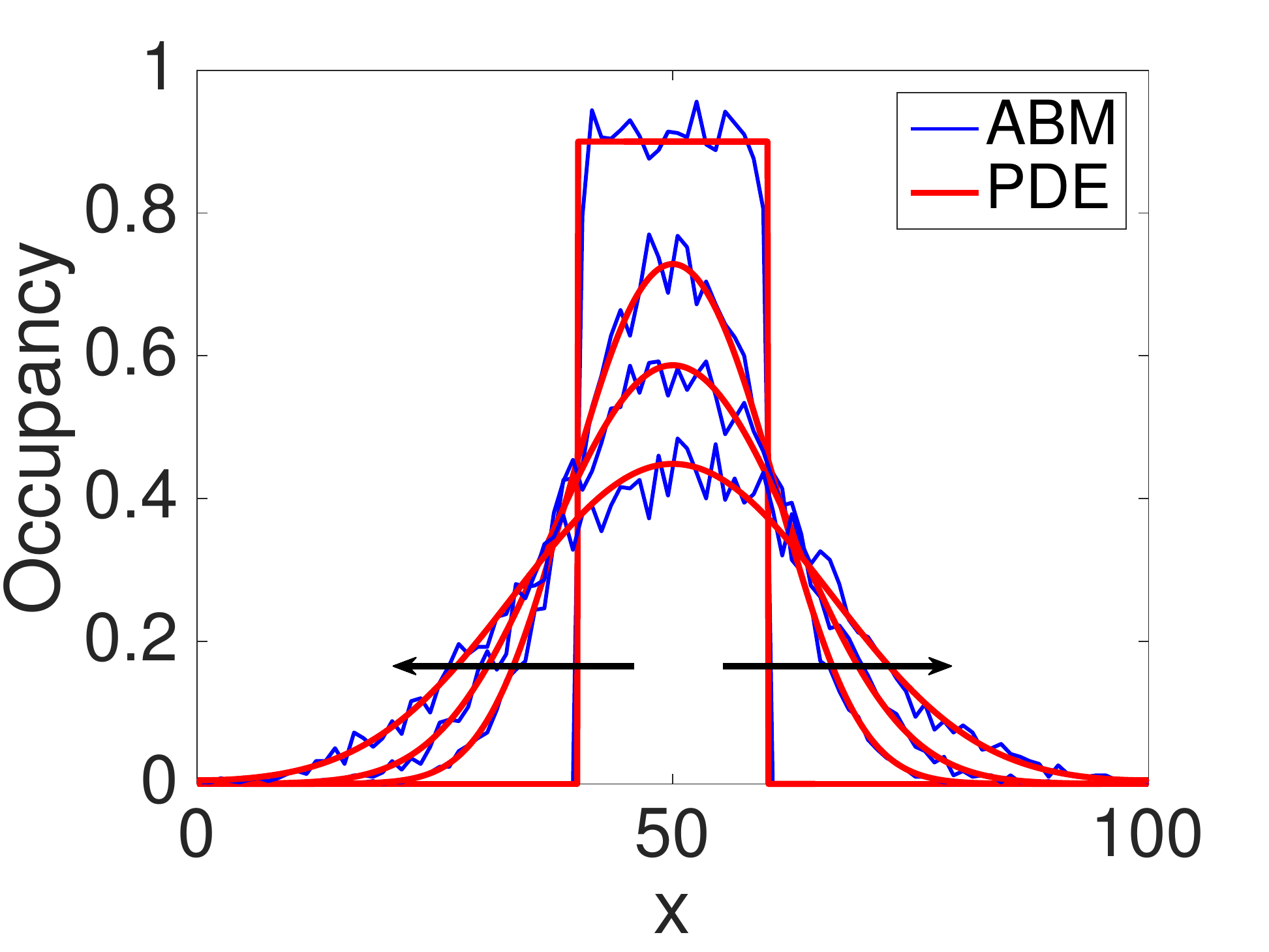}}
\end{center}
\caption{Comparison between stochastic and deterministic models of cell motility on a one-dimensional off-lattice domain. Panels (a) to (h) show four snapshots of 30  simulations of the ABM with non-interacting agents as described in Section \ref{sec:MF} (panels (a), (c), (e) and (g)) and with excluding agents ((panels (b), (d), (f) and (h)). The parameters of the model are $\alpha=4$, $d=0.5$, $R=0.2$ and the simulations are shown at times $T=0,50,100$ and $200$. In each panel $30$ independent repeats of the simulations are displayed on top each other. Agents are represented by a blue ball with radius $R$. In panels (i) and (j) we show a comparison between the occupancy of the ABM (blue line), averaged over $500$ repeats, and the numerical solution of corresponding deterministic PDE (red line). Panel (i) is for the non-interacting model and panel (j) refers to the model with volume exclusion. In both cases, the profiles are displayed at the times $T = 0, 50, 100, 200$, with the direction of the black arrows indicating increasing time.}
\label{fig:simulations_off_lattice}
\end{figure}

We consider an ABM on a one-dimensional domain, $[0,L]$, as in Figure \ref{fig:schematics_off_lattice}, with periodic boundary conditions. Each agent, $n$, is defined by its position at time $t$, $c_n(t)\in \LRs{0,L}$, and it occupies the interval $\LR{c_n(t)-R,c_n(t)+R}$, where $R$ is the analogue of the cell's radius. The model is defined in continuous time and agents perform unbiased jumps of fixed distance in one of the two directions. The rate and the distance of these jumps are denoted by $\alpha$ and $d$ respectively. See the schematics in Figure \ref{fig:schematics_off_lattice} for an illustration of the model. A set of $\cN$ agents are initially located uniformly at random in the interval $\LRs{40, 60}$ (see Figure \ref{fig:simulations_off_lattice} (a)).

For now, we assume agents move independently of other agents' positions, as in \citet{othmer1988mdb}. Consequently, a given point, $x\in \LRs{0, L}$, can be occupied by more than one agent simultaneously. In this case, obtaining a macroscopic description of the model can be achieved by employing a similar method to the corresponding on-lattice case (Section \ref{sec:on_lattice}). We consider $M$ identically prepared simulations of the model and aim to write down the master equation for the average occupancy of position $x$ at time $t$, $\Cb{x}{t}$, which is defined as 
\begin{equation}
	\label{eq:C_bar_def_off_lattice}
	\Cb{x}{t}=\frac{1}{ M}\sum_{m=1}^{M}\C{x}{t} \, ,
\end{equation}
where 
\begin{equation}
\label{eq:C_def_off_lattice}	
C^m(x,t)=|\LRb{n \st c^m_n(t)=x}| \, .
\end{equation}
The master equation for $\Cb{x}{t+\delta t}$ then reads
\begin{equation}
	\label{eq:MA_simple_off_lattice}
	\Cb{x}{t+\delta t}=\Cb{x}{t}-\overbrace{\alpha \delta t\Cb{x}{t}}^{\text{moving out of $x$}} + \underbrace{\frac{\alpha \delta t}{2} \LRs{\Cb{x -d }{t}+\Cb{x+d}{t}}}_{\text{moving into $x$}} +\mathcal{O}(\delta t ^2)\, ,
\end{equation}
where $\delta t$ is chosen sufficiently small that at most one movement event can take place in $[t,t+\delta t)$. By Taylor expanding the terms $\Cb{x \pm d}{t}$ to the second order about the point $x$ and taking the limit $\delta t,d \rightarrow 0$, while keeping $\alpha d^2$ a non zero constant, we recover the diffusion equation \eqref{eq:diffusion}, with the diffusion coefficient defined as 
\begin{equation}
	\label{eq:D_def_off_lattice}
	D=\lim_{d\rightarrow 0} \frac{\alpha d^2}{2} \, .
\end{equation}
In other words, when agents are not interacting, the off-lattice framework of the ABM does not change the resulting population-level representation which was obtained for the on-lattice case. In Figure \ref{fig:simulations_off_lattice} (i) we show a comparison of the average agent density and the corresponding diffusion equation as time evolves. The results confirm a good agreement between the stochastic and the deterministic models. 

%--------------------------------------------------------------------

\paragraph{Volume-excluding cells\\}

Incorporating volume exclusion in an on-lattice model is a natural extension of the simple multiple occupancy model. However, in an off-lattice framework there are several ways that the effects of volume exclusion can be incorporated. In this section we follow the approach of \citet{dyson2012mli}, however we highlight that other approaches can lead to slightly different results both at the individual- and population-levels.

\begin{figure}
\begin{center} 
\vspace{0.5 cm}
\includegraphics[width=0.7 \columnwidth]{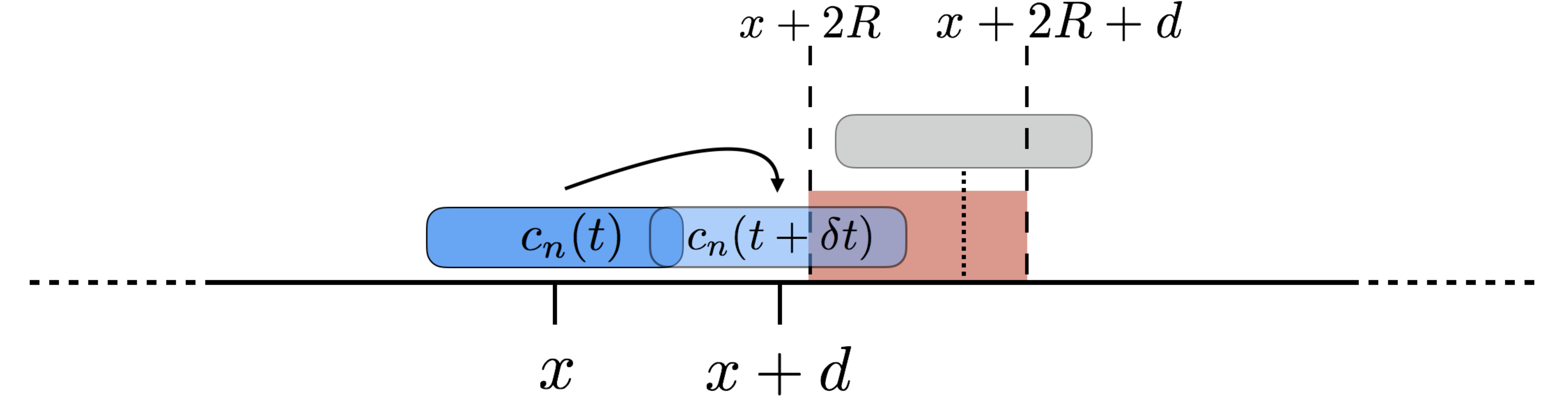}
\end{center}
\caption{Illustration of the exclusion property for the one-dimensional off-lattice model. The moving agent (blue interval) attempts to move in the right direction into position $x+d$ (light blue interval). In order for the movement to succeed, none of the other agents' centres can occupy the exclusion zone $[x+2R, x+2R+d)$ (highlighted in pink). An example of agent that would obstruct the movement is shown (light grey interval).}
\label{fig:exclusion_off_lattice}
\end{figure}

\citet{dyson2012mli} consider an exclusion mechanism in which an attempted move is aborted if it would lead to the overlap of two agents, \textit{i.e.} the corresponding centres are closer than $2R$. A schematic in Figure \ref{fig:exclusion_off_lattice} shows an example in which an agent (blue) attempts to move in the rightwards direction and an examples of an agents (grey) which would obstruct the movement. In order to write down the occupancy master equation for the average number of agents, we need to compute the transition rates, $\cT^{\pm}(x,t)$. In particular, we need to determine the probabilities that an agent located at position $x$ at time $t$, moves to the right- and left-directions, at time $t+\delta t$. By using the continuous form of the mean-field assumption \eqref{eq:MFA}, we can write
\begin{subequations}
	\label{eq:T_off_lattice}
	\begin{align}
		\cT^+(x,t)&=\frac{\alpha \delta t}{2} \LR{1-\cI^+(x,t)} \, ,\\
		\cT^-(x,t)&=\frac{\alpha \delta t}{2} \LR{1-\cI^-(x,t)} \, ,
	\end{align}
\end{subequations}
where $\cI^{+}(x,t)$ and $\cI^{-}(x,t)$ are the number agents in the exclusion zones $[x+2R, x+2R+d)$ and $[x-2R-d, x-2R)$, respectively, which impede the movement. See the interval highlighted in pink in Figure \ref{fig:exclusion_off_lattice} for an illustration right exclusion zone. 

Hence the master equation for the occupancy, $\C{x}{t+\delta t}$, reads
\begin{equation}
	\label{eq:MA_excluding_off_lattice}
	\begin{split}
	\C{x}{t+\delta t}=&\C{x}{t}-\overbrace{\frac{\alpha \delta t}{2} \C{x}{t} \LRs{\LR{1-\cI^-(x,t)}+\LR{1-\cI^+(x,t)}}}^{\text{moving out of $x$}} \\ & \underbrace{+\frac{\alpha \delta t}{2} (1-\C{x}{t})\LRs{\cI^-(x,t)+\cI^+(x,t)}}_{\text{moving into $x$}} +\mathcal{O}(\delta t ^2)\,,
	\end{split}
\end{equation}
where $C^{(m)}$ is defined as in equation \eqref{eq:C_def_off_lattice}. In order to compute $\cI^{\pm}(x,t)$, we reduce to the case in which there is \textit{at most} one other agent obstructing the movement by assuming $d<2R$. If we consider a number of agents, $\cN$, sufficiently large and we use the continuous version of the  mean-field approximation \eqref{eq:MFA}, we can write  
\begin{subequations}
\label{eq:P_R_L}
	\begin{align}
		\cI^+(x,t)&=\int_{2R}^{2R+d} \C{x+y}{t} d y \, , \\
		\cI^-(x,t)&=\int_{-2R-d}^{-2R} \C{x+y}{t} d y \, \, .
	\end{align}
\end{subequations} 
Notice that, due to the assumption on $d$, the two integrals on the right-hand side of equations \eqref{eq:P_R_L} assume values in $\LRb{0,1}$ and so the two transition probabilities defined in equations \eqref{eq:T_off_lattice} are meaningful.  We can Taylor expand $\cI^+(x,t)$ and $\cI^-(x,t)$ to second order to obtain
\begin{subequations}
	\label{eq:taylor_off_lattice}
	\begin{align}
	\begin{split}
		\cI^- (x,t)=d \C{x}{t} &+ \frac{d}{2} (4R+d) \D{\Cb{x}{t}}{x}\\&+\frac{d}{6} (12 R^2+6R d +d^2) \DD{\C{x}{t}}{x}+\mathcal{O}\LRs{(2R+d)^4} \, ,
		\end{split}\\
	\begin{split}
		\cI^- (x,t)=d \C{x}{t} &- \frac{d}{2} (4R+d) \D{\Cb{x}{t}}{x}\\&+\frac{d}{6} (12 R^2+6R d +d^2) \DD{\C{x}{t}}{x}+\mathcal{O}\LRs{(2R+d)^4}  \, .
		\end{split}
	\end{align}
	\end{subequations}
By substituting equations \eqref{eq:taylor_off_lattice} into equation \eqref{eq:MA_excluding_off_lattice} and taking the limit as $\delta t \rightarrow 0$, we obtain
\begin{equation}
	\label{eq:diffusion_off_lattice_excluding}
	\D{C^{(m)}}{t}=\frac{\alpha d^2 }{2} \D{}{x}\LRs{\LR{1+\LR{4R-d}C^{(m)}}\D{C^{(m)}}{x}} +\mathcal{O}\LR{(R+d)^4}\, ,
\end{equation}
for $m=1,\dots, M$. We can now sum equations \eqref{eq:diffusion_off_lattice_excluding} over all values of $m$ and divide by the total number of realisations, $M$, to obtain an equivalent equation for the average occupancy $\Cb{x,t}$ using the assumption that all agents are identically uniformly distributed initially we find
\begin{equation}
\label{eq:diffusion_off_lattice_excluding4}
	%\label{eq:diffusion_off_lattice_excluding2}
	\D{\bar{C}}{t}=\frac{\alpha d^2 }{2}\D{}{x}\LRs{\LR{1+\LR{4R-d)\bar{C}}}\D{\bar{C}}{x}} +\mathcal{O}\LR{(R+d)^4}\, .
\end{equation}
 The dependence on the agents' size can be explained by noting that larger agents will collide more often. \citet{dyson2012mli} identify that when $d$ is large compared to $R$, the diffusion coefficient decreases to the point at which it may be negative, which might suggest the occurrence of cell aggregation. It is also important to notice that, if the agents are initialised on a lattice with step $\Delta=2R$, and we choose $d=2R$, the ABM is equivalent to the on-lattice ABM with excluding agents. However, for such choice of $d$, the derivation of equation \eqref{eq:diffusion_off_lattice_excluding4} breaks down, which explains why simply setting $d=2R$ in equation \eqref{eq:diffusion_off_lattice_excluding4} does not recover the simple diffusion equation as might be expected. By taking the limit as $d \rightarrow 0$, while keeping $\alpha d^2$ a non-zero constant, we arrive at a non-linear diffusion equation:
\begin{equation}
	\label{eq:diffusion_off_lattice_excluding3}
	\D{\bar{C}}{t}=\D{}{x}\LRs{D(\bar{C})\D{\bar{C}}{x}} \, ,
\end{equation}
with \begin{equation*}
	D(\bar{C})=D\LRs{1+4R\bar{C} }
\end{equation*}
and $D$ is defined as in equation \eqref{eq:D_def_off_lattice}.

Notice that the exclusion property results in a non-linearity in the diffusion coefficient of the population-level equation. In particular, the term $1+4R\bar{C}$ in equation \eqref{eq:diffusion_off_lattice_excluding3} leads to faster diffusion where the average occupancy is large and  slower diffusion where the average occupancy decreases to zero, in which case the diffusion coefficient reaches its minimum value, $D$. In Figure \ref{fig:simulations_off_lattice} (j) we compare the average agent density with the numerical solution of equation \eqref{eq:diffusion_off_lattice_excluding4} at four subsequent times. The results confirm the good agreement between the ABM and the corresponding continuum equation.\\

\subsection{Higher dimensions}
\label{sec:HD}

Although the detailed derivations of the previous Section are carried out for a one-dimensional interval domain, discrete-continuum equivalence frameworks can be extended to higher dimensions.  

In fact, for models which do not account for crowding effects, either on- or off-lattice, the resulting macroscopic description can be obtained in a similar manner to the one-dimensional case \citep{othmer1988mdb,deutsch2007cam,codling2008rwm}. For the regular-square-lattice and the off-lattice models,  the resulting deterministic description is the natural generalisation of equation \eqref{eq:diffusion} which is given by 
\begin{equation}
	\label{eq:diffusion_HD}
	\D{\bar{C}}{t}=D \nabla^2 \bar{C} \, ,
\end{equation}
where $\nabla^2$ represents the Laplacian operator and the diffusion coefficient is given by $$D=\lim_{\Delta\rightarrow 0} \frac{\alpha \Delta^2}{2r} \, ,$$
where $r$ is the dimension. Equation \eqref{eq:diffusion_HD} is an isotropic PDE, \textit{i.e.} it is not biased in any spatial direction. That such isotropic equation can be derived from an  ABM which is defined on a regular, anisotropic lattice is perhaps surprising.  In other words, the intrinsic individual-level anisotropy of the ABM in the lattice directions vanishes in the diffusive macroscopic description \citep{othmer1988mdb,deutsch2007cam,codling2008rwm}. This same property has been demonstrated to not hold for models of cell movement which are based on velocity-jump processes \citep{gavagnin2018mpm}. We refer the reader to Section \ref{sec:persistence_of_motion} for a more detailed discussion of such models.

\begin{figure}
\begin{center} {\includegraphics[scale=0.2]{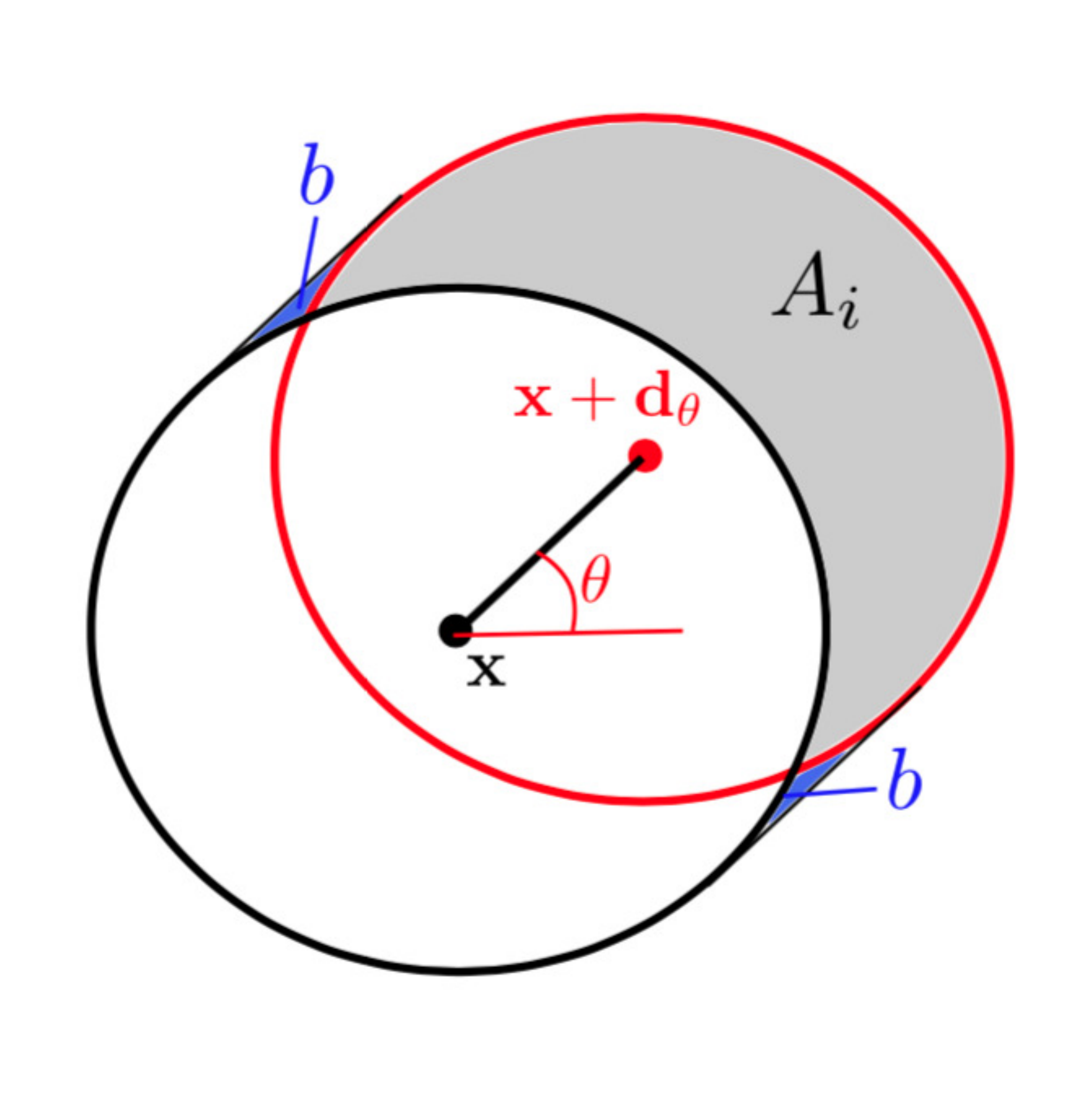}
\caption{Schematic representation of the volume exclusion property of the model of \citet{dyson2014ive} in two dimensions. A moving agent (black circle) is attempting a movement to a new position (red circle). The grey shaded region, $A_i$, is the area in which agent overlap must be avoided for a successful movement event. To simplify the algebra involved, the authors extend the integration region to include the blue shaded regions, $b$. Reproduced from \citet{dyson2014ive} with the permission of Journal of Mathematical Biology.}\label{fig:dyson2014}}
\end{center}
\end{figure}

Incorporating volume exclusion in higher dimensions does not lead to substantial changes for the on-lattice models in which case equation \eqref{eq:diffusion_HD} still holds \citep{simpson2009mss}. However, higher dimensions  significantly increase the complexity of off-lattice ABMs with crowding effects. For example, 
\citet{dyson2014ive} study an extension of their previous one-dimensional model in two and three dimensions. In  their ABMs, cells are represented as circular or spherical agents of radius $R$. To include volume exclusion, agents movements which would lead to an overlap of agents are aborted. In Figure \ref{fig:dyson2014} we reproduce a schematic of the volume exclusion property for the ABM of \citet{dyson2014ive} in two dimensions. Notice that, to compute the probability of finding obstructing agents, it is necessary to integrate the occupancy function over the grey shaded region, $A_i$, of Figure \ref{fig:dyson2014}. This makes the computation complicated and intractable from a mathematical point of view. To overcome this problem, \citet{dyson2014ive} suggest a simplification of the calculation by extending the integral to the two blue regions, $b$. Clearly, if the distance of each jump, $d_\theta$, is sufficiently small, this is a reasonable approximation which significantly  simplifies the analytical calculation.

By adopting this simplification, \citet{dyson2014ive} carry out the  derivation of diffusive PDEs for the average agent occupancy, leading to equations of the form of \eqref{eq:diffusion_off_lattice_excluding4}. In all cases, the non-linear diffusion coefficients are  increasing functions of the agents' radii, $R$, and decreasing functions of the jump distance, $d$.

 Finally, the authors consider a system comprising multiple species of agents in which the size and the movement rates of the agents depend on the species to which they belong. The authors consider an example with two species, one comprising agents that are large and slow moving, and the other smaller but quicker. A system of PDEs for the densities of the two species is derived and the results show that the species with smaller agents is less affected by volume exclusion. Interestingly, the effect is reduced in the area where the two species coexist, since the majority of the area is occupied by smaller agents.

%--------------------------------------------------------------------
% Higher orders approximaitons
\subsection{Higher-order closure approximations}
\label{sec:HOA}
In general, when agent-agent interactions are included in an ABM, as in the example of volume-exclusion above, the positions of the agents are not independent. When this happens, the evolution of the average agent  density depends on the distribution of agent pairs, which also depends on the distribution of agents triples and so on. Therefore, if we aim to derive a deterministic representation of the model, we have to deal with an infinite system of unclosed equations for each of the spatial moments of the agents' distribution. In order to overcome this problem, a moment closure approximation is necessary in order to make the system amenable to mathematical analysis. 

The mean-field closure, given by equation \eqref{eq:MFA}, represents the easiest form of moment closure which is typically used \citep{simpson2009mss,cheeseman2014std,dyson2012mli,dyson2014ive}. Although such rudimentary approximation provides good results in a wide range of scenarios, when the spatial distribution of the agents is strongly correlated, the mean-field closure can lead to under- or overestimations of the total agent density, resulting in a poor agreement between the stochastic ABM and its deterministic representation.

%Baker 2010

\citet{baker2010cmf} investigate the role of spatial correlation in exclusion processes and its effect on the agreement between the averaged agent-based dynamics and the mean-field approximation.
%details

They consider an excluding ABM on two- and three-dimensional lattices. Agents can move to neighbouring lattice sites, with rate $\alpha_m$, and proliferate by placing daughter agents into neighbouring lattice sites, with rate $\alpha_p$. They derive a general set of master equations for the $k$-point distribution functions, $\rho^{(k)}$, (of which the 1-point distribution function is simply the density, the 2-point distribution function is the pairwise occupancy etc). This infinite hierarchy of master equations is unclosed: the differential equations for the $k-1$-point distribution functions depend on the $k$-point distribution functions. Therefore, a closure approximation is required in order to solve for the lower-order distribution functions. The authors compare the averaged density for the discrete model with the first-level moment closure (the \textit{mean-field approximation}), in which neighbouring sites are assumed independent (see equation \eqref{eq:MFA}). They also compare to a second-level moment closure (the Kirkwood superposition approximation), which takes into account pairwise spatial correlations. On the square lattice, the distance between two lattice sites increases irregularly ($\Delta$, $\sqrt{2}\Delta$, $2\Delta$, $\sqrt{5}\Delta$) and the neighbours for each site, therefore, have to be calculated separately for each distance. This fact means that the number of ODEs for the correlation functions becomes intractable very quickly. The system of ODEs, therefore, needs to be truncated at a maximum distance $r_{\text{max}}$, beyond which the sites are considered independent. Different values of $r_{\text{max}}$ are compared and the results suggest that, in two dimensions, the system can be truncated at $r_{\text{max}}=3$ without losing accuracy. For the three-dimensional case, the cut-off can be reduced to $r_{\text{max}}=2$. The inclusion of correlations in the model, even if closed at level two, provides a significant improvement in the approximation to the agent-based model.  

\citeauthor{baker2010cmf} also investigate the effects of motility, birth and death events on spatial correlations. As the proliferation parameter increases with respect to the (fixed) motility parameter, spatial correlations play a more important role and the mean-field prediction appears to overestimate the growth of the population. This is due to the fact that cell motility can not break up the correlations caused by the appearance of new agents close to their parents sufficiently quickly. This leads to cluster formation which, in turn, reduces the number of successful proliferation events, slowing population growth. When agent death is included in the model, a counterintuitive effect appears. We may naively expect that deaths would decrease spatial correlation and so increase the agreement between the mean-field and the agent-based models. Instead the opposite happens. 
A high death rate has a similar effect to having a more sparsely populated initial seed. It provides opportunities for correlations to build up where previously sites occupancies were uncorrelated.

%Another consequence of including death in the model is to further decrease the steady state population density from its predicted mean-field (logistic) value. Baker et al. provide an intuitive explanation of this disagreement, arguing that, since the volume exclusion affects only the proliferation rates (abortions) and not the death rates, ``by increasing the proliferation rate, and hence the neighbour site correlations, we are effectively increasing the contribution of death relative to that of proliferation, leading to a suppression in steady state population numbers.'' \citep{baker2010cmf}

\citet{markham2013smi} continue the work of  \citet{baker2010cmf} by deriving a deterministic, continuum analogue of a spatially extended, on-lattice, agent-based model. As in the previous paper, they look at the influence of including spatial correlations in the continuum model and determine how this improves the agreement, in comparison with the mean-field model, which assumes absence of spatial correlations.

The ABM considered is the same as in \citet{baker2010cmf}, \textit{i.e.} a volume-exclusion process on a lattice on which individuals can move, proliferate and die with rates $\alpha_m$, $\alpha_p$ and $\alpha_d$, respectively.

\citet{baker2010cmf}, truncate the pairwise correlation functions at a maximum distance, $r_{\text{max}}$, resulting in a system of ODEs. The aim of \citet{markham2013smi}, however, is to derive a tractable deterministic PDE for the  evolution of the pairwise spatial correlation function. Specifically, as the lattice step $\Delta$ is small relative to the size of the domain, one can Taylor expand the correlation functions up to second order in $\Delta$ and then move to radial coordinates. Finally, assuming that $\alpha \Delta^2$ remains constant as $\Delta\rightarrow 0$, one obtains a reaction-diffusion PDE.

A trivial analysis shows that, as the motility parameter, $\alpha_m$, increases, the diffusion coefficient of the PDE for the correlation function increases. This agrees with the intuitive prediction that higher rates of movement break up clusters of agents more effectively. On the other hand, the reaction term of the PDE is negative and decreases as the proliferation rate, $\alpha_p$, increases. This means that increasing proliferation leads to a decrease in correlations. This fact appears to contradict the results of \citet{baker2010cmf}. Nevertheless, \citet{markham2013smi} provide an informal explanation for this phenomenon.
%: while a proliferation event at site $l$ clearly increases the total density of the system, it does not necessarily increase the pair distribution function for lattice sites at distance further than $\Delta$ apart. Since the correlation function is proportional to the pair correlation function and inversely proportional to the squared density, we expect the correlation function at distance decrease, as the density distribution will definitely increase.

The results of spatially extended simulations show good agreement between the solution of the PDE and the discrete model. In particular, the PDE approximation behaves similarly to the ABM which, in the case of non-zero death rate, is significantly lower than the mean-field prediction. Moreover, an improvement in the agreement between the PDE and the ABM is obtained if the movement rate increases or if higher-dimensional domains are considered. 

\citet{markham2013smi} find that there exists a region of parameter space for which the deterministic models (either with or without correlations) can not replicate the ABM dynamics. This phenomenon corresponds to high values of the death rate which leads the population of the discrete model to eventually go extinct.

%markham 2012isc
%Markham 2013smi

\citet{markham2013isc} extend their previous spatial correlation model \citep{markham2013smi} to an ABM with heterogeneous agents. They provide other examples in which spatial correlations play a crucial role in predicting the population-level behaviour correctly. Agents are divided into $U$ species and can move, proliferate and die with rates $\alpha^I_m$, $\alpha^I_p$ and $\alpha^I_d$ respectively, where $I\in \lbrace 0,\dots , U\rbrace$ denotes the specie of the selected agent. With the same idea as their previous work, \citet{markham2013isc} focus on the evolution of the pairwise correlation, closing the system at the level above. Firstly they derive a set of ODEs for the system in which the pairwise correlation is divided into \textit{auto-correlation}, i.e. between agents of the same species, and \textit{cross-correlation}, if the agents belong to different species. Subsequently they obtain a set of PDEs by Taylor expanding and taking the limit as the lattice step goes to $0$. A key assumption throughout the paper, is that the pairwise correlation depends only on the distance between the agents, i.e. it is translationally invariant and isotropic. For this reason, the agents are always assumed to be spread uniformly across the domain initially.

The results for two species show a good agreement between the PDE model and the ABM in comparison to the agreement between the mean-field approximation and the ABM. The authors assume agent death is negligible and consider two species of cells with the same proliferation rates and different movement rates. While the logistic dynamics of the mean-field model predicts that the densities of the two species will converge to the same steady state, in the ABM, the species with the greater movement rate reaches a higher density at the equilibrium. They also show that with specific values of proliferation rates, the results can be the reversed, \textit{i.e.} in the mean-field approximation, the density of one species at the equilibrium is greater than the other specie's density, while the ABM predicts that, at equilibrium, the two species reach the same density. Remarkably, the PDE model incorporating agent-agent correlation shows a good agreement with the ABM in all these scenarios. 

%--------------------------------------------------------------------
% Models
\section{Model extensions}
\label{sec:models}
The models summarised in the previous section, represent the fundamental basis for the vast majority of spatially extended models of cell migration. However, during the process of migration, cells perform a variety of other actions and interactions whose role can dramatically impact upon the macroscopic behaviour of the system \citep{niessen2007tja,carmona2008cil,ward2003dbd,keynes1992rca,trepat2009pfd,tambe2011ccg}. Here we provide a brief overview of some of the most important extensions and modification of the standard ABMs and demonstrate their effects on the corresponding macroscopic models.

%--------------------------------------------------------------------
% Proliferation
\subsection{Cell Proliferation}
\label{sec:proliferation}

In all the models considered so far, the ability of cells to divide and produce daughter cells was not included. In reality, in certain circumstances (e.g. tumour invasion and wound healing) the role of cell proliferation is crucial for the dynamics of the system \citep{sherratt2001nmm, maini2004twm}.

%Simpson 2007

\citet{simpson2007sic} propose an ABM which takes into account cells' motility and proliferation. The ABM is defined on a two-dimensional lattice with a simple exclusion property, meaning that each lattice site can be occupied by at most one agent at the time. The model advances in  discrete time. At each time step, every agent can move and proliferate with probabilities $P_m$ and $P_p$, respectively. Agents move according to a simple, unbiased random walk to one of the their four nearest sites (von Neumann neighbours). When a proliferation event occurs, the proliferating agent moves to one of its eight surrounding sites  (Moore neighbours) and the new offspring is displaced to the site diametrically opposite. If either of the two sites is already occupied, the proliferation event is aborted. Both motility and proliferation events are limited by a carrying capacity $\kappa\in \lbrace 1, \dots , 8 \rbrace$. In particular, if an agent attempts to move or proliferate into a site with a number of surrounding neighbours greater than $\kappa$, the event is aborted.

Firstly, the authors simulate an invasion wave and investigate some of the features of the ABM. They relate and compare the behaviour of their ABM with the traditional deterministic Fisher equation \citep{fisher1937waa}:
\begin{equation}
	\label{eq:fisher}
	\D{C}{t}=D \DD{C}{x} +\nu C\LR{1-\frac{C}{K}} \, ,
\end{equation}  
where $D$ is the diffusion coefficient, $\nu$ is the growth rate and $K$ is the carrying capacity. Note that the average speed of invasion of the deterministic Fisher wave is known to be $v^*=2\sqrt{D\nu}$.

The results of the ABM simulations show that, consistent with the continuum equation, the wave speed of the invasion increases with motility parameter, $P_m$, and proliferation probability, $P_p$. However, the speed in the ABM is more sensitive to variation in proliferation than in motility, in contrast to the corresponding deterministic description which predicts equal sensitivity. As in the continuum setting, the speed of invasion is found to be independent of the carrying capacity.  

To establish a connection between the ABM and the traditional continuum model given by equation \eqref{eq:fisher}, \citeauthor{simpson2007sic} compare the behaviour of the two levels of description for scenarios in which either proliferation or motility are active, but not both. When agents can only proliferate, the total density evolves in a logistic manner and the authors identify a linear relationship between the parameters of the agent-based and population-level models. When agents can move but not proliferate, the results show that the diffusivity of the model is a decreasing function of the background density. Nevertheless, a linear dependence is found between the motility probability of the ABM, $P_m$, and the diffusion coefficient at zero agent density.

Finally, the ABM is used to investigate the individual cells' trajectories within the invasion wave. The authors use synthetic data to obtain statistical information on the average direction of movement. The results compare the  behaviour of the most advanced cell in the invasion with a second cell close behind the wave front. Despite the cells moving according to an unbiased random walk, the crowding effects inhibit the motility  into highly populated regions and this induces a bias towards the direction of the invasion in both the cells considered.  By using the individual agent trajectories, this bias can be quantified. Consistent with experimental data \citep{druckenbrod2007ben}, the observations on the ABM show a more evident bias for cells near the wave front in comparison to cells behind the wave front.

\begin{figure}
\begin{center} 
\subfigure[][]{\includegraphics[scale=0.26]{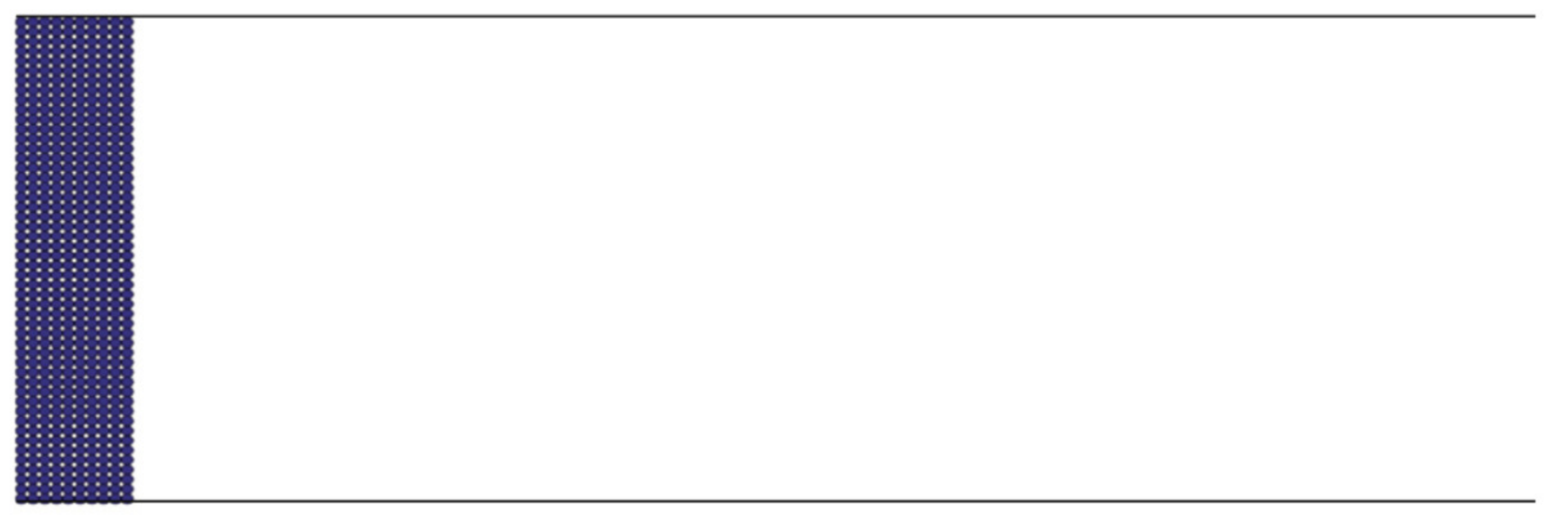}}\\
\subfigure[][]{\includegraphics[scale=0.26]{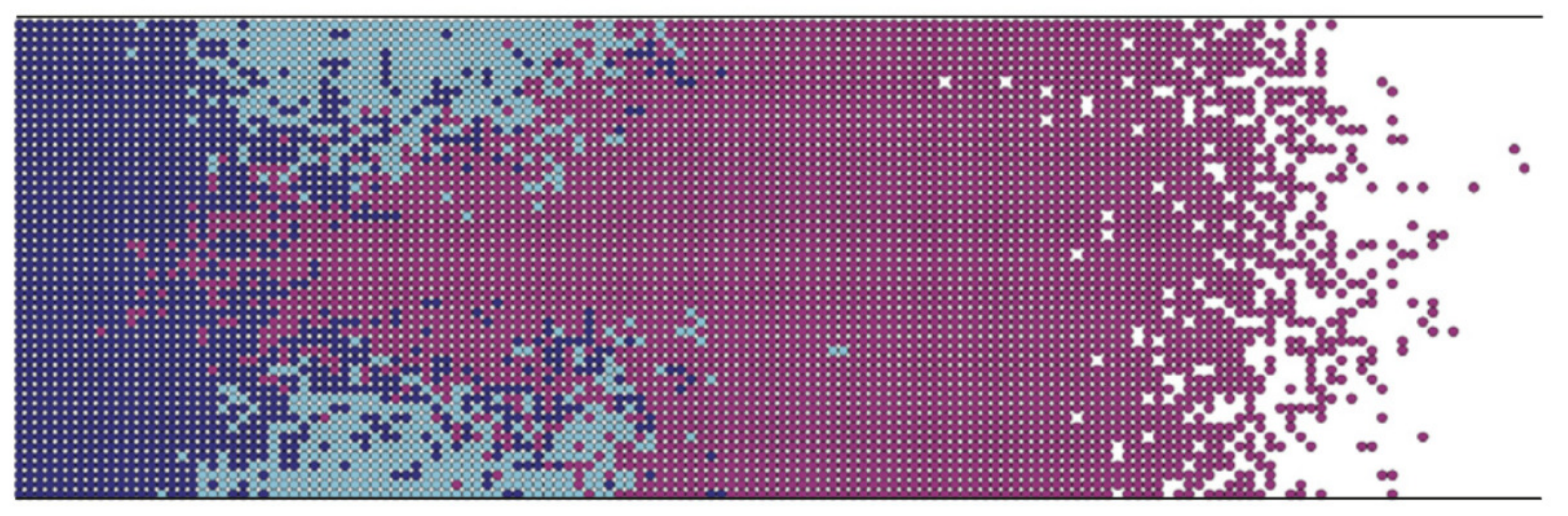}}\\
\subfigure[][]{\includegraphics[scale=0.26]{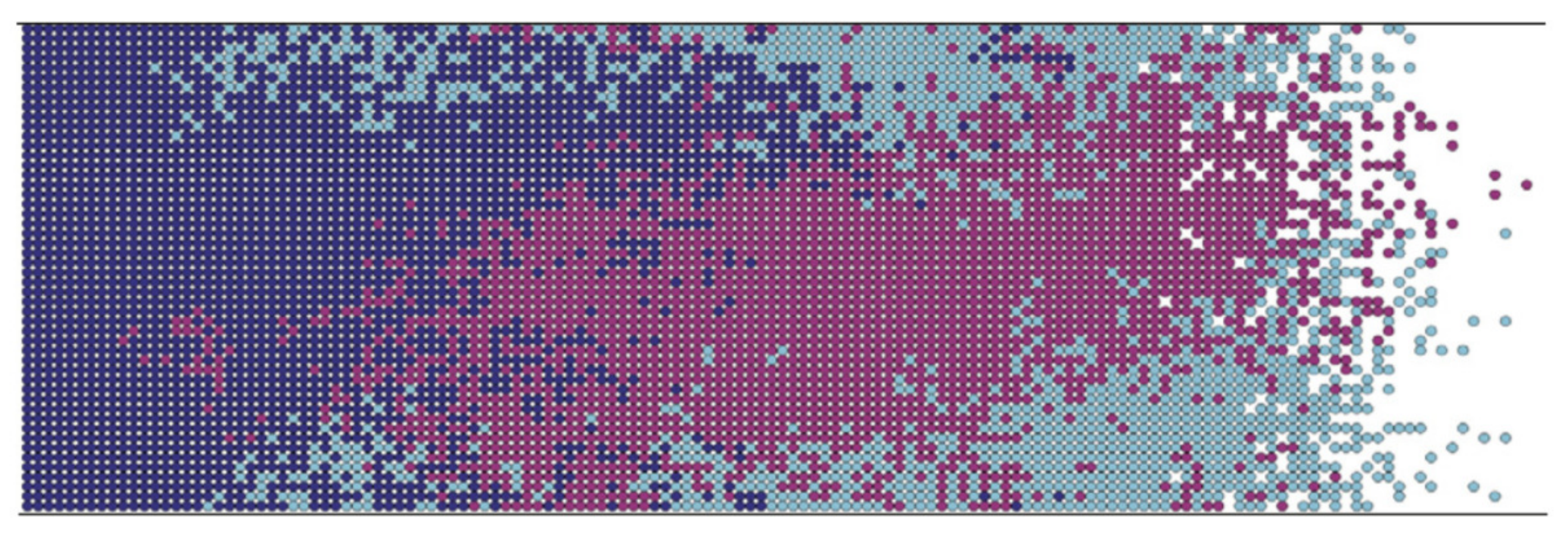}}
\end{center}
{\caption{Two simulations of the ABM of \citet{cheeseman2014std} for an invasion wave with agent lineage tracings. Panel (a) shows the initial condition for all simulations. Panels (b) and (c) show two realisations of the ABM in which the agents populate the domain through movement and proliferation. The largest and second largest single agent lineage tracings (pink and turquoise respectively) and the 498 other agent lineage tracings (all collected together in blue) are illustrated. Reproduced from \citet{cheeseman2014std} with permission of Journal of Theoretical Biology.}\label{fig:cheesman2014}}
\end{figure}

%cheesman2014 
\citet{cheeseman2014std} continued the study of cell proliferation by modelling the spatial and temporal dynamics of different cellular lineages within an invasion wave. The authors defined an ABM model on a two-dimensional lattice with a volume exclusion property. The model is initialised with a population of cells located on one end of the domain (see Figure \ref{fig:cheesman2014} (a)). Agents are labelled according either to their generation number or the lineage to which they belong. The results for the ABM simulation show a clear spatial organisation in the distribution of the agent generation number. However, there is a large individual variability in the spatial distribution of a single agent's linage tracing (see Figure \ref{fig:cheesman2014}). 
In order to reproduce the dynamics of the agent generation number, a set of non-linear diffusion equations are derived from the ABM and are studied in one dimension. In particular, by using the same approach as \citet{simpson2009mss}, \citeauthor{cheeseman2014std} derive a system of conservation of mass equations describing the density profile for each generation $\eta_i(x,t)$. In order to investigate the lineage tracings, they develop a Generation-Dependent Galton-Watson (GDGW) process and they look at the Lorenz curves, which describe the proportional contributions of each of the initial seed cells' lineages to the final population \citep{lorenz1905mmc}.

%------------
The results from agent-based and PDE models show that the invasion wave is composed of spatially regular and predictable generations. As the invasion occurs, the older generations reach a steady state behind the travelling wave. The structure is more apparent when averaging over more realisations of the ABM model. Indeed, both the mean and the variance of the generation number increase linearly with the distance from the location of the initial group of cells. Conversely, the individual agent lineages exhibit a clear asymmetry in their contribution to the final population. A small group of cells (\textit{superstars}) generate the majority of the total population. The GDGW model shows a good agreement with the ABM results. This fact is interesting, since the GDGW process ignores some of the spatial and temporal correlations between individual lineages. Nevertheless this phenomenon can be partially explained by noticing that the competition between agents affects the generation densities, $\eta_i$, which are used in the Galton-Watson process.\\

%Plank2012-2013

As highlighted in Section \ref{sec:methods}, choosing between on- and off-lattice frameworks can lead to differing macroscopic descriptions for excluding ABMs incorporating motility only. The same can be said for models of cell proliferation.  \citet{plank2012mcc,plank2013lfm} provide a pioneering comparison between on-lattice and off-lattice models of proliferating and migrating cells. 

The authors develop a new off-lattice discrete-time ABM for migration and proliferation of cells in a two-dimensional domain. Agents are represented as incompressible circles of diameter $\Delta$ and the total number of agents at time $t$ is denoted $\cN(t)$. At each time step, $\cN(t)$ agents are selected independently and are given the chance to move with probability $P_m$. A selected agent attempts to move a fixed distance, $\Delta$, in a given direction, $\theta$, which is chosen uniformly at random in $[0, 2\pi)$. To include crowding effects, the movement is aborted if any of the other agents lie within a distance $\Delta$ of the line segment which connects the current location to its potential new location. Once all the motility events have been attempted, another $\cN(t)$ agents are selected independently to attempt proliferation events with probability $P_p$. A proliferating agent attempts to divide into two daughter agents whose positions are chosen diametrically opposite each other at distance $\Delta/2$ from the original agent. The chosen proliferation event takes place only if it does not cause overlap between agents.

The authors derive a mean-field approximate ordinary differential equation for the spatially averaged agent density, $C_m(t)$. In doing this, they assume that the population of agents is homogeneous in the domain, which is known to be a poor assumption for large values of proliferation \citep{baker2010cmf, markham2013isc}. 

\citet{plank2012mcc} show a comparison between their off-lattice model and a standard on-lattice model \citep{simpson2010cip} for a spatially uniform initial condition. For small values of agent density, the absence of significant agent-agent interactions leads to similar behaviours for the on- and off-lattice models. However, the two types of approach show a substantial difference for high densities. The on-lattice model leads to a faster growth of the population density. The density eventually reaches the maximum value of unity when all the lattice sites are occupied. Conversely, it is impossible for agents in the off-lattice model to reach the theoretical maximum agent density. This because the agents are not perfectly aligned, and at high densities become jammed in more realistic, yet irregular configurations. The appearance of a natural carrying capacity in the off-lattice model, as opposed to the artificial value induced by a on-lattice approach, suggests the off-lattice model is a more suitable representation for biological applications.

\citet{plank2013lfm} further investigate the behaviour of their on-lattice and off-lattice models in scenarios in which the spatial variability of the cell density profile plays an important role. In particular, they initialise the domain by uniformly populating only the left-hand side region with a relatively low density. They then proceed to study the speed and the shape of the resulting invasion waves in the two models. 

Their results highlight that the two models behave differently behind the invasion front. However, the models' behaviours are similar at the leading edge. These findings are consistent with the previous observation, that the effects of crowding are more evident in the off-lattice model than in the on-lattice model \citep{plank2012mcc}. Specifically, in the invaded region, where the agent density is higher, the population size of the off-lattice model approaches carrying capacity more slowly than the analogous on-lattice model. However, at the front of the invasion, the low value of agent density makes the two models almost indistinguishable. This implies that, in the long-term, the speeds of the invasion fronts in the on-lattice and off-lattice models are the same.

Finally, \citet{plank2013lfm} carry out a least-squares parameter estimation in order to fit the two continuum representations to density profile data simulated using the off-lattice ABM. The solutions for both the on-lattice and off-lattice PDEs are in good agreement with the simulated data and it is hard to distinguish between the two fitted curves. The authors present this as an example that highlights the difficulty of choosing the correct mathematical framework when modelling real experimental data. \\
%This suggests that a robust parameter estimation should be based on aspects of the model which are unaffected by the presence, or absence, of the lattice, such as the speed of the invasion front.

% Yates 2017msr

Typically, cell proliferation is represented in ABMs as a Poisson process with a certain rate $\alpha_p$, which means that each agent's inter-division times are exponentially distributed \citep{mort2016rdm,simpson2007sic,treloar2012vjp,turner2009cbc}. One of the advantages of this approach is that, due to the memoryless property of the exponential distribution, the process can be efficiently simulated using the popular Gillespie algorithm \citep{gillespie1977ess}.

However, assuming exponentially distributed cell inter-division times is not biologically realistic \citep{golubev2016aie}. In particular, the monotonicity of the exponential distribution implies that the most likely time for a cell to divide is immediately after its own creation. Recently, \citet{yates2017msr} have proposed a novel model for incorporating non-exponentially distributed  cell-cycle times, based on a multi-stage scheme. The authors divide the cell cycle into $s$ stages. The waiting time for an agent to pass from the $i$-th phase to the $(i+1)$-th phase is exponentially distributed with rate $\lambda_i$. When an agent exits from the $s$-th stage, it divides into two daughter agents which are initialised in the first phase. This can be summarised by the following chain of reactions:
\begin{equation*}
	X_1\xrightarrow{\lambda_1}X_2 \xrightarrow{\lambda_2} \cdots \xrightarrow{\lambda_{k-1}} X_s \xrightarrow{\lambda_k} 2X_1\; .
\end{equation*}    
In general, the resulting probability distribution of the total cell cycle is a hypoexponential distribution. Note that this general implementation involves $k$ independent parameters ($\lambda_1\dots, \lambda_k$), one for each stage transition and, if $k$ is large, then this may lead to issues of parameter identifiability. In order to reduce the number of free parameters while maintaining the advantage of using a multi-stage representation, \citet{yates2017msr} focus on two cases: the case in which all the transition rates are identical, $\lambda_i= \lambda$ for $i=1, \dots , k$, and the case in which all the transition rates are identical apart from one, $\lambda_i = \lambda$ for $i=1, \dots , k-1$. The corresponding distribution for the total cell cycle in these two cases are the Erlang distribution and exponentially modified Erlang distribution, respectively. The authors show these distributions to be both biologically plausible and computational feasible.

In order to investigate how the multi-stage representation of the cell-cycle affects the total population growth, \citet{yates2017msr} implemented two models (one spatial and one non-spatial). Firstly, they modified the model of \citet{turner2009cbc} in which the spatial position of the cells is considered unimportant. The authors investigate the stage distribution of agents at the steady state for large times.  By writing down a system of ODEs describing the average proportion of agents at each stage $j$, $M_j$, for $j=1,\dots, s$, \citet{yates2017msr} show that the number of agents in each stage is not proportional to the average length of that stage. In particular, for identical transition rates, as the total number of stages becomes large, the proportion of agents at the first stage approaches twice that of agents at the last stage.

The authors also modify the spatially extended ABM of \citet{baker2010fmm} to include their multi-stage proliferation scheme. The agents are initialised uniformly on a lattice with periodic boundary conditions and a standard volume exclusion property. Agents attempt movements with rate $\alpha_m$. To facilitate the comparison with the traditional model with exponentially distributed waiting time of rate $\alpha_p$, the authors consider the case of identical transition rates $\lambda_i=s \alpha_p$ for $i=1, \dots , k$. Hence the average waiting time required for progress through all the $s$ stages is  independent of the number of stages. In particular, the two models, with and without multi-stage scheme, have the same average attempted proliferation waiting time.

\begin{figure}
\begin{center} 
\subfigure[][]{\includegraphics[scale=0.26]{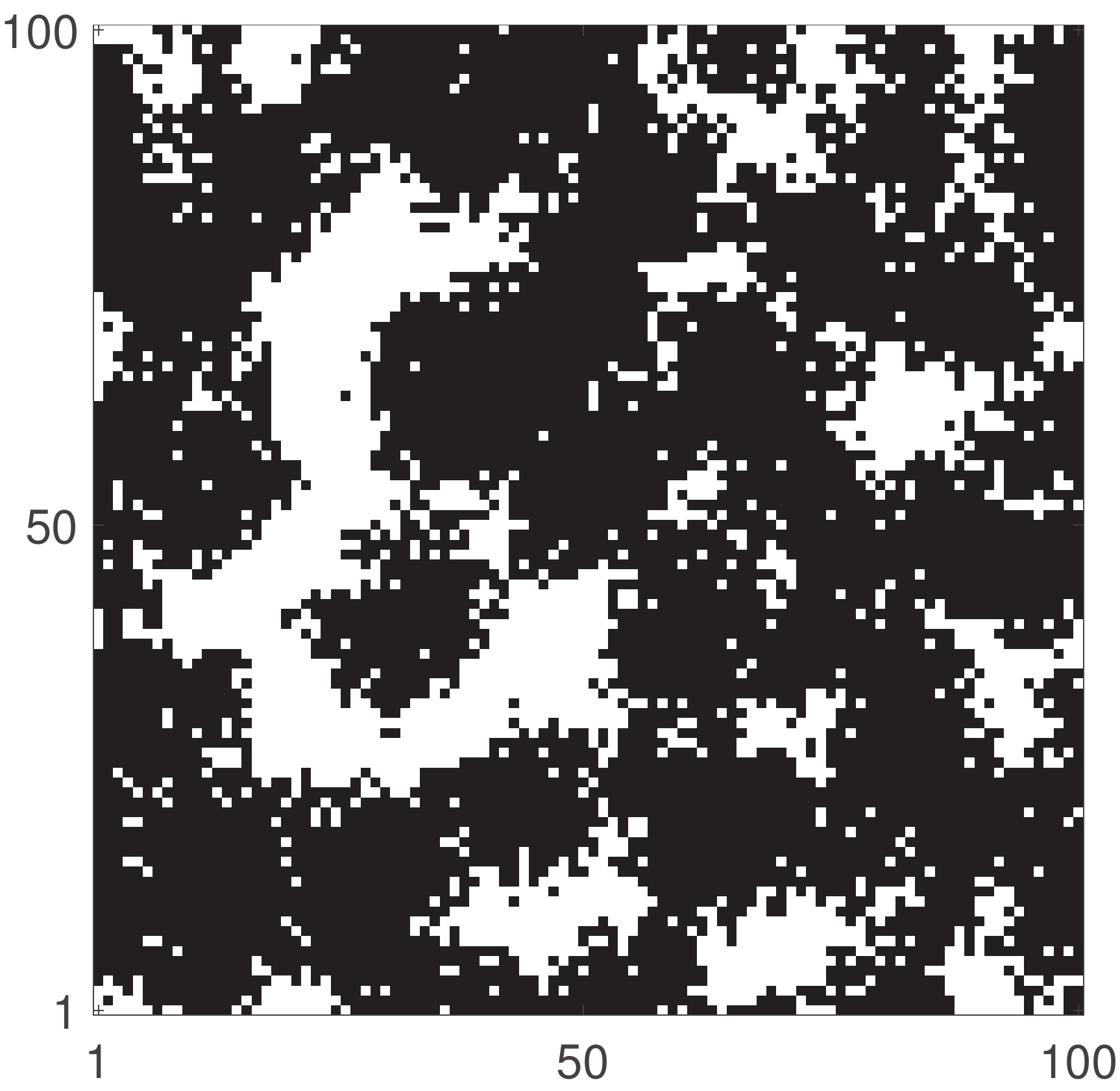}}
\subfigure[][]{\includegraphics[scale=0.26]{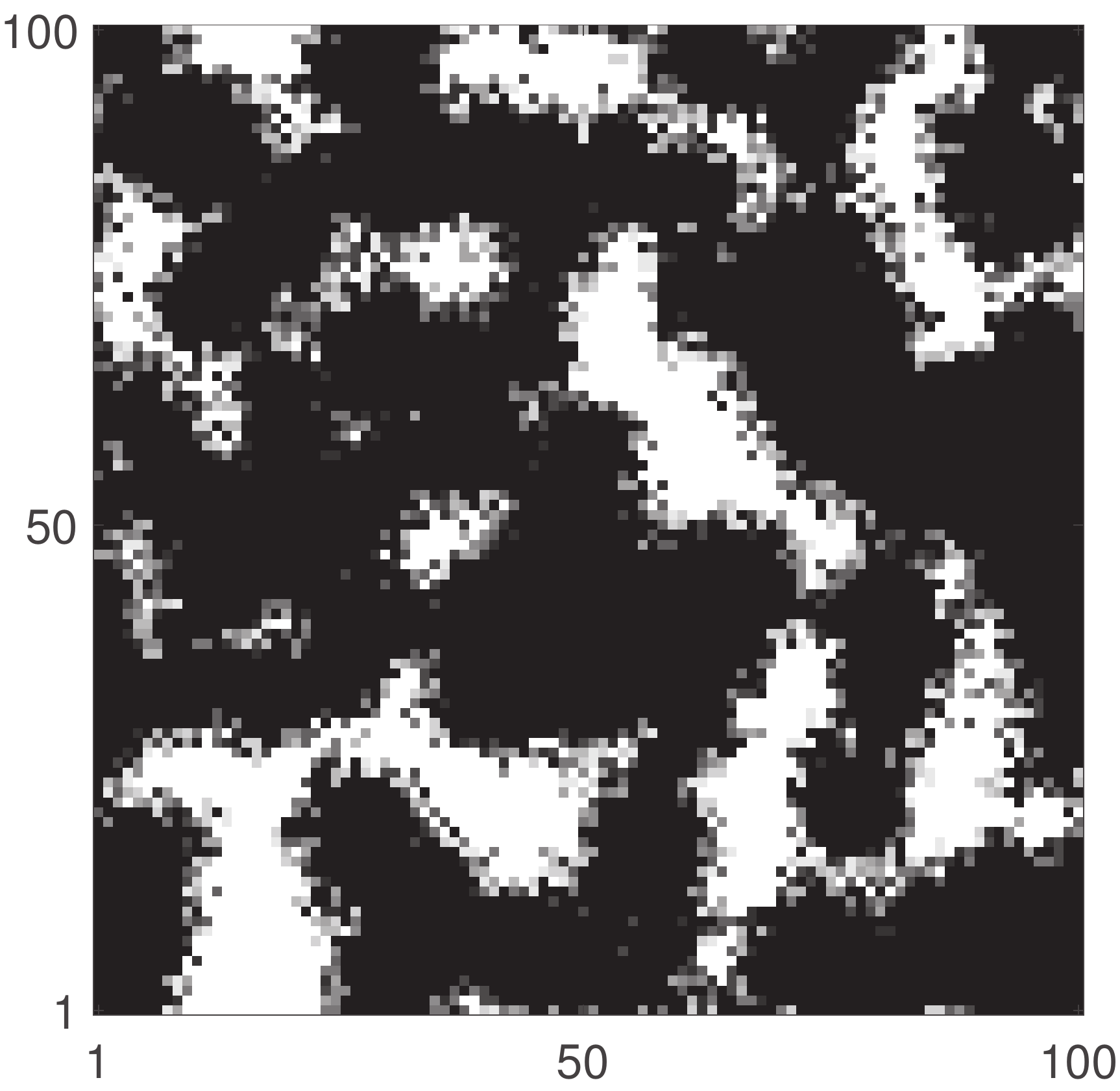}}
\subfigure[][]{\includegraphics[scale=0.26]{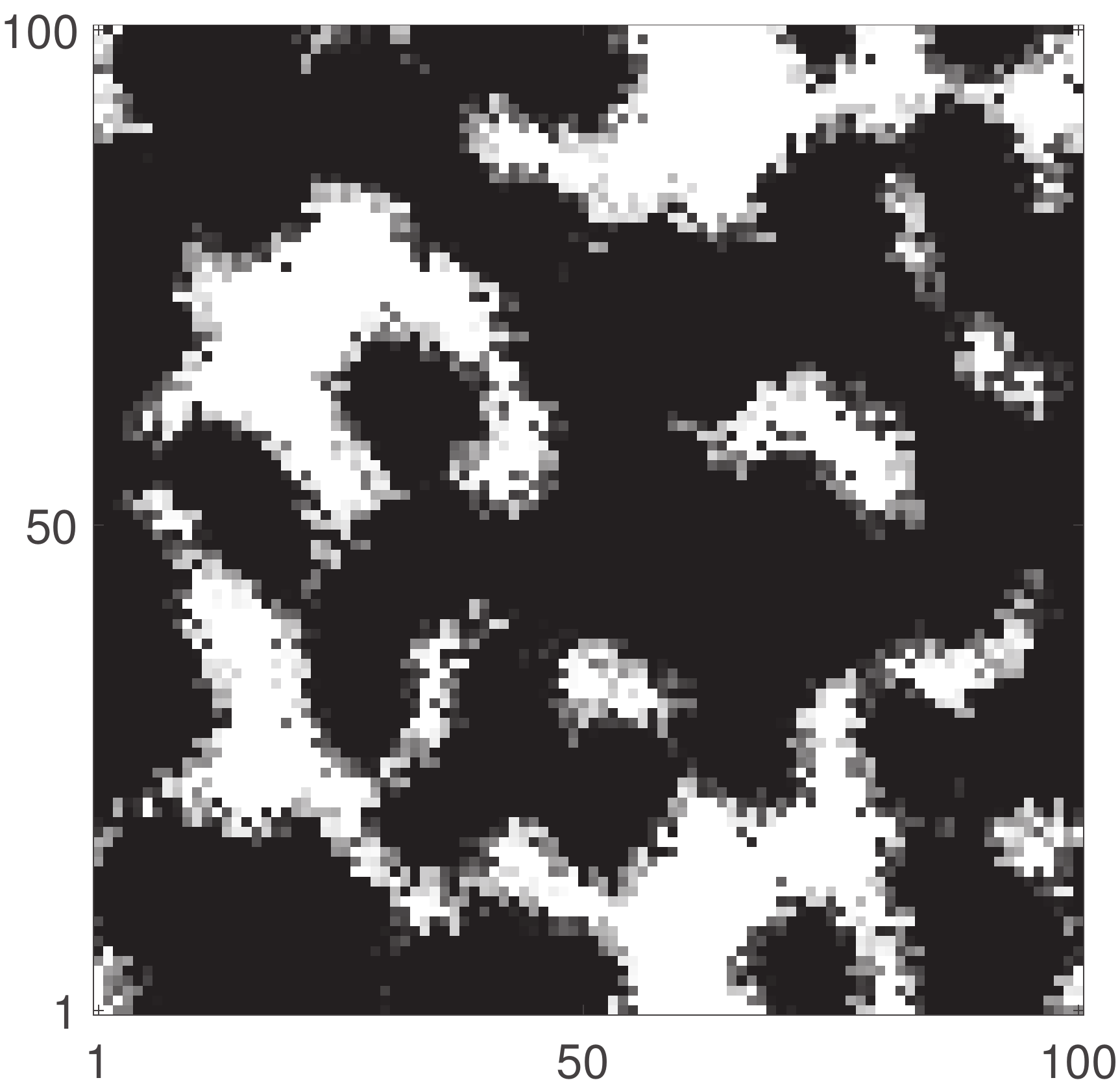}}
\end{center}
{\caption{Influence of multi-stage representation of the cell-cycle on the spatial coverage of cells populating the domain at $t = 10$. Agents in different stages are represented by different shades of grey. Darker shading corresponds to later stages, and white indicates empty sites. Parameters are $\alpha_m=1$, $\alpha_p=1$. The initial condition consists of 100 agents uniformly distributed on the lattice. Panel (a), (b) and (c) correspond to the cases $s = 1, 10, 100$, respectively. Increasing $s$ causes an increase in the total cell density in this scenario at time $t=10$. Reproduced from \citet{yates2017msr} with permission of Bulletin of Mathematical Biology.}
\label{fig:yates2017msr}}
\end{figure}

The simulations of the ABM highlight that, for low-density colonies, the multi-stage scheme leads to a slower population growth compared to the model with exponentially distributed proliferation. However, under the realistic assumption that agents that fail to proliferate due to crowding remain at the last stage (as opposed to returning back to the first stage of the cycle), a clear proliferating rim of (grey) cells can be seen with the bulk of cells being kept at stage $s$ (see Figure \ref{fig:yates2017msr}). Agents located in this rim, re-attempt  division after aborted events more quickly than in the single-stage cell cycle model since they only have to wait a time $\alpha_p /k$ on average. Therefore, the effective average inter-division time for cells with a multi-stage cell cycle at the proliferating rim of a cluster decreases in comparison with cells to a single-stage cell cycle. 

--------------------------------------------
% Cell-cell interaction
\subsection{Cell interactions}
\label{sec:cell-cell_interaction}
Cells can undergo a great variety of interactions in response to their environment and more specifically, their neighbouring cells \citep{cai2006mdg,dworkin1985cim,trepat2009pfd,tambe2011ccg}. Typically, these can be divided in two broad categories: \textit{indirect} and \textit{direct} \citep{othmer1997abc}. 

When cells have the ability to detect the presence of an external signal and change their behaviour according to its concentration, we call these indirect interactions. This can affect the movement speed or the turning rate (\textit{kinesis}) \citep{cai2006mdg}, it may induce a directional bias (\textit{taxis}) \citep{painter2002vfq} or it may involve a combination of these \citep{erban2004fic}. In general, the  external signal can depend on the cells themselves. For example, it can be directly produced by a moving cell as it moves, forming a trail behind it, as it has been shown for Myxobacteria \citep{dworkin1985cim}. Notice that this type of interaction does not involve direct cell-cell contact, communication is mediated only through the external signal. Direct interactions are based on the ability of cells to detect and interact with surrounding cells. These include contact interactions such as volume exclusion \citep{abercrombie1979cim} and adhesion-repulsion \citep{trepat2009pfd,tambe2011ccg}.

The introduction of these interactions in models of cell migration is crucial, since they often constitute the driving forces that generate spatial structure at the population scale. In this Section we focus our attention on a set of important cell interactions: the indirect response to a chemical signal, direct adhesion-repulsion forces between cells and the ability of cells to push and pull each other.

%---------------------------------------------------
\subsubsection{Chemotaxis}
\label{sec:external_signalling}

%othmer1997abc
\citet{othmer1997abc} derive a general class of PDEs from a set of ABMs incorporating a simple signal following mechanism. The aim of their paper is to determine whether stable bacterial aggregation can arise as result of a simple chemotactic response. 

The main ABM considered in the paper consists of a system of non-interacting agents which move on a one-dimensional lattice with unit step size, according to a continuous-time, nearest-neighbour random walk. As time evolves, agents produce an attracting signal, $w$, which sits on an embedded lattice of half the step size. 

The authors distinguish three types of model depending on the information used to compute the transition rates. These are the \textit{local}  model, if the transitions depend only on the signal concentration at the current site, $w_n$; the \textit{barrier} model, in which cells only sense the value of the signal at the short-range, nearest-neighbour sites, $w_{n\pm 1/2}$; and the \textit{gradient-based} model, in which the transition rates depend on the long-range, nearest-neighbour differences, $w_n-w_{n\pm 1}$. In the barrier and gradient-based cases an additional distinction is made between normalised and unnormalised transition rates. For unnormalised rates, larger values of $w$ lead to a faster total movement rate, whereas, for normalised rates, the average waiting time at each site does not depend on the signal concentration. In all cases, by using a limiting argument, the authors derived the corresponding continuous diffusive approximations for the average agent density. 

\citeauthor{othmer1997abc} carry out a stability analysis of the solutions of the continuous barrier model with normalised transition rates. The analysis is repeated with three different models for the evolution of the signal: linear growth, exponential growth and saturating growth. The results show that linear growth can only lead to uniform agent density profiles, even when starting from a single peak in the initial distribution of agents (\textit{collapse}). On the other hand, when the signal grows exponentially, the system tends to converge to a single peaked distribution (\textit{blow-up}). Finally, when a saturation in the production of $w$ is assumed, stable aggregation can occur as a result of the interplay between the production of the signal and the short-range chemotactic response. The formation of such aggregates strongly depends on the decay rate of the signal and on the initial condition. In particular, a large decay rate or a small initial value of $w$ can lead to aggregation or blow-up, whereas, in the absence of signal decay, the agent density will eventually collapse to uniformity.

%%Erban 2004
%\citet{erban2004fic} continue the study migrating cells which respond to an external signal. The authors provide a series of  The analysis of \citeauthor{erban2004fic} is based on velocity jump processes for describing the motion of individuals such as bacteria, wherein each individual carries an internal state that evolves according to a system of ordinary differential equations forced by a time- and/or space-dependent external signal. In the problem treated here the turning rate of individuals is a functional of the internal state, which in turn depends on the external signal. Using moment closure techniques in one space dimension, we derive and analyse a macroscopic system of hyperbolic differential equations describing this velocity jump process. Using a hyperbolic scaling of space and time, we obtain a single second-order hyperbolic equation for the population density, and using a parabolic scaling, we obtain the classical chemotaxis equation, wherein the chemotactic sensitivity is now a known function of parameters of the internal dynamics. Numerical simulations show that the solutions of the macroscopic equations agree very well with the results of Monte Carlo simulations of individual movement.

%-----------------
% Adhesion-repulsion
\subsubsection{Adhesion-repulsion}\label{section:adhesion_repulsion}

%Anguige 2009

\citet{anguige2009odm} investigate the emergence of aggregation through the mechanisms of cell-cell adhesion and volume exclusion. They predict, under their model, that the spontaneous aggregation of cells is not possible if the initial cell density throughout the domain is too low, regardless of the intensity of the adhesion force.

To describe cell migration, \citet{anguige2009odm} consider a discrete-space, continuous-time random walk model on the unit interval. Multiple agents can occupy the same compartment up to a carrying capacity, $S$, and they move at random with given rates to one of the two nearest neighbour compartments. In order to include volume exclusion in the model, the transition rates in both directions are decreased linearly with the density of the target site. If a site is fully occupied, no transitions into that site can occur. To incorporate adhesive forces, the rate of moving in a particular direction is decreased linearly with the density of the adjacent site in the opposite direction, in proportion to an \textit{adhesion parameter}, $\beta$.

By Taylor expanding and neglecting terms of third (or higher) order it is possible to derive the diffusive limit of the discrete model as the number of lattice sites goes to infinity. This macroscopic model is a non-linear diffusion equation as equation \eqref{eq:diffusion_off_lattice_excluding3} with 
\begin{equation*}
	D(\bar{C})=D\LRs{3 \beta\LR{\bar{C}-\frac{2}{3}}^2+1-\frac{4}{3}\beta } \, ,
\end{equation*}
for which the homogenous density profile is the only steady state. \citet{anguige2009odm} find that there is a critical value of the adhesion parameter $\beta_c=0.75$. For the low-adhesion regime ($\beta<\beta_c$), pattern formation is not possible in either the discrete or continuum descriptions. 

When the adhesion force is large ($\beta>\beta_c$), the results of the discrete model show complex behaviours such as pattern formation and spatial oscillations in density. Patterns in the discrete model are found to be transient and metastable. All clusters which arise eventually coalesce to single stable aggregate or to two aggregates separated by a single trough. For the macroscopic PDE representation, the authors identify an interval of unstable values of density for which the diffusion coefficient of the non-linear PDE takes negative values making the continuum model ill-posed. 

To overcome the issue of the ill-posedness in the PDE and to obtain a reasonable continuum description, \citet{anguige2009odm} propose to include more terms from the Taylor expansion of their discrete model. These higher order corrections result in a fourth-order diffusion equation reminiscent of the Cahn-Hilliard equation \citep{sun2000dci}. The presence of a viscosity terms in the revised equation allows the authors to prove it has a well-posed initial value problem.

%Thompson 2012

\begin{figure}
\begin{center} {\includegraphics[scale=0.43]{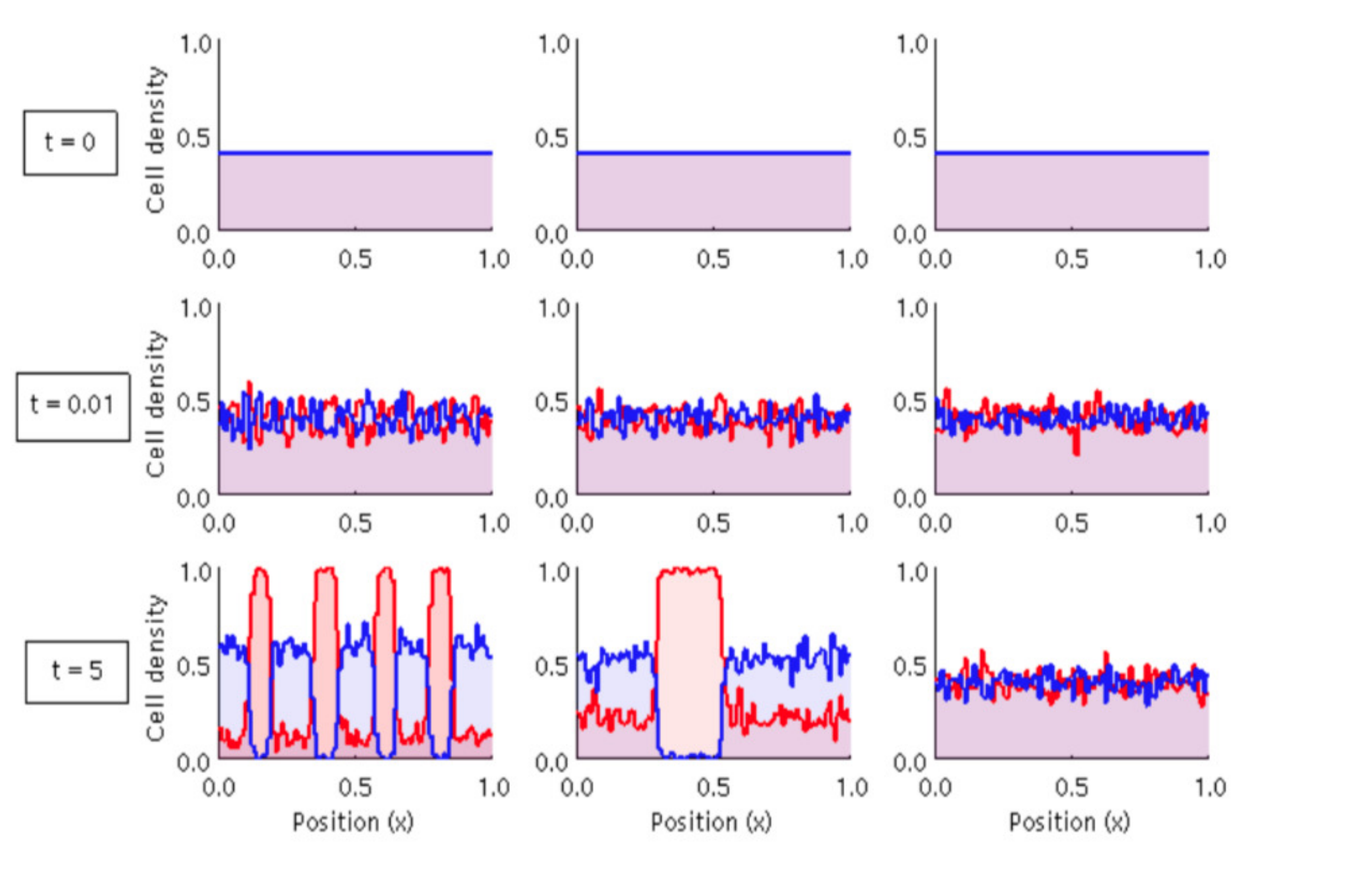}
\caption{Simulations of the two-species ABM of \citet{thompson2012mcm} in one dimension. Blue lines denote the density of the type $A$ agents and red lines the density of type $B$ agents. The domain is initialised with homogeneously distributed agents of type $A$ with density $0.4$. Depending on the intra- and inter- adhesion forces, three qualitative configurations are possible: $\beta_{A,A}=0.1$, $\beta_{B,B}=0.9$ and $\beta_{A,B}=\beta_{B,A}=0$ lead to cell sorting (left), $\beta_{A,A}=0.1$, $\beta_{B,B}=0.9$ and $\beta_{A,B}=\beta_{B,A}=0.2$ lead to engulfment (middle) and $\beta_{A,A}=0.1$, $\beta_{B,B}=0.9$ and $\beta_{A,B}=\beta_{B,A}=0.6$ produce a mixing configuration (right). Reproduced from \citet{thompson2012mcm} with permission of Bulletin of Mathematical Biology.}\label{fig:thompson2014}}
\end{center}
\end{figure}

\citet{thompson2012mcm} continue the work of  \citet{anguige2009odm} on the modelling of cell-cell adhesion at multiple scales. The authors consider a modification of the ABM of \citet{anguige2009odm} in order to study the interactions with a second species of cell. In this new model, agents are of two types, $A$ and $B$, and the there are four different coefficients ($\beta_{A,A},\beta_{B,B},\beta_{A,B}$ and $\beta_{B,A}$) governing the intra- and inter-species adhesion forces, respectively. \citet{thompson2012mcm} show that the model is capable of reproducing three configurations (\textit{complete sorting}, \textit{engulfment} and \textit{mixing}),
which have been observed previously both in experiments and continuous models \citep{armstrong2006cam}. In Figure \ref{fig:thompson2014} we report an example of these three configurations depending on the choice of the adhesion parameters. 
Specifically, if inter-species adhesion is small and intra-species adhesion is large the system reaches a structured configuration (cell sorting) in which agents of the same species tend to cluster together (see left column of Figure \ref{fig:thompson2014}). If inter-species adhesion is small and one of the intra-species adhesions is larger than the other, then engulfment of the species with the larger adhesion occurs (see middle column of Figure \ref{fig:thompson2014}). Finally, when inter-species adhesion is large and the intra-species adhesion is small, then mixing occurs, \textit{i.e.} agents of the two species are uniformly distributed across the domain (see right column of Figure \ref{fig:thompson2014}).

% Binny2015 
Recently, \citet{binny2015smd} have studied a one-dimensional off-lattice ABM (later extended to two-dimensions \citep{binny2016ccb}) in which both the rate and the direction of the movement of each agent are influenced by the configuration of the neighbouring agents. In their model an agent, $n$, moves with a rate, $\alpha_n$, which is computed by
\begin{equation*}
\label{eq:motility_binny}
	\alpha_n(t)=\max \LRb{0, \tilde{\alpha} +\sum_{m\ne n}w \LR{c_m(t)-c_n(t)}}\, ,
\end{equation*}
where $\tilde{\alpha}$ is an intrinsic motility rate, independent of the other agents' positions, and $w(z)$ is a kernel function weighting the strength of interaction between agents positioned a distance $z$ from each other. \citet{binny2015smd} consider the case in which the kernel is Gaussian:
\begin{equation}
\label{eq:kernel_binny}
	w(z)=\gamma_r \exp\LR{-\frac{z^2}{\sigma_r^2}}\, ,
\end{equation} 
where $\gamma_r$ and $\sigma_r$ determine the intensity and the range of the interaction.  Notice that if $\gamma_r>0$, agents tend to move more often when they are surrounded by other close  agents. Conversely, for $\gamma_r<0$, the presence of close neighbours inhibits the motility. When a movement even takes place, the moving agent performs a jump of random length and direction. The length of jump is drawn at random from a Laplace distribution (\textit{i.e.} short steps are more likely to be taken than long jumps), independent of the other agents' positions. For the direction of the movement, the authors incorporate a directional bias, $b_n(t)$, such that the presence of neighbouring agents can affect the final direction. They defined the directional bias for an agent, $n$, as
\begin{equation*}
	b_n(t)=\sum_{m\ne n}v' \LR{c_m(t)-c_n(t)} \, ,
\end{equation*} 
where $v(z)$ is a Gaussian kernel function of the same form of equation \eqref{eq:kernel_binny} with intensity and rage given by $\gamma_b$ and $\sigma_b$, respectively. If $\gamma_b>0$, agents are biased to move away from highly concentrated regions, whereas if $\gamma_b<0$, agents tend to move towards one another.
% as may occur for slime-following scenarios \citep{othmer1997abc}.

To study the resulting spatial structure from a deterministic prospective, \citet{binny2015smd} develop a population-level model in terms of the first two spatial moments of the ABM. By using the Kirkwood superposition approximation (see Section \ref{sec:HOA} for a discussion about alternative approaches to the mean-field approximation), the authors derive a closed system of two ordinary differential equations (ODEs) describing the evolution of the average density of single agents, $Z_1$ and of pairs of agents separated by a distance $\xi$, $Z_2(\xi)$. 

In order to test the accuracy of their deterministic representation against the numerically simulated individual-based model, the authors compare the pair correlation functions (PCFs) \citep{illian2008sam} of the ABM with the PCF predicted by the spatial moment model. The pair correlation function can be expressed in terms of the first two spatial moments as 
\begin{equation}
	PCF_{SM}(\xi)=\frac{Z(\xi)}{Z_1^2}\, .
\end{equation} 
Notice that $PCF_{SM}\equiv 1$ corresponds to the case of absence of spatial structure, whereas $PCF_{SM}(\xi)> 1$ denotes evidence of spatial correlation and  $PCF_{SM}(\xi)< 1$ implies anticorrelation, at length $\xi$. 

Their findings show that the moment model provides a good approximation of the spatial structure predicted by the ABM for a wide range of parameter choices. The moment model underestimates or overestimates the spatial structure only for large values of intensity $\gamma_m$ and $\gamma_b$. \citet{binny2015smd} suggests that this is due to higher order interactions which are not incorporated in the deterministic representation.

\citet{binny2016ccb} continue the study of agent-agent interactions by extending the previous model to two dimensions and including agent proliferation and death. In the new model, agents are allowed to proliferate with a rate which is dependent on their neighbourhood and defined similarly to $\alpha_n(t)$ of equation \eqref{eq:kernel_binny}. The rate of agent death is assumed independent of the location of other agents. Following \citet{binny2015smd}, the authors derive a deterministic model for the first two spatial moments of the ABM. After comparing the performance of four different types of moment closure, the authors select an asymmetric power-2 closure as the most appropriate \citep{murrell2004mcp,raghib2011mme} and they adopt it throughout the paper. The results show that the good agreement between the stochastic and deterministic models, found in \citep{binny2015smd} is preserved when proliferation and death are incorporated in the model. Unsurprisingly, the presence of non-uniform spatial structure leads to both population growth and the average cell density differing significantly from the predictions of a mean-field model that ignores agent-agent interactions.

Finally, \citet{binny2015smd, binny2016ccb} investigate the role of neighbour-dependent movement and proliferation in the formation of spatial structure. In general, movement tends to break up spatial structure because new agents can move out of clusters generated by short-range proliferation \citep{baker2010cmf}. When movement is biased away from neighbouring agent, $\gamma_b<0$, this effect appears more clearly, since agents undergo directed movement out of clusters. The same dispersal of spatial structure is found for the neighbour-dependent-inhibition of the motility rate, but it occurs in a less efficient way. The authors also highlight that the crowding-induced inhibition of proliferation is capable of counteracting the formation of clusters due to short-range proliferation.  

%--------------------------------------------------------------------
% Pulling-pushing
\subsubsection{Pushing and pulling}

\citet{yates2015ipe} study how cell-cell pushing affects cells' dispersal and proliferation by considering on- and off-lattice ABMs. The authors begin by considering a simple ABM incorporating exclusion on a two-dimensional lattice, as in \citet{simpson2007sic}. They modify the basic model in order to incorporate the ability of cells to push their neighbours \citep{ewald2008cem,schmidt2010icm}. Pushing is implemented according to four different mechanisms with variable degrees of complexity and realism. 

Consider an agent at position $(i,j)$ which attempts a movement into an occupied site to its right. In the most basic pushing mechanism, the authors let the moving agent push the agent at position $(i+1,j)$ into the site $(i+2,j)$, with a given probability, $H$, with the pushing agent taking the pushed agents' original position, $(i+1,j)$. The pushing is effective only if the target site of the pushed agent, $(i+2,j)$, is empty. The entire movement is aborted otherwise. For the next two mechanisms, the assumption on the location of the target site of the pushed agent is relaxed. For one of these mechanisms the pushed agent is allowed to move into each of its empty neighbouring sites with equal probability. If all neighbouring sites of the pushed agent are occupied, then the whole event is aborted. In the other mechanism, the target site is chosen uniformly at random from amongst the pushed agent's three nearest neighbours, but if the chosen site is found to be occupied then the entire event is aborted. In the final case considered, the moving agent can attempt to push up to $K$ other agents in a straight line in a chosen direction. 

In all the cases considered, the authors derive diffusive PDE approximations for the total agent density. For the basic pushing mechanism, for example, the resulting non-linear PDE takes the form of equation \eqref{eq:diffusion_off_lattice_excluding3} with density dependent coefficient given by
\begin{equation}
\label{eq:push}
	D(\bar{C})=D\LR{1+4 H \bar{C} }\, .
\end{equation}
  In most scenarios, the comparison between the ABM and the corresponding PDE is good. However, the more complicated pushing mechanisms, such as the linear pushing of multiple agents, are found to introduce strong spatial correlation in the occupancies of adjacent sites. Since the derivation of the continuum models is based on a mean-field approximation, the presence of such spatial correlation leads to a divergence between the ABM and its deterministic counterpart.

\citet{yates2015ipe} also investigate the effect of introducing pushing in an off-lattice set-up by modifying the one-dimensional model of \citet{dyson2012mli} (see Section \ref{sec:MF}). In this model, a pushing event is attempted with probability $H$ when a movement would lead to an overlap of agents, which are represented as interval of length $2R$. The moving agent can displace the pushed adjacent agent far enough for it to complete its movement, but only if this will not produce an overlap with a third agent. In this case the entire movement would be aborted. By using a similar approach to that of \citet{dyson2012mli}, a non-linear diffusive PDE of the form \eqref{eq:diffusion_off_lattice_excluding3} is derived with
\begin{equation}
D(\bar{C})= D \LR{1+2R(2-H) \bar{C}}\, ,	
\end{equation}
 and the numerical solution of the PDE shows good agreement with the averaged density of the ABM. 

In all the cases considered, both on- and off-lattice, \citet{yates2015ipe} find that the introduction of pushing behaviour leads to density-dependent diffusion coefficients in the macroscopic descriptions. For the on-lattice models, pushing leads to a faster diffusion, due to the larger number of possible movements for each agent in comparison to the simple exclusion model. However, in the off-lattice scenario, increasing the tendency to push leads to a decreasing diffusivity. To explain this behaviour, \citet{yates2015ipe} underline that when a pushing event occurs in the off-lattice model, the moving agent and the pushed agent end up being adjacent, even if they were not in contact prior the movement. This aggregation effect leads to a slower dispersal of agents corresponding to a lower diffusion in comparison to the non-pushing case.\\

There is much evidence in the literature to suggest that cells can pull each other while they are undergoing collective migration 
\citep{bianco2007tdm,ghysen2007llm}. \citet{chappelle2018pmc} explore the effect of allowing cell-cell pulling in both on- and off-lattice models of cell migration.

Firstly, the authors specify a simple on-lattice excluding ABM in which agents are capable of pulling each other. For example, if an agent is moving to the right and another agent is occupying the neighbouring site to its left, the moving agent will pull the neighbouring agent rightwards, with probability $\omega$. 

By writing down the occupancy master equation of the model and taking appropriate limits, the authors obtain a non-linear diffusion equation for the average agent density, which takes the form of equation \eqref{eq:diffusion_off_lattice_excluding3} with
\begin{equation*}
	D(\bar{C})=D\LR{1+3 \omega \bar{C}^2 }\, .
\end{equation*}Pulling is found produce a quadratic density-dependence in the diffusion coefficient of the macroscopic PDE. A comparison with the corresponding model for agents which are able to \textit{push} each other (see equation \eqref{eq:push}) highlights that the density dependence of diffusivity due to pushing is larger than in the pulling case. 

In order to study the effect of pulling in off-lattice models, \citet{chappelle2018pmc} use a similar approach to that of \citet{dyson2012mli} (see Section \ref{sec:MF}).  The one-dimensional model in \citet{dyson2012mli} is extended by introducing a \textit{pulling distance} proportional to the moving distance, $l=kd$, such that, if an agent is chosen to move in a given direction and there is another agent whose centre is within distance $l+2R$  in the opposite direction, both agents move in the given direction. The corresponding PDE is of the form of equation \eqref{eq:diffusion_off_lattice_excluding3} with 
\begin{equation*}
	D(\bar{C})=D\LRs{1+4R\bar{C}\LR{1-\omega k} } \, ,
\end{equation*} 
which predicts that the effect of pulling is to decrease the effective diffusivity in the off-lattice model, leading to a slower dispersion at the macroscopic level. The apparent contradiction with the on-lattice counterpart is explained in the paper by using a similar argument to \citet{dyson2012mli}.

%--------------------------------------------------------------------
% Growing domain
\subsection{Growing domains}
\label{sec:growing_domain}
Considering cell motility on a purely stationary domain may often be unrealistic for biological processes: cell growth, division and movement itself will cause the size of the tissue to change dynamically in many biologically plausible situations \citep{rogulja2005rcp,wolpert2015pdev}. Surprisingly, it is only in the last decade that attention has been given to investigating how this phenomenon affects models at multiple scales. \citet{baker2010fmm} represents a pioneering example of the incorporation of domain growth into models of cell migration.

% Baker 2010

For the motility scheme, \citet{baker2010fmm} follow the approach of \citet{othmer1997abc}. At the individual-cell-level they consider a continuous-time, discrete-space ABM. Agents are initialised on a one-dimensional lattice with zero-flux boundary conditions. They have the ability to sense the concentration of a signal profile, $w$, at the their current lattice site and at their two nearest neighbour sites. Agents can jump to immediately neighbouring lattice sites. These jumps are regulated by two transition rates that are linear combinations of the signalling molecule concentration at the agent's current and immediately sites. The authors consider four types of transition rate comprising different linear combinations of these concentrations (\textit{local, non-local, average} and \textit{difference}). If no signal profile exists then agents can either diffuse randomly (jumping with constant rates independent of their position) or implement density-dependent transition rates. For each case, using a master equation, the authors derive population-level descriptions, which comprise an advection-diffusion PDE for the average agent density, $C(x,t)$, as a function of position, $x$, and time, $t$.

The inclusion of domain growth in the model is carried out in two phases. Firstly, \citet{baker2010fmm} consider the case of exponential growth. On the individual-level this implies that each lattice site divides at a constant rate per unit time. When the site divides, a new daughter site is added adjacent to the parent site. The agents in the parent site are divided between these two sites according to a symmetric distribution and the sites to the right of the daughter site (and their contents) are all shifted one site's width to the right. A PDE description of cell density is derived both from a phenomenological, continuum perspective, using a conservation of matter argument \citep{crampin1999rdg}, and from the master equation of the ABM. In order to derive the corresponding PDE from the master equation, the authors use a moment closure approximation on the number of agents in each lattice site. This approximation is based on the assumption that movement occurs on a faster time-scale than domain growth.

Secondly, the authors consider a more general type of domain growth which is density-dependent. As  before, upon being chosen to undergo a growth event, a lattice site is divided into two daughter sites, but now with a density-dependent rate, $f\left(C(x,t)\right)$. A continuous approximation is derived using a further moment closure assumption - that the mean of a non-linear function of the agent density can be expressed as the same non-linear function of the mean agent density. The authors prove that domain growth is linear for the case of linear density-dependence on domain growth, while it must be evaluated for numerically in the general case. 

The paper shows comparisons between the simulations of the individual-based models and the numerical solution of the PDEs. In addition, the predicted macroscopic domain length is compared with the average value from simulations. Results are displayed for the cases of the non-growing domain, constant growth, linearly density-dependent growth and quadratically density-dependent growth. In each case, the results show a good agreement between the discrete model density and the continuous approximation. However, the predicted value of the domain length in the continuum approximation is underestimated in comparison to the true value from the stochastic model in the case of the non-linear growth model. 

The last part of the paper incorporates a signalling profile which agents can sense and respond to. Specifically, an exponentially decreasing concentration of signalling molecule, $w$, is placed in the domain and agents are assumed to interact with it, both via the local and non-local schemes. Results show good agreement between the simulation and the deterministic prediction for both linear and logistic domain growth. As a last application, \citet{baker2010fmm} consider linearly density-dependent growth with both local and non-local sensing again demonstrating good agreement between the two modelling regimes. 

%Yates 2012

\citet{yates2012gfm} generalise the existing ABMs of \citet{baker2010fmm} to a non-uniform lattice allowing the process of cell division on the underlying domain to be modelled in a more realistic manner. The domain is allowed to extend in a manner which can be made arbitrarily close to continuous, as opposed to the discrete increments by which the domain is extended in \citet{baker2010fmm}. 

The authors consider two distinct ways to discretise a one-dimensional domain in a non-uniform way: the \textit{Voronoi partition method} and the \textit{interval-centred method}. In the first case, the agent positions, $c(n,t)$ for $n=1, \dots , \cN$, are chosen first and the edges of the intervals are the bisectors of these points. In the second case, the edges of the intervals are specified first and the agents lie at the centre of these intervals. As in \citet{baker2010fmm}, no exclusion property is implemented and multiple agents can occupy the same lattice site. 

In the first part of the paper the domain is fixed and agents move following a position-jump process on the irregular lattice. The transition rates of an agent in interval $i$ are computed from the mean first passage times that a  Brownian particle, initialised at $x_i$, takes to hit one of the neighbouring particle positions ($x_{i-1}$ or $x_{i+1}$). Using the master equation for agent densities, \citet{yates2012gfm} derive a macro-scale description of the model in the diffusive limit. For the Voronoi partition, the density evolves according to the diffusion equation and the comparison between the simulations and the PDE shows a good agreement. In the case of the interval-centred partition, the agreement between the individual-level behaviour and the population-level diffusion equation is poor since the transition rates are not inherently linked to the size of the intervals. 

The remainder of the paper focuses on the introduction of domain growth to the models of agent migration using the Voronoi partition. \citet{yates2012gfm} start by defining a deterministic scheme for domain growth. Every time an agent moves, every interval grows an amount proportional to its length and proportional to the time step between movement events. When an interval reaches a threshold length, it splits into two daughter intervals. The boundaries of such daughter intervals are chosen in order to preserve the Voronoi property of the domain partition. Each agent is redistributed to a new interval with probability proportional to the overlap between the new intervals and the old intervals. The authors also consider an alternative growth mechanism in which, with a give rate, an interval is chosen at random to grow with probability proportionally to its length. When a \textit{growth event} occurs, both the selected interval and an adjacent interval grow in order to preserve the Voronoi property. For both growth schemes, a population-level description can be obtained through the master equation for the densities. The resulting PDEs are of the same form as in \citep{baker2010fmm}. 

Finally, a series of comparisons between simulations and PDEs confirm the good agreement between the microscopic and macroscopic models for the Voronoi domain partition, but a poorer agreement for the interval-centred domain partition, as expected. \\

%Thompson

In addition to their investigations into cell-cell adhesion and volume exclusion on a static domain (see Section \ref{section:adhesion_repulsion}) \citet{thompson2012mcm} study the interplay of the same properties on a growing domain using a modification of the ABMs of \citet{baker2010fmm} and \citet{yates2012gfm}. The authors incorporate a partially excluding property into the model of domain growth of \cite{baker2010fmm} by allowing the carrying capacity of each compartment to be proportional to its size (see Figure \ref{fig:schematics_compartment} for a schematic illustration). When a growth event occurs, a compartment is chosen at random and its carrying capacity is increased by unity. Commensurately, its size it is also increased. When a compartment reaches a pre-defined threshold, it is split into two compartments each with carrying capacity set to half of the previous value. To compensate for the unequal compartment sizes generated by growth, the transition rates are amended as in \citep{yates2012gfm}. A population-level description of the model is derived, but it is only valid for short time-scales, specifically until the first splitting event occurs.

The results demonstrate that domain growth decreases the chance of cell clustering occurring by cell-cell interactions. For high values of adhesion and with small growth rates, clusters can still appear over short time-sales. However, as the growth rate increases, the probability of clusters appearing decreases and all initial cell clusters are eventually destroyed. \\

%Yates 2014dcm

\begin{figure}
\begin{center} 
\vspace{0.5cm}
\subfigure[][]{\includegraphics[width=0.45 \columnwidth]{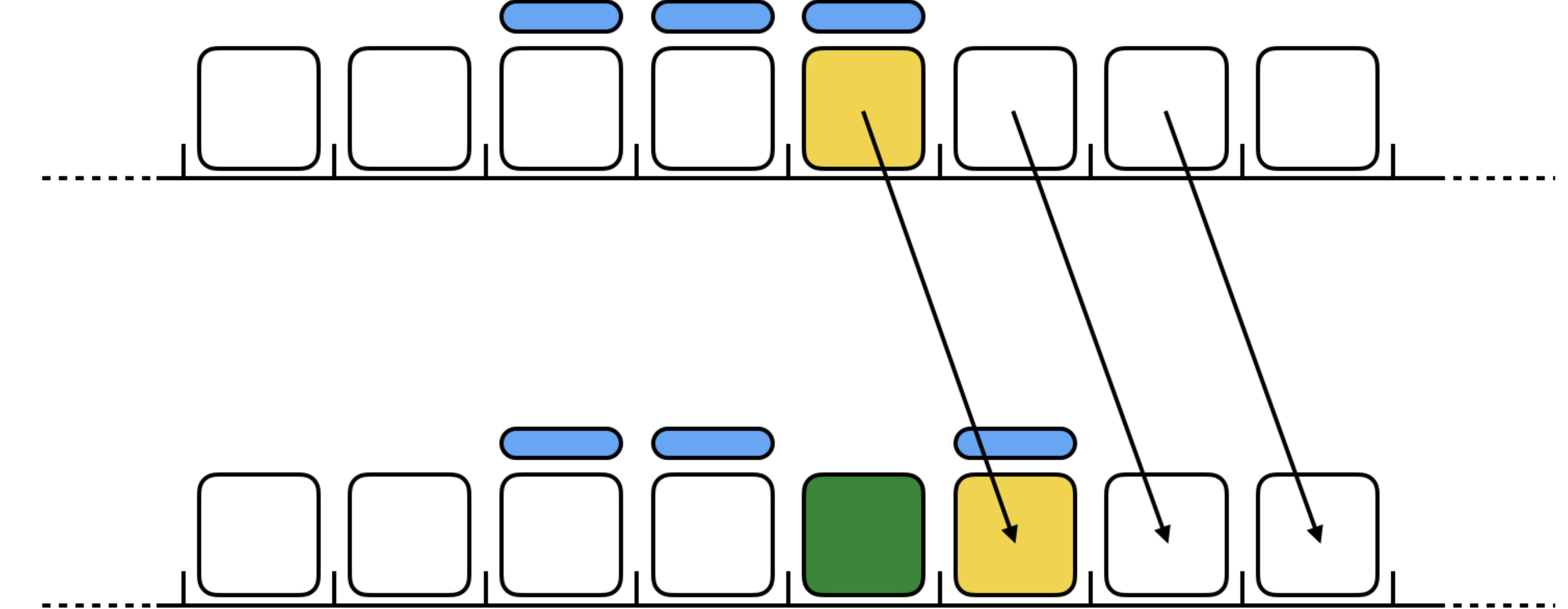}}
\subfigure[][]{\includegraphics[width=0.45 \columnwidth]{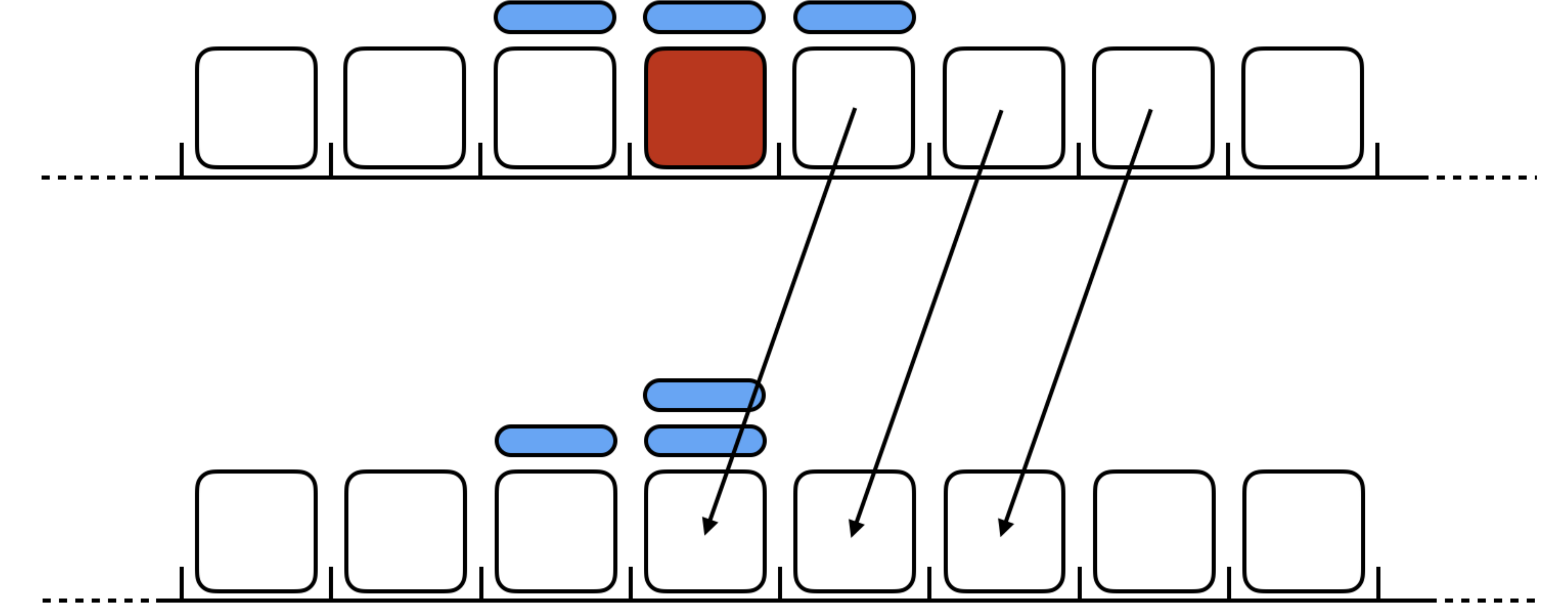}}
\caption{Schematics of the ABM of \citet{yates2014dcm}. White squares represent empty lattice spaces and blue markers represent the tracer agents. Panel (a) shows an example of site division in which the dividing site (yellow) is also marked by a tracer agent. The tracer agent moves with the dividing agent to the right and a new site (green) is added in the vacant space. Panel (b) shows an example of site death. A marked site (red) is removed from the lattice, its tracer agent remains in its position and causes an overlap of two tracer agents.}
\label{fig:schematics_growing}
\end{center}
\end{figure}

\citet{hywood2013mbt} suggest a modified version of the model of \citet{baker2010fmm} to represent tissue growth. The authors initialise a one-dimensional lattice with a set of contiguous non-overlapping agents (\textit{tracers} agents). These agents are inactive (\text{i.e} they do not perform any  jumping movement between sites) and their role is only to mark the position of the site in which they are located as the underlying lattice grows. The authors focus on the case of exponential growth with constant rate, $b$, which is implemented similarly to \citet{baker2010fmm}. If a marked site splits into two daughter sites, the corresponding tracer agent moves to the right daughter site (see Figure \ref{fig:schematics_growing} for an illustration). The usage of tracer agents follows from the previous work of \cite{binder2009epg} in which, contrastingly, domain growth is implemented in a deterministic manner. 

By coupling the position of the tracers to a system of non-interacting random walkers and writing down the corresponding occupancy master equation for the tracers, \citet{hywood2013mbt} derive a formula for the infinitesimal mean and variance of the underlying stochastic process, $\mu(x,t)$ and $\sigma^2(x,t)$, respectively. These expressions can then be used as coefficients for an advection-diffusion PDE (Fokker-Planck equation (FPE)) describing the spatio-temporal evolution of the average occupancy of the tracer agents, $\bar{C}(x,t)$: 

\begin{equation}
	\label{eq:FPE}
\D{\Cb{x}{t}}{t}=\DD{}{x}\LRs{\frac{\sigma^2(x,t)}{2} \Cb{x}{t}}-\D{}{x}\LRs{\mu(x,t) \Cb{x}{t}}	 \; .
\end{equation}

\citet{yates2014dcm} extends the work of \citet{hywood2013mbt} to more general scenarios in which the rate at which sites divide is time-dependent, $b(t)$. This extension allows the incorporation of a variety of  biologically realistic mechanisms of domain growth. By using similar steps, \citet{yates2014dcm} obtains a series of PDEs for the average occupancy of the tracer agents of the same form as equation \eqref{eq:FPE}. In particular, the author provides the expressions of $\mu(x,t)$ and $\sigma^2(x,t)$ for the case of linear growth, $\frac{dL}{dt}=r$, generalised logistic growth, $\frac{dL}{dt}=r L \LRs{1- \LR{\frac{L}{R}}^\nu} $, and Gompertzian growth, $\frac{dL}{dt}=r L \LR{\,\text{ln} \LR{\frac{L}{R}} }$.

Finally, \citet{yates2014dcm} studies the implications of implementing site death in the model which can be done by removing sites with a given rate, $d(t)$. When this occurs, all the site to the right of the removed site are moved to the left in order to fill the vacant space created by the site's removal. If a tracer agent occupied the removed site, it remains in its position. Notice that this new mechanism can lead to multiple tracer agents occupying the same site. See Figure \ref{fig:schematics_growing} for an illustration. The comparison between the ABM simulations and the corresponding PDE preserves a good agreement when site death is introduced, even for cases of net domain shrinkage, \textit{i.e} $b(t)<d(t)$.

The results of \citet{yates2014dcm} highlight the danger of neglecting death events in the model, even when the net growth rate is positive, $b(t)-d(t)>0$. In other words, although the mean growth rate can be estimated correctly by using purely growing dynamics (with growth rate given by $\tilde{b}(t)=b(t)-d(t)$), the second and higher moments of the process will be incorrect. For example, the higher value of variance leads to faster diffusivity of the agents when site death is explicitly incorporated in the model.

%--------------------------------------------------------------------
% Persistence of motion
\subsection{Persistence of motion}
\label{sec:persistence_of_motion}

A fundamental assumption of modelling cell movement with a position-jump process (such as those implemented in Section \ref{sec:methods}), is that the cells' positions are subject to a series of Markovian jumps in space and, therefore, the direction of motion between sequential jumps is totally uncorrelated. In reality, experimental observations suggest that many types of cell show a tendency to preserve their direction of motion for some time before reorienting \citep{berg1972cec, hall1977amc, gail1970lmf, wright2008def}, even when the movement remains unbiased in the long term. This behaviour is known as \textit{persistence of motion} or \textit{of direction} \citep{patlak1953rwp}. The standard framework to model persistence of motion in an ABM based is to employ a \textit{velocity-jump} process \citep{othmer1988mdb,othmer2000dlt, othmer2002dlt2,codling2008rwm,campos2010prm} in which the variable which is performing markovian jumps is the velocity rather than the position of the agents \citep{othmer1988mdb}.

A simple example of an velocity-jump process ABM in one dimension was analysed by \citet{goldstein1951ddm} and later by \citet{kac1974smr} and \citet{othmer1988mdb}. The model can be formulated as follows. Consider an ABM in which agents have an assigned direction of motion, either right or left. As time evolves they can either move, with rate $\alpha$, or change their direction, with rate $\lambda$. The two events are assumed to occur independently. When a movement event takes place, the moving agent performs a jump of distance, $v$, in its assigned direction. We  can interpret $\pm v \alpha$ as the two possible values of the velocity of an agent, depending whether it is moving in the right- or left- direction, respectively. Notice that this formulation is valid for both on- and off-lattice models. However, in an on-lattice framework, the distance, $v$, has to be an integer multiple of the lattice step $\Delta$. 

Let $\bar{R}(x,t)$ and $\bar{L}(x,t)$ denote the average occupancy of agents in position $x$ at time $t$ with associated direction to the right and left, respectively. For a system of non-interacting agents, the corresponding occupancy master equations can be written as
\begin{subequations}
\label{eq:velocity_MA}
	\begin{align}[left = \empheqlbrace\,]
		\bar{R}(x,t+\delta t)&=\bar{R}(x,t)+\alpha \delta t \LRs{\bar{R}(x-v,t) -\bar{R}(x,t)} +\lambda \delta t\LR{ \bar{L} -\bar{R}}\, ,\\[6pt]
		\bar{L}(x,t+\delta t)&=\bar{L}(x,t)+\alpha \delta t\LRs{\bar{L}(x+v,t) -\bar{L}(x,t)} + \lambda \delta t\LR{ \bar{R} -\bar{L}}\, .
	\end{align}
\end{subequations}
By Taylor expanding the two terms $\bar{R}(x-v,t)$ and $\bar{L}(x+v,t)$ about $x$ to first order and taking the limit $\delta t, v \rightarrow 0$ with $v \alpha $ kept constant, one obtains the system of advective PDEs given by
\vspace{0.2cm}
\begin{subequations}
\label{eq:velocity_PDE}
	\begin{align}[left = \empheqlbrace\,]		\D{\bar{R}}{t}&=-V \D{\bar{R}}{x}+\lambda \LR{\bar{L} -\bar{R}}\, , \label{eq:velocity_PDEa}
\\[7pt]
		\D{\bar{L}}{t}&=\, \, V \D{\bar{L}}{x}+\lambda \LR{\bar{R} -\bar{L}}\, , \label{eq:velocity_PDEb}
	\end{align} 
\end{subequations}
where $V=\lim_{v\rightarrow 0}v\alpha$.  By adding equations \eqref{eq:velocity_PDE} and differentiating with respect to $t$ we can write
\begin{equation}
		\DD{(\bar{R}+\bar{L})}{t}=V \DDdiff{(\bar{L}-\bar{R})}{t}{x}\, .\label{eq:velocity_sum_diff_PDEa}
	\end{equation} 
Similarly, by subtracting equation \eqref{eq:velocity_PDEa} from equation \eqref{eq:velocity_PDEb} and differentiating with respect to $x$, we obtain
	\begin{equation}
		\DDdiff{(\bar{L}-\bar{R})}{x}{t}=V \DD{(\bar{R}+\bar{L})}{x}-2 \lambda \D{(\bar{L}-\bar{R})}{x}\, .\label{eq:velocity_sum_diff_PDEb}
	\end{equation} 
Finally, substituting \eqref{eq:velocity_sum_diff_PDEa} into equation \eqref{eq:velocity_sum_diff_PDEb} and recalling equations \eqref{eq:velocity_PDE} we can write 
\begin{equation}
	\label{eq:telegraph}
	\DD{\bar{C}}{t}+2 \lambda \D{\bar{C}}{t}= V^2 \DD{\bar{C}}{x}\, ,
\end{equation}
where $\bar{C}=\bar{R}+\bar{L}$ represents the total average occupancy. Equation \eqref{eq:telegraph} is also known as a \textit{telegraph equation} since it was originally derived to describe the propagation of signal waves travelling trough a telegraph transmission wire \citep{goldstein1951ddm}. \citet{othmer1988mdb} was the first to obtain such equations from a system of non-interacting agents. \citet{othmer2000dlt} demonstrated that it is possible to recover the canonical diffusion equation as the parabolic limit of the telegraph equation by taking the limit as $V$ and $\lambda$ to infinity simultaneously, such that $V^2/\lambda$ is constant. In other words, the two ABMs with and without persistence are governed by continuum models of the same form in the long time limit. This can be understood by noticing that the short-term directional bias due to the presence of persistence becomes less evident at the spatial scale which are much larger than the average distance moved by an agent before reorienting and temporal scales which are much larger than the average reorientation time \citep{othmer2000dlt,othmer2002dlt2,codling2008rwm}.

Notice that the derivation of the equation \eqref{eq:telegraph} from the equations \eqref{eq:velocity_PDEa} and \eqref{eq:velocity_PDEb} is possible only for the case of non-interacting agents in one dimension. When the model is defined on a two-dimensional lattice, for example, we can still write down a system of four advective PDEs for the average occupancies of agents moving in the four direction of the lattice \citep{gavagnin2018mpm}. However, a closed formula for the total average occupancy is no longer obtainable.

A similar problem occurs when agent-agent interactions are added to the model, in which case deriving an analogue of the telegraph equation and, consequently, its diffusive limit, is not possible. \citet{treloar2011vjm} have studied the implication of incorporating crowding effects in a model of persistence. At the stochastic-level, they consider a modification of the ABM of \citet{othmer1988mdb} on a one dimensional lattice with three different volume exclusion properties each of increasing complexity. For each case, the authors derive the a system of advective PDEs describing the evolution of the average occupancy of the two subpopulations of agents depending on their direction of movement. 

The results of \citet{treloar2011vjm} show that the details of the crowding interactions lead to differences in the corresponding continuum models. This highlights a substantial difference from the analogous persistence-free position-jump process (see Section \ref{sec:MF}), for which crowding effects do not change the corresponding macroscopic representation. \\

The continuum models derived by \citet{treloar2011vjm} do not admit a diffusive interpretation. A diffusive interpretation is desirable, since it allows a direct comparison with other diffusive models and establishes a direct connection with commonly used statistical tools of movement analysis. Recently, \citet{gavagnin2018mpm} have studied a generalisation of the ABM of \citet{treloar2011vjm} in two dimensions for which is possible to obtain a faithful diffusive description at the population-level. The authors suggest the introduction of an additional parameter, $\varphi\in [0, 1]$, in the ABMs of \citet{treloar2011vjm} which modulates the intensity of the short-term directional bias and hence the intensity of the persistence. 

By rescaling the parameter with the size of the lattice step, $\varphi\sim \mathcal{O}(\Delta)$, \citeauthor{gavagnin2018mpm} obtain a general set of advective-diffusive PDEs. A comparison of the column-averaged density shows a good agreement between the stochastic and the deterministic models for a wide range of parameters. However, when the agent jump length becomes large, the density profile of the ABM presents regular peaks in density which are not captured by the corresponding continuum models. 

\begin{figure}
\begin{center}
\subfigure[][]{\includegraphics[width=0.4 \columnwidth]{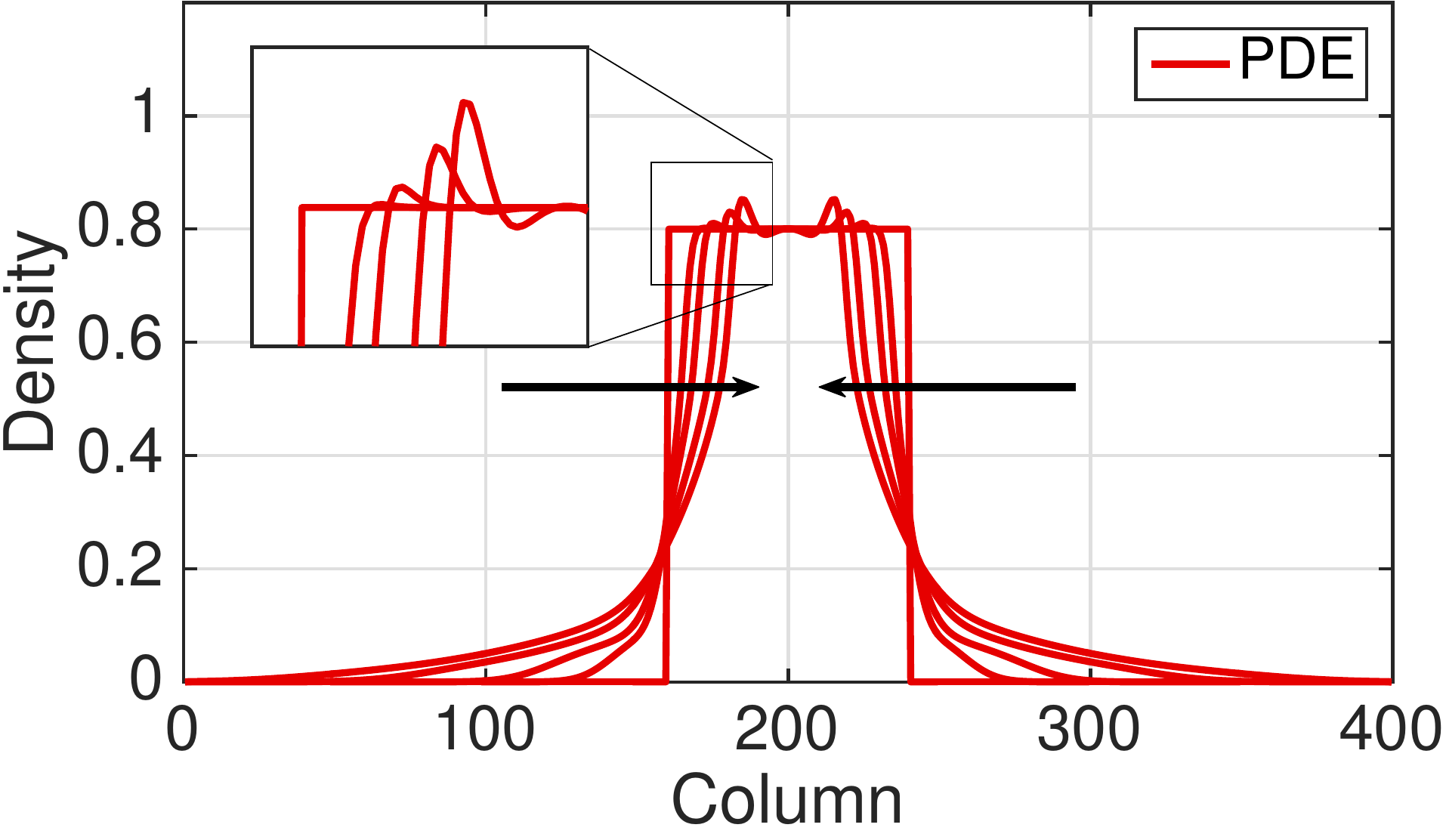} }
\subfigure[][]{\includegraphics[width=0.4 \columnwidth]{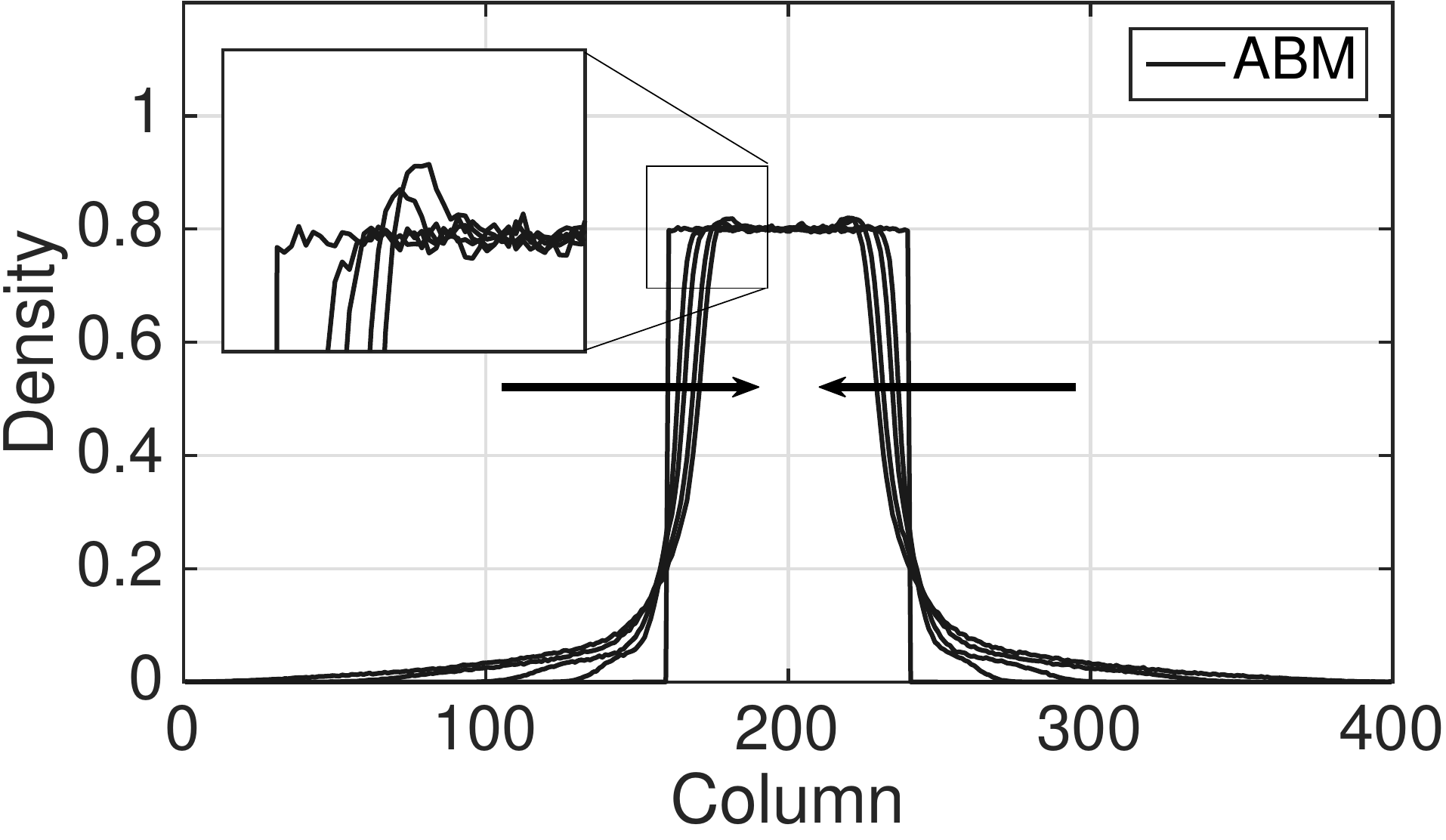} }
\end{center}
\caption{Spontaneous aggregation induced by persistence and volume exclusion. The top panels show (a) the numerical solution of the PDE for the column-averaged total density and (b) column-averaged density of the ABM, averaged over $M=100$ repeats. The profiles are shown at time $T=0,50,100,200,300$ with the direction of the black arrows indicating increasing time. Reproduced from \citet{gavagnin2018mpm} with permission of Physical Review E.}
\label{fig:gavagnin2018} 
\end{figure}

For strong values of persistence, \citet{gavagnin2018mpm} find evidence of a spontaneous form of agent aggregation as result of the interplay of persistence and crowding effects in highly populated regions (see Figure \ref{fig:gavagnin2018}). Notice that such a form of agent aggregation is not possible in ABMs which do not incorporate persistence such, since the overall behaviour is governed by simple diffusion \citep{simpson2009mss}. Their deterministic model is capable of providing a qualitatively matching description of such aggregation phenomenon, even though the presence of short-range correlations affects the quality of the agreement (see Figure \ref{fig:gavagnin2018}).

Finally, the two dimensional framework of \citet{gavagnin2018mpm} highlights a positive anisotropy of the model in the axial directions of the lattice. This  anisotropy appears as an intrinsic feature of the persistent model combined with the lattice environment. Off-lattice models should have the advantage that they are not afflicted by such anisotropy. However, the derivation of a corresponding macroscopic description becomes more complicated and sometimes intractable.\\

\section{Conclusion}
\label{sec:conclusion}
Cell migration is essential in a wide range of biological contexts, including many developmental and homeostatic mechanisms in the human body \citep{gilbert2003med, keller2005cmd, maini2004twm, deng2006rma}. As a result, a great deal of attention has been given to the study of such a complex phenomenon in the last few decades. However, when an experimental approach becomes difficult or even impossible, mathematical modelling provides an avenue through which investigations can continue. The power of modelling relies on its ability to test and verify experimental hypotheses, as well as making predictions which can indicate appropriate experimental directions.

In general there is a dichotomy between the types of models used for the mathematical representation of cell migration. On one side of the divide are continuum models capable of representing the population-level characteristics of a group of cells and often amenable to mathematical analysis. When required, such models are fast to simulate numerically and can often be linked explicitly to the model parameters. However, these models are less useful for capturing individual-level detail and are consequently more difficult to link directly to experimental data; abilities which are inherent to the discrete models, on the other side of the divide. Discrete models are often intuitive to formulate and are able to incorporate stochasticity in a natural manner. However, it is often difficult to link the results of such models directly to the model parameters in order to gain a population-level overview and their simulation can be computationally expensive. Building equivalence frameworks allows the exploitation of  their complementary strengths and the circumvention their complementary weaknesses. 
 
In the first part of this review we have outlined a class of methodologies designed to bridge the divide between the individual-level and population-level regimes. As fundamental examples, we have considered a series of simple ABMs representing cell migrating on a one-dimensional domain. We have recalled explicitly how the canonical diffusion equation \eqref{eq:diffusion} can be derived from these models when agents are not interacting with each other. The same equation is obtained for both on- and off-lattice systems. By repeating the computation for ABMs which incorporate crowding effects, we have underlined a substantial difference between on-lattice ABMs and off-lattice ABMs. Although the on-lattice models for diffusion give the same PDE irrespective of whether volume exclusion is implemented or not, the same can not be said in almost any other case, for example the off-lattice ABM that we considered.

In the second part of this chapter, we have provided a brief review of some of the important features which can influence cell behaviour at the population level, either by suppressing or enhancing the collective movement. For each case, we have summarised the relevant studies devoted to incorporating such specific features into mathematical models both in the stochastic and deterministic regimes. \\

In conclusion, although significant progress has been made towards the understanding of cell migratory behaviours through mathematical modelling,  many questions concerning the direct relationship between single cell behaviours and collective invasion remain unclear. For example, there is still little understanding of the role of heterogeneity in cell behaviour, such as leader-follower mechanisms \citep{mclennan2012mmc, mclennan2015vsi,schumacher2017shc} or heterogeneitic cell proliferation \citep{smadbeck2016cmd}.  In this context, developing mathematical models capable of incorporating both these aspects represents an important avenue for future research.

%--------------------------------------------------------------------
% Acknowledgments
%\ack

%--------------------------------------------------------------------
% Bibliography
\newpage
\bibliography{database}{}
\bibliographystyle{unsrtnat}

\end{document}